\documentclass[times]{qjrms4}

\newcommand\BibTeX{{\rmfamily B\kern-.05em \textsc{i\kern-.025em b}\kern-.08em
T\kern-.1667em\lower.7ex\hbox{E}\kern-.125emX}}

\usepackage{moreverb}
\usepackage{natbib}
\usepackage{balance}
\usepackage[normalem]{ulem}
\usepackage{color}
\usepackage{subfig}
\def\volumeyear{2018}
\def\volumenumber{144}
\begin{document}

\markboth{Y. Wang et al.}{Atmospheric circulation regimes}

\runningheads{Y. Wang et al.}{Atmospheric circulation regimes: I.}

\title{Comparative terrestrial atmospheric circulation regimes in simplified global circulation models: I. from cyclostrophic super-rotation to geostrophic turbulence}

\author{Y. Wang\affilnum{a}, \corrauth P.~L.~Read\affilnum{a},   F. Tabataba-Vakili\affilnum{a,b}, R. M. B. Young\affilnum{a,c}}

\address{\affilnum{a}Atmospheric, Oceanic \& Planetary Physics, Department of Physics, University of Oxford, Clarendon Laboratory, Parks Road, Oxford, OX1~3PU, UK \\
\affilnum{b} Jet Propulsion Laboratory, Pasadena, USA \\
\affilnum{c} Now at Laboratoire de M\'et\'eorologie Dynamique/Institut Pierre-Simon Laplace (LMD/IPSL), Sorbonne Universit\'es, UPMC Univ Paris 06, PSL Research University, \'Ecole Normale Sup\'erieure, Universit\'e Paris-Saclay, \'Ecole Polytechnique, Centre National de la Recherche Scientifique, 75005 Paris, France
}

\corraddr{Atmospheric, Oceanic \& Planetary Physics, Clarendon Laboratory, Parks Road, Oxford, OX1~3PU, UK.}



\begin{abstract}
The regimes of possible global atmospheric circulation patterns in an Earth-like atmosphere are explored using a simplified GCM based on the University of Hamburg's Portable University Model for the Atmosphere (PUMA) --- with simplified (linear) boundary layer friction, a
Newtonian cooling scheme and dry convective 
adjustment (designated here as PUMA-S). A series of controlled experiments are conducted by varying planetary rotation rate and imposed
equator-to-pole temperature difference. These defining parameters are further combined with each other into dimensionless forms to 
establish a parameter space, in which the occurrences of different circulation regimes are mapped and classified.
Clear, coherent trends are found when varying planetary rotation rate (thermal Rossby number) and frictional and thermal relaxation timescales. The sequence of circulation regimes as a function of parameters, such as the planetary rotation rate, strongly resembles that obtained in laboratory experiments on rotating, stratified flows, especially if a topographic $\beta$-effect is included in those experiments to emulate the planetary vorticity gradients in an atmosphere induced by the spherical curvature of the planet. 
A regular baroclinic wave regime is also obtained at intermediate values of thermal Rossby number and its characteristics and dominant zonal wavenumber depend strongly on the strength of radiative and frictional damping. These regular waves exhibit some strong similarities to baroclinic storms observed on Mars under some conditions. Multiple jets are found at the highest rotation rates, when the Rossby deformation radius and other eddy-related length scales are much smaller than the radius of the planet. These exhibit some similarity to the multiple zonal jets observed on gas giant planets. Jets form on a scale comparable to the most energetic eddies and the Rhines scale poleward of the supercritical latitude. The balance of heat transport varies strongly with $\Omega^\ast$ between eddies and zonally symmetric flows, becoming weak with fast rotation. 
\end{abstract}


\maketitle

\section{Introduction} \label{sec:introduction}
An understanding of the dominant factors and mechanisms controlling the general circulation of a planetary atmosphere is a 
prerequisite to our understanding of the climate variability of the Earth, both in the past and in the 
future. This goal has stimulated an immense body of research during the past century, focussing particularly on the details of the Earth's atmosphere, surface and planetary environment and the parameters which define them. From one viewpoint this is entirely understandable, since many aspects of the changing climate of the present day are manifested in relatively subtle ways through small but systematic changes in the statistical moments of key climate variables (temperature, wind, precipitation etc.) in amongst the highly chaotic fluctuations associated with normal weather and seasonal changes. Too much focus on such details, however, can sometimes obscure much more fundamental principles that may be made more readily apparent by exploring a much broader parameter space, one that might embrace not only the Earth but also the atmospheres of other planetary bodies. 

Within our solar system, substantial atmospheres are found on a series of planetary bodies: Venus, Mars, the gas/ice giants (Jupiter, Saturn, Uranus and Neptune), and Titan---the largest moon of Saturn (see 
e.g. \cite{sanchez-lavega2011,mackwell2013}). 
Even beyond the Sun, more than 3600 extrasolar planets (a.k.a. exoplanets) have now been detected 
using various astronomical observation techniques, and many of them are believed to host substantial atmospheres (e.g. see reviews by \cite{Seager2010}, \cite{showman2010} and \cite{showman2014}). Many differences in their climate and styles of circulation amongst 
these planetary atmospheres are clearly seen and demand to be understood. 


From a physicist's viewpoint, all of these planetary atmospheres can be abstracted as rotating, stratified
fluids moving under mechanical/electromagnetic and other forcings from the host star and the underlying planet. 
It is thus natural to ask the question as to what characteristic factors most fundamentally determine the 
various circulation patterns that we observe in those exotic worlds, and also our own.

The general methodology commonly employed for this sort of problem is to establish a model
atmosphere founded upon various reasonable idealisations, and then to gain insights by 
studying the behaviour of this model atmosphere under different conditions. In most cases, simple models incorporating only the most significant dynamical 
constraints and key physical processes are used rather than the highly complex and \lq\lq realistic\rq\rq\
models which seek to simulate the climate and atmospheric circulation of a given planet in great quantitative detail. This is, of course, partly due to the convenience of handling a simpler model, but 
more importantly, because it tells us more about fundamental physical processes and causal relationships,
which are usually not explicitly expressed or straightforward to extract in the results of complex models.

Laboratory analogues of the circulation of planetary atmosphere have been explored for many years, and usually take the form of viscous stratified fluids confined
in a rotating annulus (or tank) heated from the boundaries (e.g. see \cite{vonLarcher2014}). Although such physical models do not typically take into account full spherical geometry or radiative processes, they are able to capture many aspects of the essential physics of the atmospheric circulation, treating it as a 
fluid in rotation under the forcing of differential stellar heating. Since the pioneering work of 
\cite{Hide1953} and \cite{Fultz1959}, much knowledge of the atmospheric general circulation as well as other
aspects of atmospheric dynamics has been gained by studying these flow patterns (see \cite{hide1975}, 
\cite{Lorenz1967}). Flow behaviours under various conditions have been thoroughly measured and classified, 
and their dependence on non-dimensional parameters can be shown by mapping the flow patterns within a 
parameter space (the resulting map is known as a \emph{regime diagram}; 
e.g. see \cite{Read1998}, \cite{Read2001}, 
\cite{Wordsworth2008}, 
\cite{vonLarcher2014} and references therein).

Another approach, adopted here, is to make use of simplified numerical models of planetary atmospheres to investigate the 
behaviour of the circulation. 
There has been a series of work taking a parametric approach to studying atmospheric circulation since the 1970s \citep{Hunt1979,Williams1982,Geisler,Williams1988a,Williams1988b,DelGenio1987,Jenkins1996}, with motivations ranging from paleoclimate simulations to GCM 
parameter optimisation. This has continued into the 21st century with a view to improving our understanding of aspects of the circulation and climate of planets not too different from Earth \citep[e.g.][]{Navarra2002,barry2002,schneider2006,Mitchell2010}, with the ultimate approach to parameter ensembles adopted by the climateprediction.net and weather@home groups \citep[e.g.][]{stainforth2005,massey2015}. The imperative to explore widely different regions of parameter space in planetary climate and circulation has also prompted the extrasolar planet community to make use of three-dimensional atmospheric circulation models \citep[e.g.][]{joshi1997,joshi2003,merlis2010,heng_vogt2011,selsis2011,wordsworth2011,yang2013,yang2014,hu2014,forget2014,kaspi2015}. Depending upon the purpose of the study, the parameters to vary within a simple GCM have included planetary rotation rate, planetary radius, gravity, obliquity etc., although many exoplanet studies have tended to focus on planets whose rotation is tidally-locked to their orbit so they permanently present the same face to their parent star.

Perhaps the closest previous work to what is presented here is found in the studies of \cite{Geisler}, \cite{Mitchell2010} and \cite{kaspi2015}. \cite{Geisler} used a simplified GCM with no hydrological cycle, Newtonian relaxation towards a prescribed thermal field in place of a radiation scheme and a parameterization of Ekman friction, but only considered rotation rates equal to or slower than that of the Earth and employed relatively crude spatial resolution (R15). \cite{Mitchell2010} also used a simplified GCM with Newtonian relaxation but at higher spatial resolution (T42), and also focussed on exploring mechanisms for super-rotation in slowly rotating, Earth-like planets. \cite{kaspi2015} explored a wider range of parameter space, including not only changes in planetary rotation but also variations in stellar radiative flux, atmospheric mass (or surface pressure) and planetary density (governing surface gravity) in a model that included a gray radiation scheme and simple hydrological cycle (including moist convection). The spatial resolution and range of rotation rate was similar to what is considered in the present study (T42 - T170). However, only one parameter was varied at  time in their study, and the distribution of absorbers was carefully tailored to emulate that of water vapour in the Earth's atmosphere (with enhanced concentrations at low levels near the equator) so as to produce a specifically Earth-like distribution of zonal mean temperature. Their main focus was also on the zonal mean component of the simulated circulation and associated meridional transports of heat, momentum and moisture. In our study, however, we have chosen to use a more generic temperature distribution for relaxation forcing which is more typical of planets without an intense hydrological cycle, though the thermal distribution is still sufficiently Earth-like for the results to be comparable with the Earth itself.

In the present work, we use the simplified Global Circulation Model---PUMA (the Portable University Model of Atmospheres) developed by the Meteorological Institute, University of Hamburg, to conduct controlled experiments and sensitivity tests on factors including planetary rotation rate, radiative and frictional timescales. Corresponding non--dimensional parameters are suitably defined to 
construct a dimensionless parameter space in which circulation regimes are mapped. This paper is the first in a series of papers reporting studies based on the results of diagnostics of systematically controlled experiments with various versions of the PUMA model, of which the present work focuses on the baseline version of PUMA using linear Newtonian relaxation forcing of the temperature field (hereafter designated PUMA-S). Future work will report on parallel studies using a more physically consistent gray radiation scheme. 

Section \ref{sec:exp-design} describes the experiment setup
and definitions of the dimensionless parameters of the regime diagram. This is 
followed by a brief overview and classification of the major circulation regimes 
observed in our experiments in Section \ref{sec:reg-class}. Sections \ref{sec:zonal}
and \ref{sec:eddies} then discuss the trends observed in the zonal mean and eddy/wave
diagnostics. The 
regular baroclinic waves regime are 
discussed in detail in Section \ref{sec:reg-rg} and the overall results are discussed in Section \ref{sec:discussion}. 

\section{Experiment design and regime diagram}\label{sec:exp-design}
The model used is the Portable University Model of the Atmosphere
(PUMA; e.g. see \citet{fraedrich1998,frisius1998,vonhardenberg2000}). PUMA represents the dynamical core
of a spectral atmospheric general circulation model (AGCM), based on the work of \citet{hoskins1975}. The dry primitive
equations on a sphere are integrated using the spectral transform method \citep{orszag1970}. Linear
terms are evaluated in the spectral domain while nonlinear products are calculated in grid point space. Temperature, divergence,
vorticity, and the logarithm of the surface pressure are the prognostic variables. The vertical is divided into a number of
equally spaced $\sigma$ levels (where $\sigma=p/p_s$, where $p$ and $p_s$ denote
the pressure and the surface pressure, respectively). The integration in time is carried out with a filtered leap-frog semi-implicit
scheme \citep{robert1966}. 

\subsection{Parameterisation of physical processes}
While the representation of atmospheric dynamics is similar to a full state-of-the-art AGCM, the representation of
diabatic and subscale processes in PUMA is very simple and linear. 

\subsubsection{Friction and (hyper-)diffusion}
Dissipation terms in the vorticity and divergence equations are parameterised as a Rayleigh 
friction:
\begin{equation}
P_\zeta=\frac{\zeta}{\tau_F(\sigma)}+H_\zeta 
\label{eq:rayleigh_zeta}
\end{equation}
\begin{equation}
P_D=\frac{D}{\tau_F(\sigma)}+H_D,
\label{eq:rayleigh_d}
\end{equation}
where $\tau_F(\sigma)$ is the characteristic timescale of momentum dissipation, $H_\zeta$ 
and $H_D$ are the hyperdiffusion terms. 
In our experiments, $\tau_F(\sigma)$ is set to about
$1$ Earth day in the lowest layers (where $\sigma > 0.8$, where $0.8$ is taken to correspond roughly to the level 
of the top of planetary boundary layer) and $\rightarrow \infty$ in the free atmosphere 
(defined to be where $\sigma\leq0.8$).

A horizontal hyperdiffusion ($\propto \nabla^8$), applied to temperature, divergence,
and vorticity, represents the effect of subgridscale horizontal mixing and local energy dissipation due to
eddies. This is applied to both vorticity/divergence and to temperature, with a hyper-diffusion timescale $\tau_{H} \sim \delta^8/K  = 0.25$ Earth days (where $\delta$ is the gridscale and $K$ the hyper-diffusion coefficient).

\subsubsection{Diabatic Heating}
Diabatic heating and cooling of the atmosphere is parameterised by a simple linear 
Newtonian relaxation formulation:
\begin{equation}
\frac{J}{c_p}
=\frac{T_R-T}{\tau_R(\sigma)}+H_T,
\label{eq:newtonian}
\end{equation}
where $\tau_R$ is the characteristic timescale for the temperature field to relax towards 
the prescribed restoration temperature field, and $H_T$ is a 
hyperdiffusion term. 
The restoration temperature field is prescribed as a function of latitude and height:
\begin{equation}
T_R(\phi,\sigma)=T_{Rz}(\sigma)+f(\sigma)T_{Rl}(\phi),
\end{equation}
in which
\begin{eqnarray}
T_{Rz}(\sigma) & = & (T_{Rz})_{tp}+\sqrt{\Bigl[\frac{L}{2}(z_{tp}-z(\sigma))\Bigr]^2+S^2} \\ \nonumber
 & & + \frac{L}{2}\Bigl(z_{tp}-z(\sigma)\Bigr),
\end{eqnarray}
where $(T_{Rz})_{tp}=(T_{Rz})_{grd}-Lz_{tp}$ is the restoration temperature at the 
tropopause, $L$  the vertical lapse rate, $z_{tp}$  the global constant height of the 
tropopause, $(T_{Rz})_{grd}$ the restoration temperature at the ground. The constant 
$S$ acts as a smoothing term at the tropopause. In the vertical direction, the 
restoration temperature field essentially consists of a troposphere with constant (moist)
adiabatic lapse rate and an isothermal stratosphere. If $S$ is set to zero, then there 
will be an abrupt discontinuous change of temperature at the transition between 
troposphere and stratosphere. In our experiments, $L=6.5 \text{K km}^{-1}$, 
$z_{tp}=12000\ \text{m}$ and $S=2\ \text{K}$. 


The meridional variation of the restoration temperature field is of the form
\begin{equation}
T_R(\phi)=(\Delta T_R)_{NS}\frac{\sin{\phi}}{2}-(\Delta T_R)_{EP} \Bigl(\sin^2{\phi}-
\frac{1}{3}\Bigr),
\end{equation}
where $(\Delta T_R)_{EP}$ is the prescribed constant restoration temperature difference 
between the equator and the poles, $(\Delta T_R)_{NS}$ is the variable part of the 
meridional temperature gradients which can be made to change with time to simulate an 
annual cycle.

The meridional variation is modulated in altitude by the function $f(\sigma)$ so that 
the variation vanishes at the isothermal tropopause, that is assumed to be:
\begin{equation}
f(\sigma)=
\begin{cases}
	\sin{\Bigl(\frac{\pi}{2}\bigl(\frac{\sigma-\sigma_{tp}}{1-\sigma_{tp}}\bigr)
	\Bigr)} & \text{if $\sigma \geq \sigma_{tp}$}\\
	0 & \text{if $\sigma < \sigma_{tp}$}
\end{cases}
\end{equation}

\begin{figure}
 \centering
 \includegraphics[bb=0 0 456 238,width=0.9\columnwidth,clip=true]{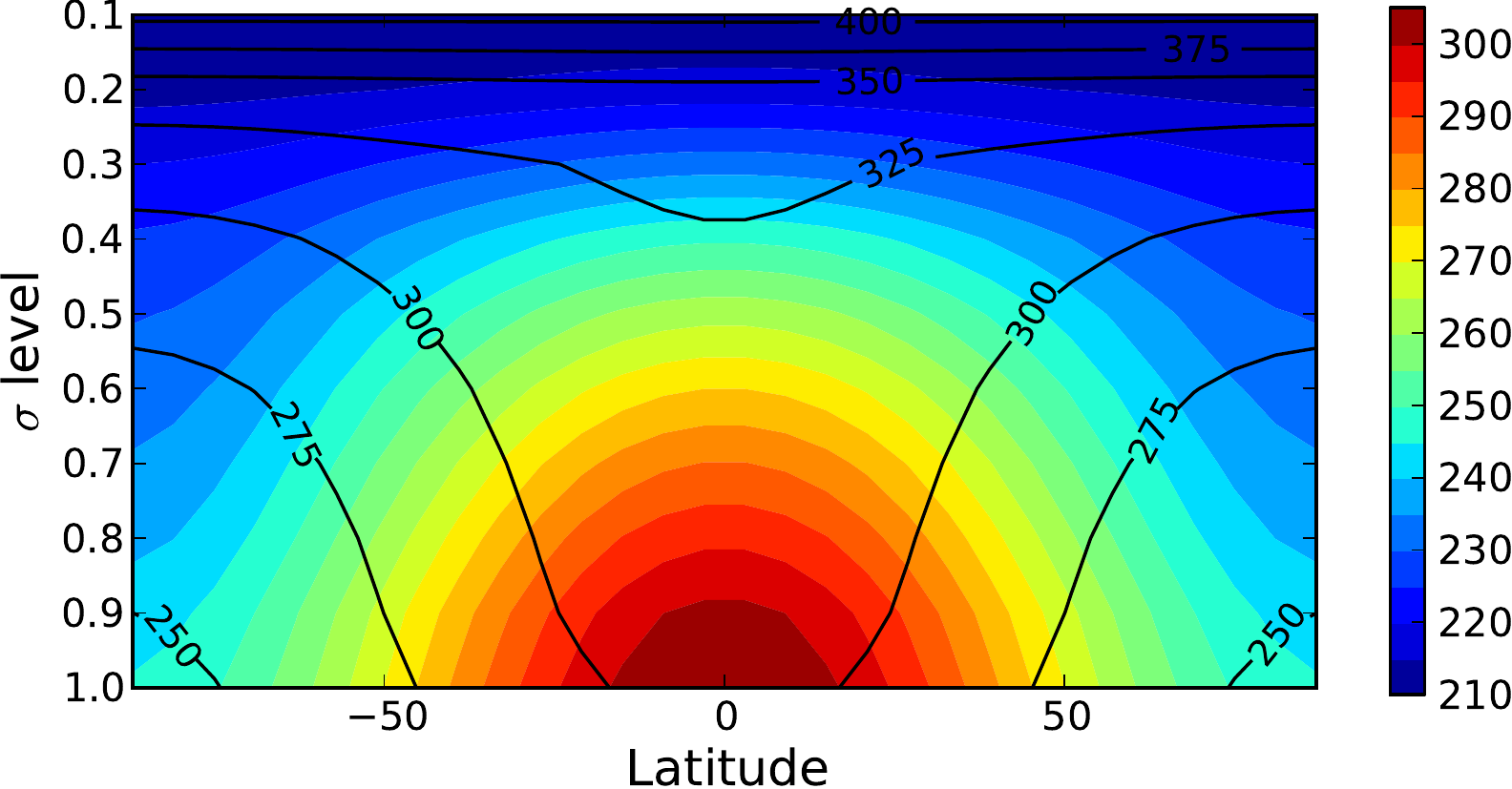}
 \caption{Restoration temperature (colour) and potential temperature (contour) field 
 with equator-to-pole temperature difference of $60\ \text{K}$.}\label{fig:4}
\end{figure}

The complete restoration temperature field with equator-to-pole temperature difference
of $60\ \text{K}$ is shown in Fig~\ref{fig:4}.


\subsection{Simulation experiment design and parameters}
PUMA-S is used here to study the behaviour of a prototype planetary 
atmosphere (albeit for which the atmospheric composition, gravitational acceleration and 
planetary radius are still set to terrestrial values). The model is set up with no 
topographical effects (e.g. normal drag, gravity waves excited by terrain, thermal 
contrast between continents and oceans...). Seasonal and diurnal cycles are suppressed and a fixed annual mean thermal forcing is applied through Newtonian relaxation.

Controlled simulation experiments were conducted by varying rotation speed $\Omega$ and the 
equator-to-pole temperature difference $\Delta T_h$. We tested three values of 
$\Delta T_h=$60 K, 10 K, and 5 K. For each value of
$\Delta T_h$, eight values of $\Omega$ were investigated ($\Omega^{\ast}=\Omega/\Omega_E=8, 4, 2, 1,
1/2, 1/4, 1/8, 1/16$, where $\Omega_E$ is the rotation rate of the Earth). This led to the series of simulations shown as a function of position in parameter space in Fig. \ref{fig:regdiag}. 
Horizontal resolution was set to T42 for slowly rotating simulations
($\Omega^{\ast}\leq 1$), T127 for faster rotating simulations with 
$\Omega^{\ast}=2, 4$, and T170 reserved for simulations with $\Omega^{\ast}=8$. For all
the simulations, there were 10 vertical levels ($\sigma=0.1, 0.2, 0.3,...,1.0$).
Frictional timescale, $\tau_f$, was set to $0.6$ Earth days at $\sigma=1.0$ and $1.6$
Earth days at $\sigma=0.9$, with zero value at all other levels above, producing an
Ekman-like planetary boundary layer. The radiative timescale, $\tau_R$, was set 
to 30 Earth days in the free atmosphere, decreasing to $2.5$ Earth days at 
$\sigma=1.0$. The initial condition in each case was an isothermal atmosphere at 
rest. Each simulation experiment was run for 10 model years (with 360 days per model year) to 
ensure that a statistically steady state was reached. All of the following
diagnostics were based on data from the tenth model year unless otherwise stated.

\subsection{Mapping circulation regimes in parameter space}
The set of simulations outlined above were designed to explore a wide range of planetary rotation rate and frictional and thermal relaxation timescales. The relevant circulation regime could then be identified, as discussed in Section \ref{sec:reg-class} below, and each experiment was then located with reference to a suitable (dimensionless) parameter space. The results from this exercise are shown as a map of identified circulation regimes in Fig. \ref{fig:regdiag}. Following precedents by \cite{Geisler}, \cite{Read2011} and others, by analogy with studies of flow regimes in laboratory experiments, this was constructed with reference to the thermal Rossby number $\mathcal{R}o_T$ (see \cite{Mitchell2010},  \cite{Read2011}) 
\begin{equation}
 \mathcal{R}o_T = \frac{R\Delta \theta_h}{\Omega^2 a^2} \label{eq:Ro}\\
\end{equation}
where $\Delta \theta_h$ is the equator-to-pole potential temperature difference,
$a$ the planetary radius and $R$ the specific gas constant. In the laboratory, frictional effects are typically represented with reference to a Taylor number, which can be represented (to within a factor O(1)) as
\begin{equation}
 \mathcal{T}a = 4(\Omega \tau_{E})^4, 
 \end{equation}
where $\tau_E$ is the Ekman spindown timescale $\tau_E \sim H/\sqrt{\nu \Omega}$, $H$ is the vertical scale of the domain and $\nu$ the kinematic viscosity. In the present models, which use a simpler linear Rayleigh friction in the boundary layer, we can define a similar spindown timescale $\tau_{ft}$ \citep{valdes1988} by $\tau_{ft} \simeq \overline{\tau_f} H/d$, where $H$ is taken to be the pressure scale height, $d$ the height of the boundary layer and $\overline{\tau_f}$ the mean value of $\tau_f$ in the boundary layer. Accordingly we can define a frictional Taylor number for our model as
\begin{equation}
 \mathcal{T}a_f = 4(\Omega \tau_{ft})^4.  \label{eq:Taf}
 \end{equation}
Our experiments are configured with values of $H$ and $d$ such that $H/d\sim5$, 
leading to a frictional Taylor number $ \mathcal{T}a_f \sim 10^4 \Omega^4\tau_f^4$, 
where $\tau_f$ is approximately 1 Earth day given the profile of $\tau_f$ as 
stated above. An equivalent thermal or radiative Taylor number can also be defined as
\begin{equation}
 \mathcal{T}a_R = 4(\Omega \overline{\tau_R})^4, \label{eq:Tarad}
 \end{equation}
where $\overline{\tau_R}$ is the mean value of $\tau_R$ throughout the atmosphere.

\section{Classification of circulation regimes}\label{sec:reg-class}

\begin{figure*}
 \centering
  \resizebox{0.8\textwidth}{!}{
  \includegraphics{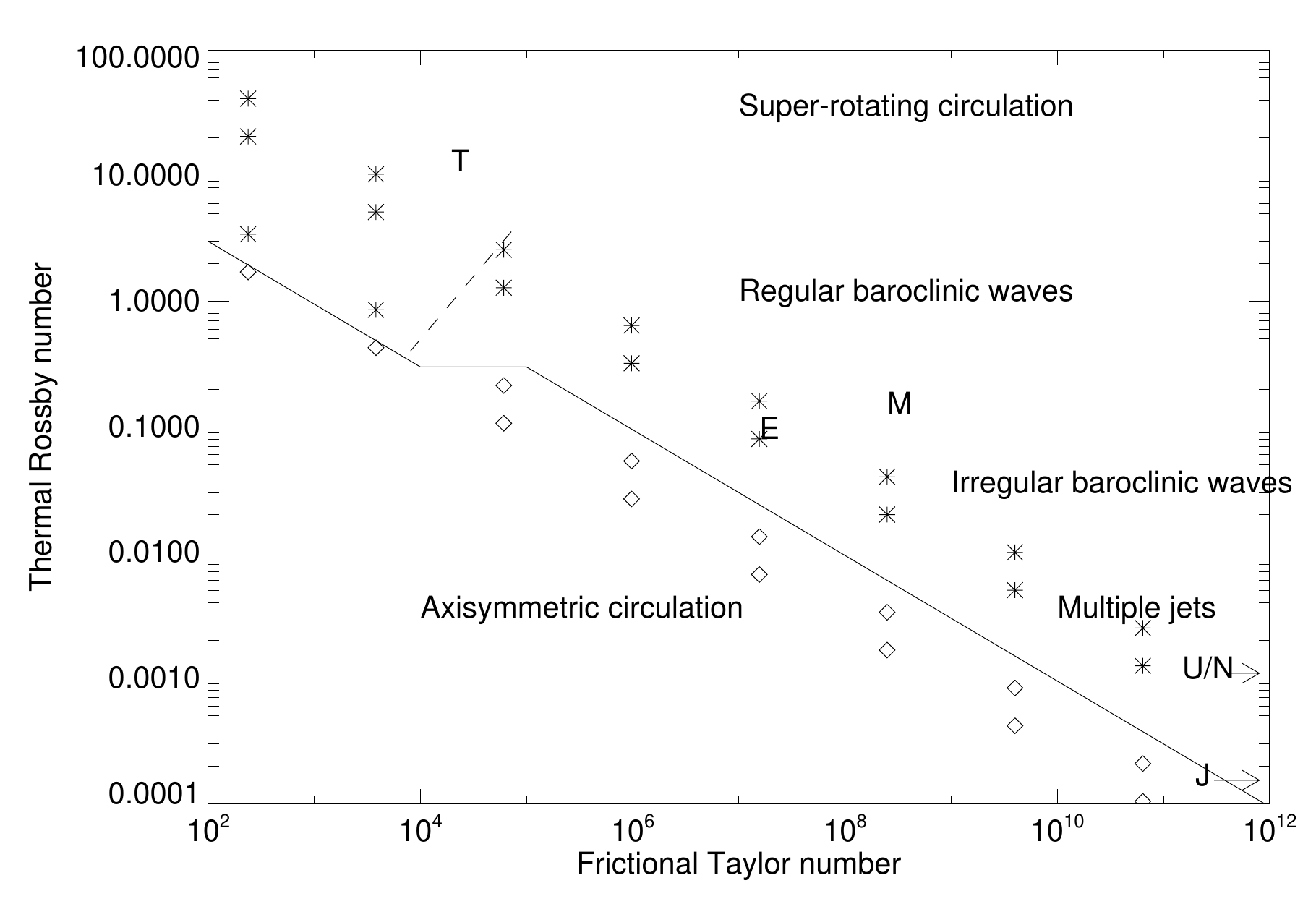}
 }
 \caption{Regime diagram showing the various circulation regimes with respect to characteristic dimensionless parameters ($ \mathcal{R}o_T$ and $\mathcal{T}a_f$). Stars refer to experiments in which wavy flows are discovered, whereas 
open diamonds indicate experiments in which axisymmetric flows were found. The approximate location in parameter space of some Solar System planets (Earth, Mars, Titan, Jupiter, Saturn, Uranus and Neptune) are labeled by their initial letters. The solid line delineates the boundary between axisymmetric circulations and circulations with wavy/turbulent flows. The dashed lines indicate the boundaries between different circulation regimes within the wavy/turbulent region.}
 \label{fig:regdiag}
\end{figure*}

As stated above, the regime diagram shown in Fig. \ref{fig:regdiag} was obtained by mapping 
the occurrences of the observed circulation regimes within a parameter space 
spanned by thermal Rossby number $\mathcal{R}o_{T}$ and frictional Taylor number
$\mathcal{T}a_{f}$. Five different types of global circulation can be identified
in this regime diagram.

\subsection{Steady, axisymmetric flow}
In our experiments, axisymmetric flows 
are found in the lower left region of the regime diagram, with sufficiently 
small equator-to-pole temperature contrast or strong enough diffusion or frictional 
damping. Axisymmetric circulations are characterised by smooth, steady, laminar flows 
encircling the axis of planetary rotation, with no wave/eddy disturbances. A similar regime was identified by \cite{Geisler} in their simplified GCM and corresponds closely to the ``Lower symmetric'' regime in the laboratory \citep[e.g.][]{hide1975}. 

\begin{figure*}
 \centering
\subfloat{\includegraphics[bb=75 85 700 510,width=0.45\textwidth,clip=true]{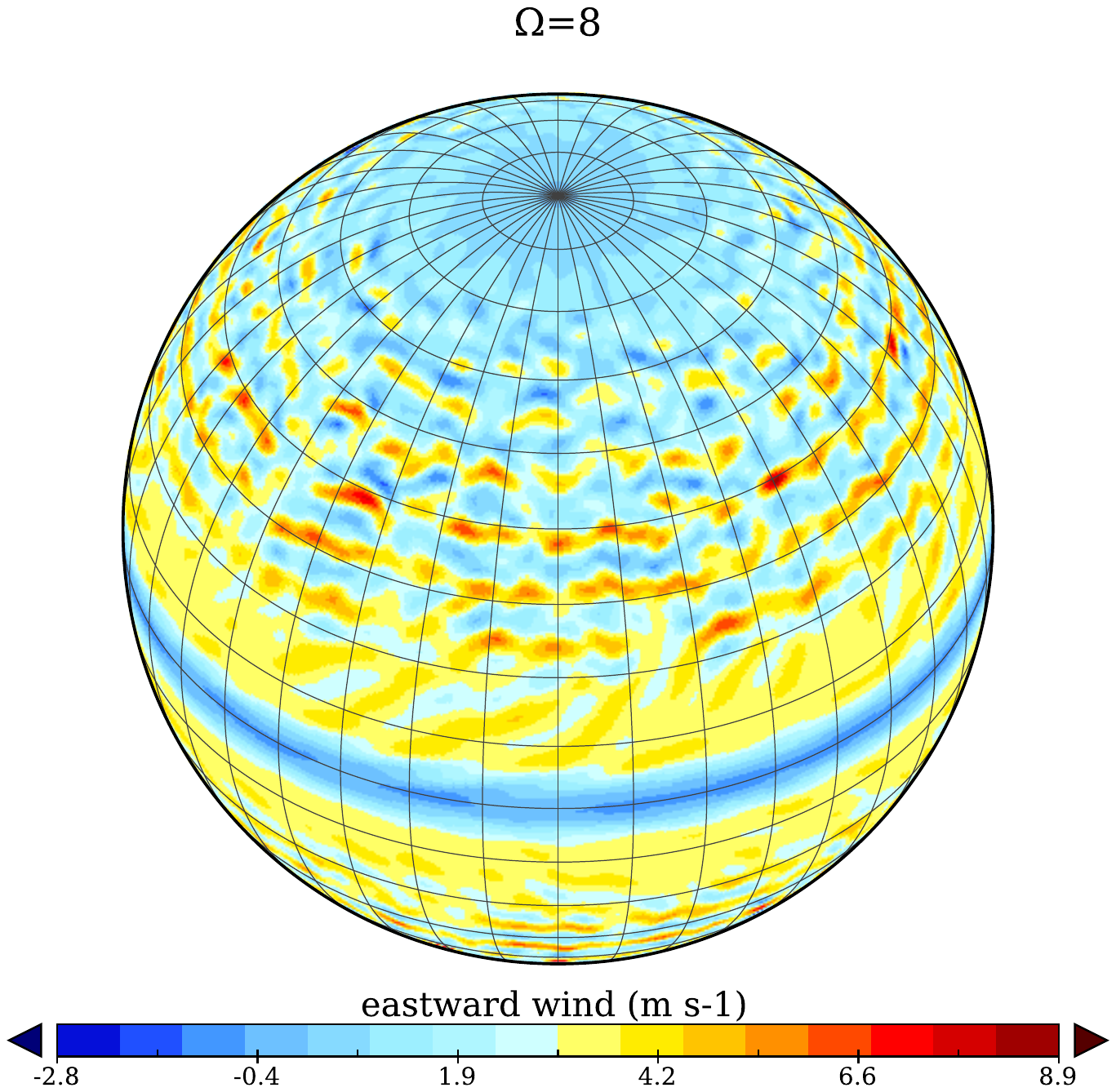}}
\subfloat{\includegraphics[bb=75 85 700 510,width=0.45\textwidth,clip=true]{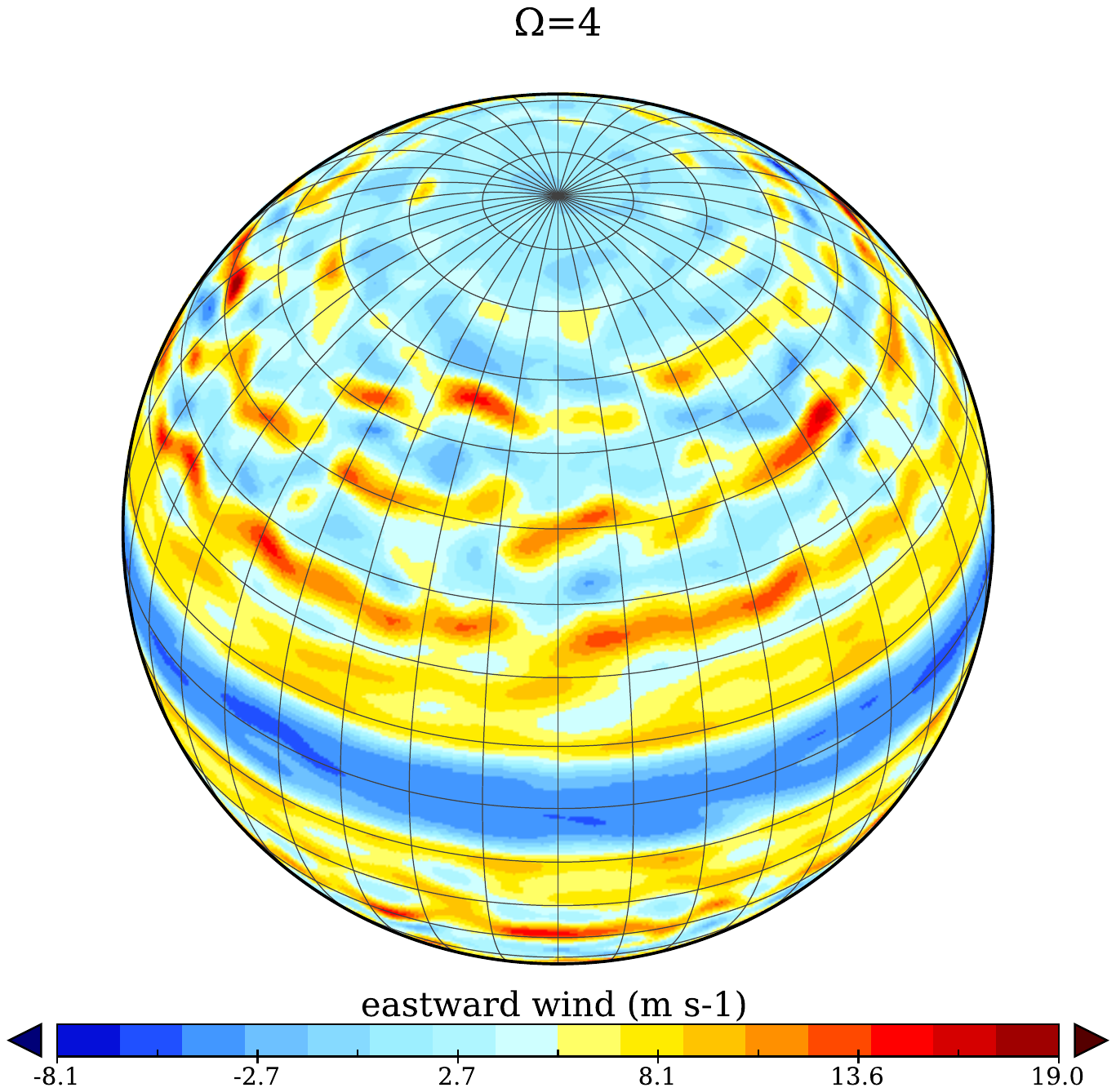}}\\
 (a) \hspace{8.5cm}(b)\\
\subfloat{\includegraphics[bb=75 85 700 510,width=0.45\textwidth,clip=true]{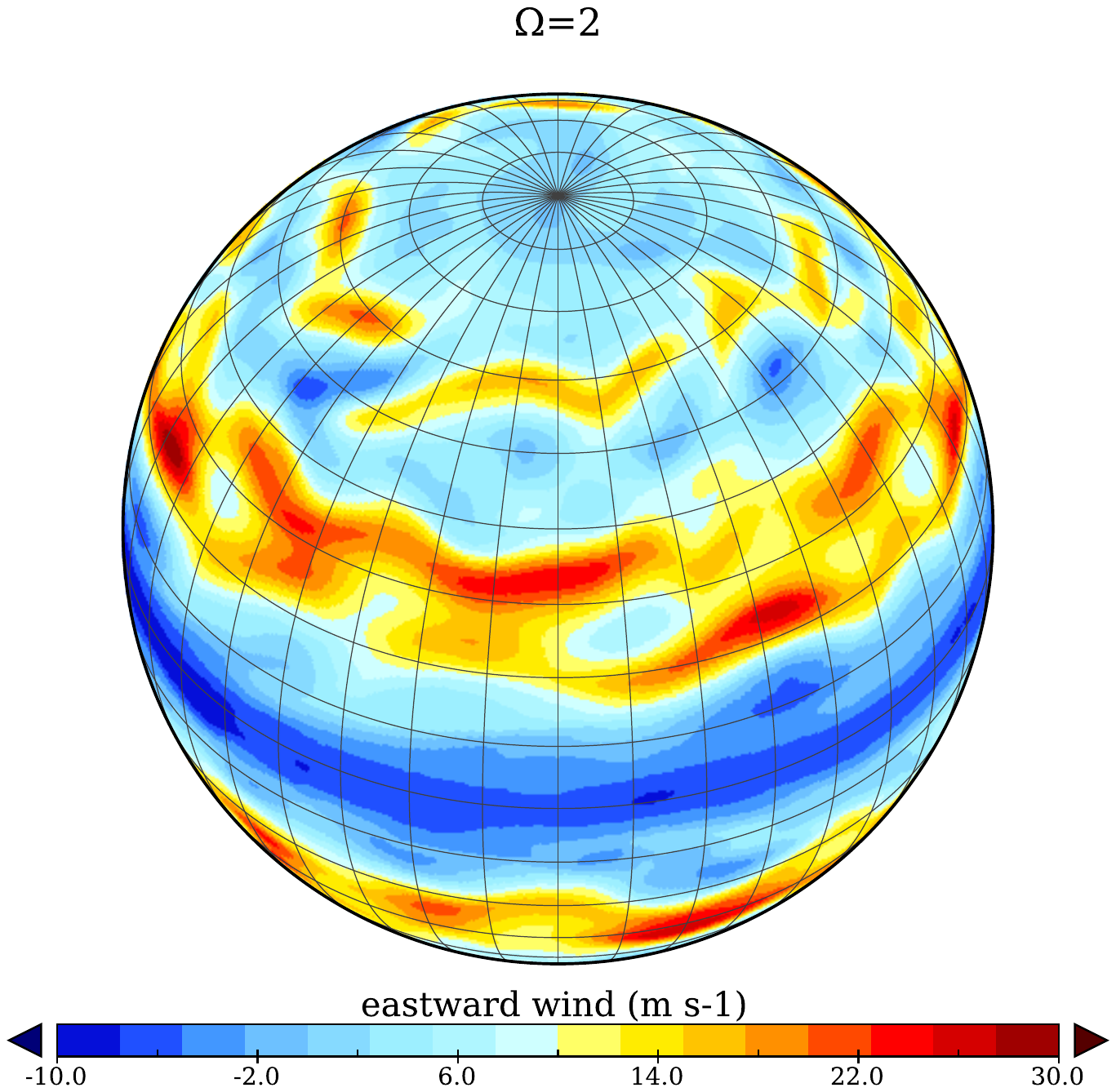}}
\subfloat{\includegraphics[bb=75 85 700 510,width=0.45\textwidth,clip=true]{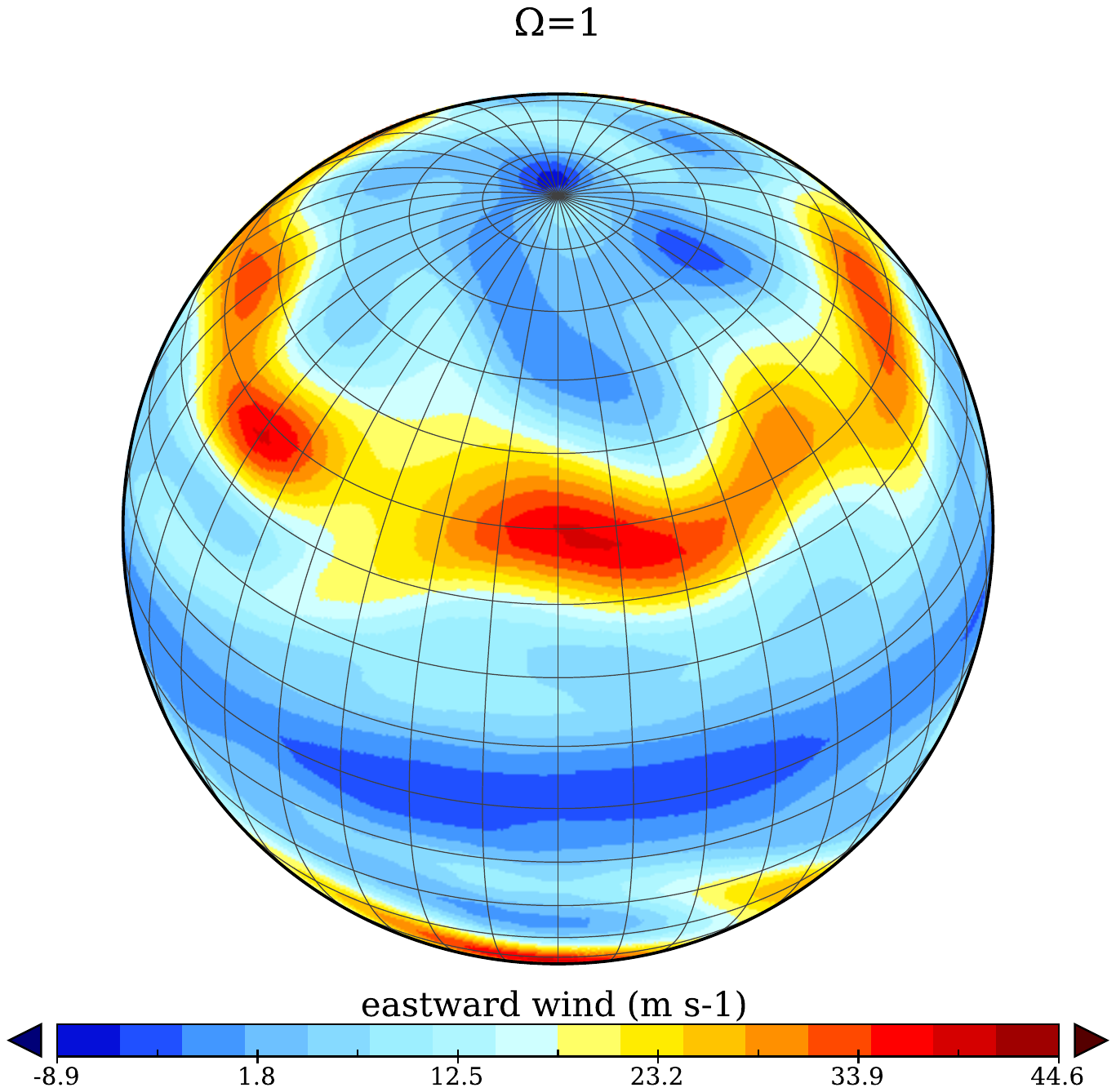}}\\
(c) \hspace{8.5cm}(d)\\
\subfloat{\includegraphics[bb=75 85 700 510,width=0.45\textwidth,clip=true]{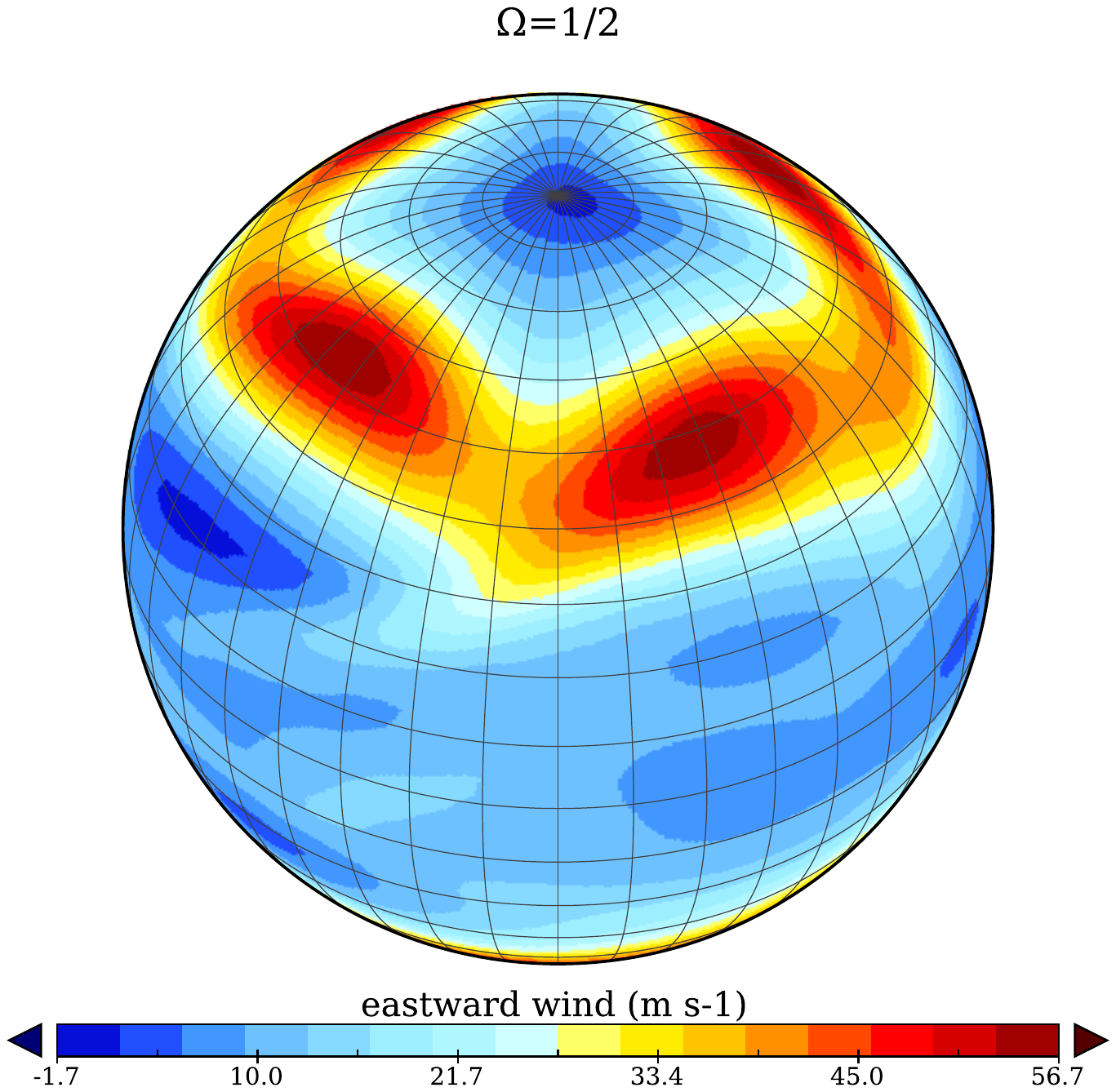}}
\subfloat{\includegraphics[bb=75 85 700 510,width=0.45\textwidth,clip=true]{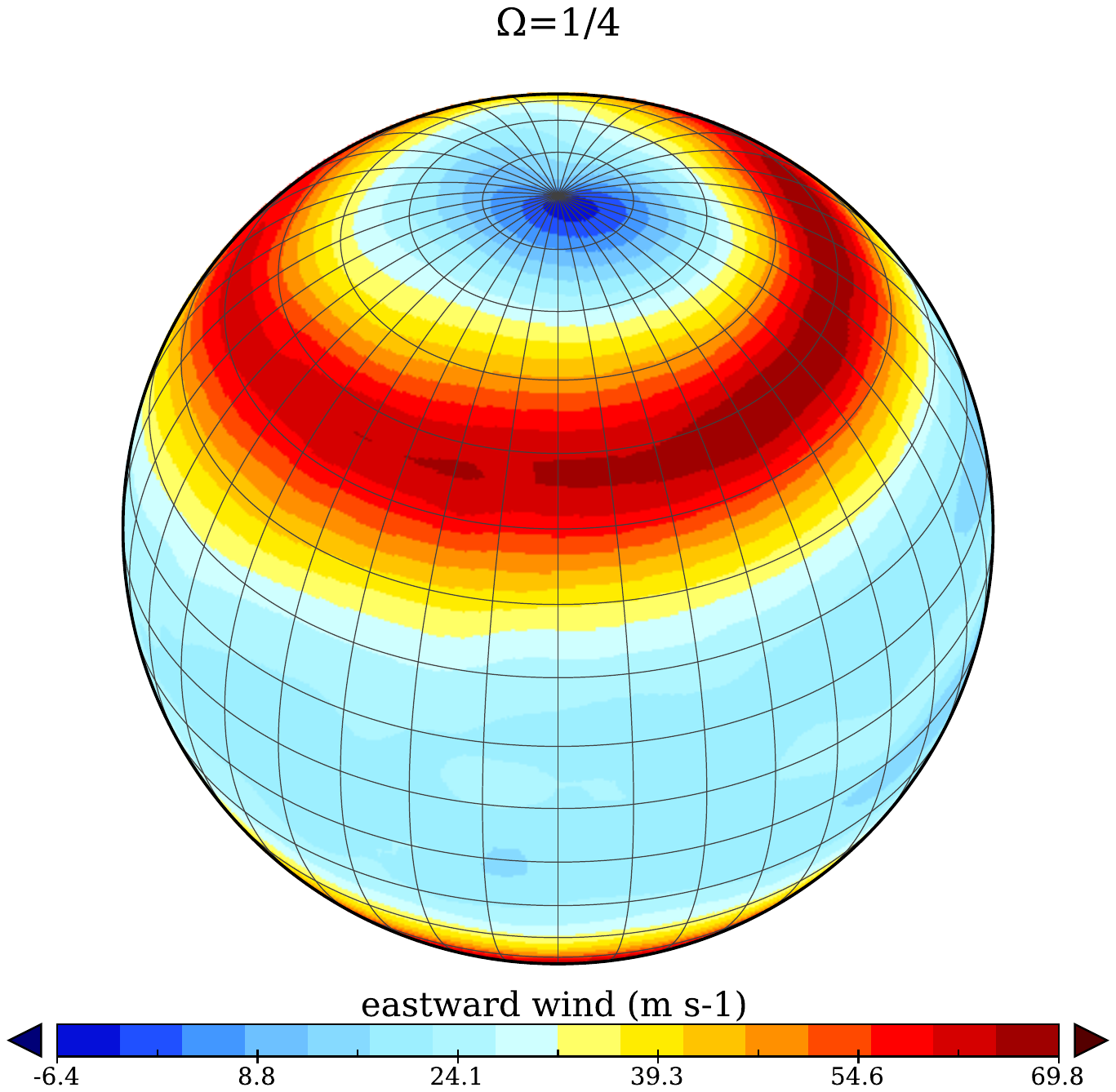}}\\
(e) \hspace{8.5cm}(f)\\
 \subfloat{\includegraphics[bb=75 85 700 510,width=0.45\textwidth,clip=true]{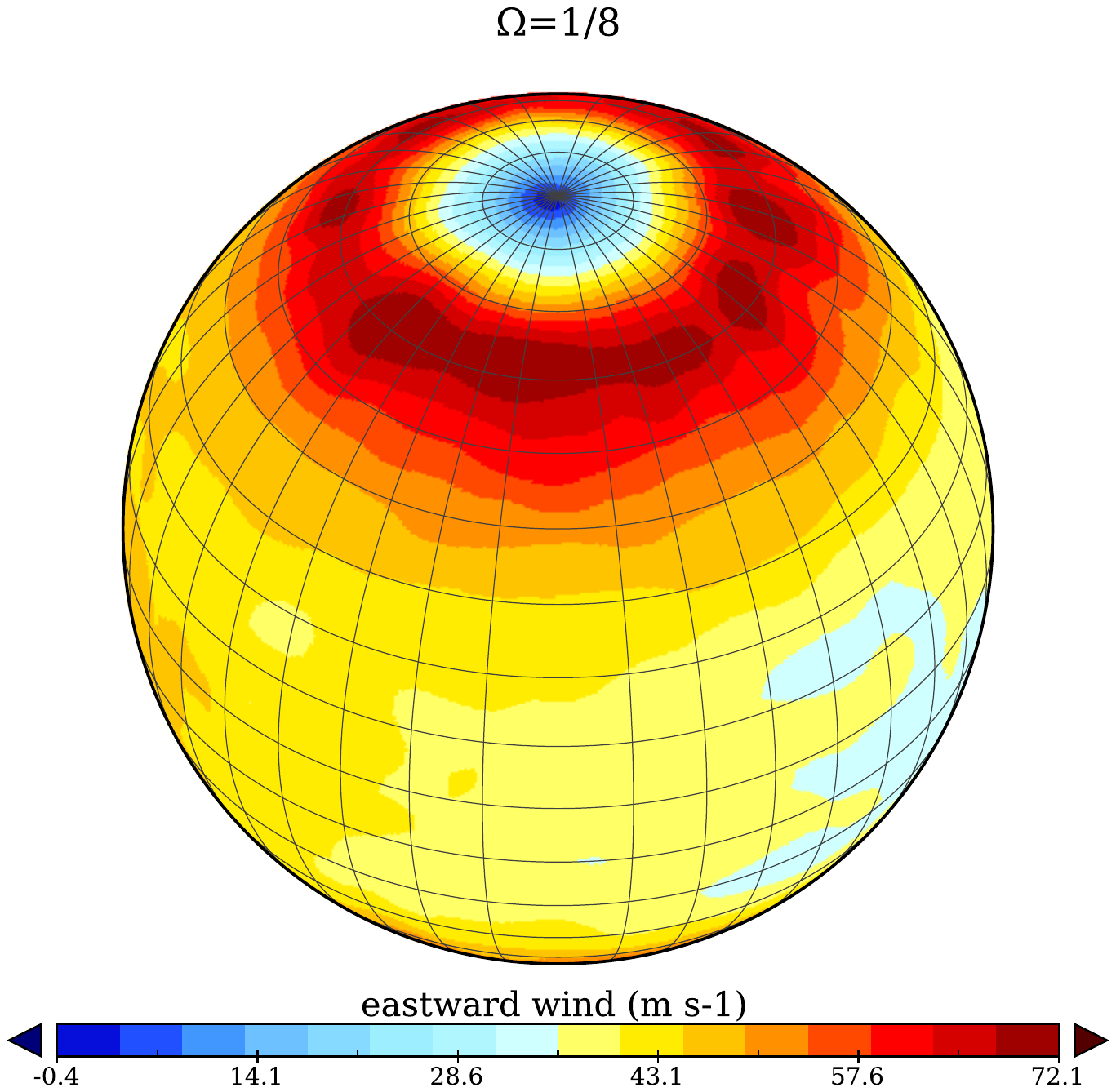}}
\subfloat{\includegraphics[bb=75 85 700 510,width=0.45\textwidth,clip=true]{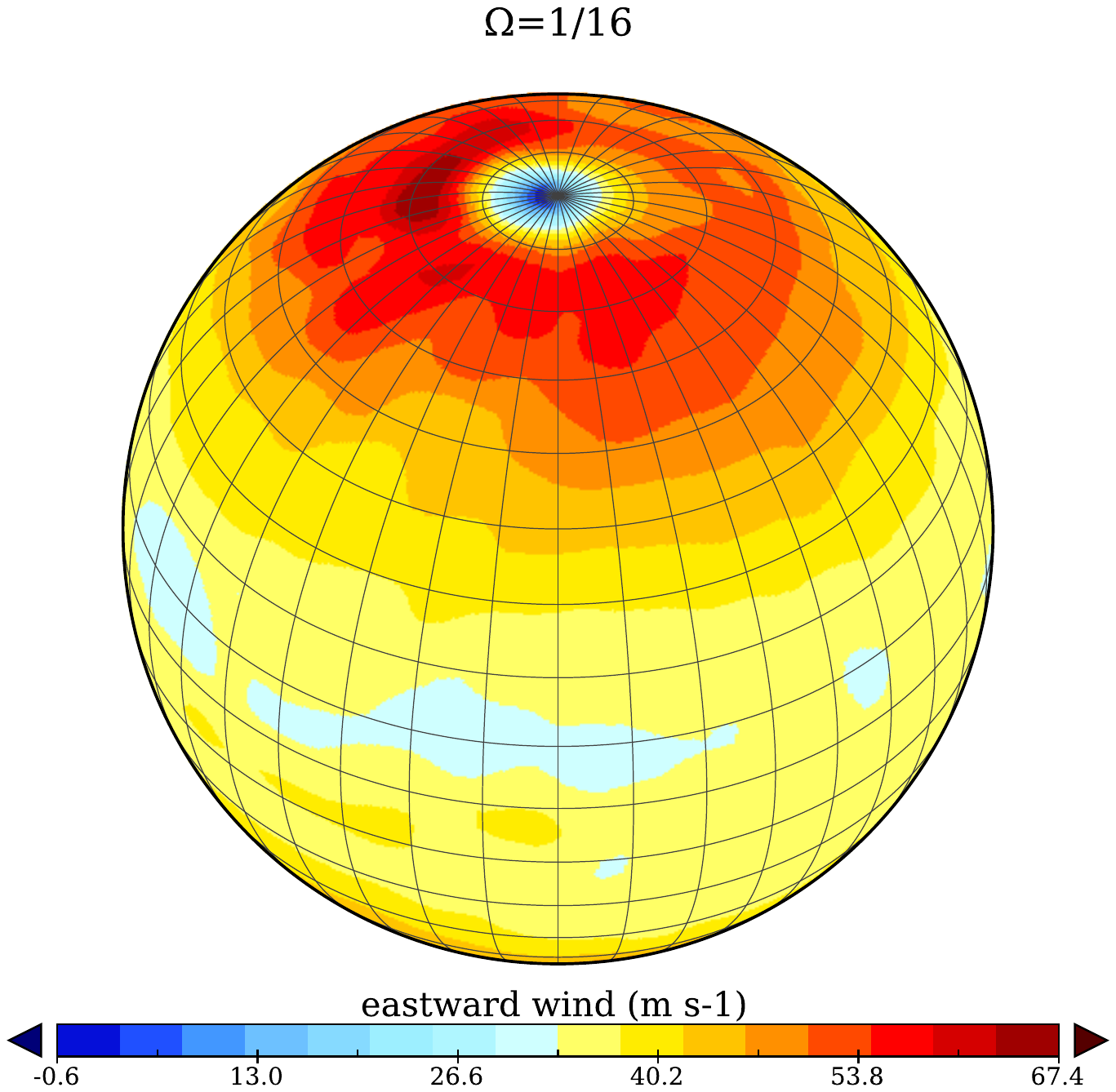}}\\
 (g) \hspace{8.5cm}(h)
\caption{Snapshots of upper level (200 hPa) zonal wind, projected onto a spherical surface at $\Omega^{\ast}=8$ [(a)], 4 [(b)], 2 [(c)], 1 [(d)], 1/2 [(e)], 1/4 [(f)], 1/8 [(g)] and  1/16 [(h)], illustrating the basic organization of winds across the planet.}
 \label{fig_spherical_pumas}
\end{figure*}

\subsection{Cyclostrophic, super-rotating flow}
This regime is found under conditions in which the planet rotates much more slowly than the Earth. Due to the weakness of
the Coriolis force on these slow rotators, the major force balance accounting
for the large-scale atmospheric motion is between the pressure gradient and
the centrifugal ``force'', known as cyclostrophic balance 
(see \cite{Holton1992}). Another distinctive feature of this kind of circulation 
is that there is typically a very strong prograde zonal wind over the equatorial 
regions, in contrast to the retrograde wind more typically seen in Earth's tropical atmosphere, 
a phenomenon known as equatorial super-rotation (e.g. see \cite{Read1986,Read2018}). These features are apparent in the
snapshots of zonal wind illustrated in Fig. \ref{fig_spherical_pumas}(f)-(h). 
A cross-section of the zonal mean zonal 
wind averaged over the last 360 model days for slowly rotating planets can be seen in Figure \ref{fig_ustrm_highrot_pumas}(n)-(p), 
which show substantial eastward (prograde) wind in 
the upper atmosphere over the equator. The extratropical jet streams, which
are usually located in mid-latitude or subtropics in Earth's atmosphere, are
pushed further polewards due to the expanded Hadley cells under weak
rotational constraints (see Section \ref{sec:zonal} below).


\subsection{Regular baroclinic wave flow}
For planets with an intermediate
thermal Rossby number (larger than Earth's, but smaller than those of Venus 
and Titan), planetary scale baroclinic waves are found to occur at mid-latitudes but tend to be very regular and 
coherent in both their spatial structure and in their time dependence. 
A perspective view of the upper level $\overline{u}$ field is 
shown in Fig. \ref{fig_spherical_pumas}(e). This simulation has a roughly similar thermal Rossby 
number to that of Mars, and a regular baroclinic wave with a dominant mode of either 
wavenumber-4 
or wavenumber-3 is found at the 500 hPa level 
(see Section \ref{sec:reg-rg} for further discussion).



\subsection{Irregular baroclinic wave flow}
This is the circulation regime that 
resembles the Earth's atmospheric circulation more closely than other regimes reported
here. It is characterised by spatially and temporally irregular 
baroclinic waves with mixed wavenumbers and aperiodic (probably chaotic) time-dependence. 
This regime typically exhibits a single, meandering jet stream in each hemisphere, in which the shape of the wavy structure changes irregularly with time in terms of both
phase and amplitude. The flow in Fig. \ref{fig_spherical_pumas}(d) is not dominated by a single zonal wavenumber mode but by a range of wavenumbers from $\sim5-8$ which fluctuates continuously (see also Fig. \ref{fig:latfft_highrot} below).

\subsection{Multiple zonal jet flow}
For planets with much smaller
thermal Rossby number than the Earth ($\mathcal{R}o_T<10^{-2}$), the development of 
geostrophic turbulence in the presence of a strong $\beta$-effect and
weak friction leads to the formation and maintenance of a flow pattern dominated by 
multiple eddy-driven zonal jets. Characterised by multiple eddy-driven jets over the 
extratropical regions, flows in this region of parameter space develop 
strongly zonal structures at planetary scale, as shown in Fig. \ref{fig_spherical_pumas}(a)-(c), 
in contrast to the rather more obviously meandering wavy structures
of other non-axisymmetric regimes. It is notable that gas giant planets in the Solar System 
(Jupiter, Saturn, Uranus and Neptune) all exhibit a set of strong, parallel, nearly rectilinear zonal 
jets (\cite{showman2010}), as reflected in their banded appearances. 
The experiment with $\Omega^{\ast}=8$ has a somewhat similar value of 
$\mathcal{R}o_T$ to Jupiter and Saturn (as can be appreciated from 
Fig. \ref{fig:regdiag}). Recent studies suggest that such banded zonal 
jets can also be found in the Earth's oceans (for example, as found 
in the eddy-resolving ocean model simulations of the North Pacific Ocean,
e.g. see \cite{Galperin2004}), but the resulting jets are much weaker and more strongly meandering. 

The diversity of circulation regimes identified in this regime diagram suggests a number
of fundamental questions regarding the general circulation of planetary atmospheres:
What determines the preferred eddy scales in a planetary atmosphere? How does the 
planetary rotation rate affect the meridional heat transport? What causes and 
maintains the equatorial superrotation when rotation rate is low? What determines
the selection of wavenumbers in steady baroclinic waves? Why do eddies organise 
themselves into jets and what insights can be obtained on the jet formation 
mechanism from modern theories of geostrophic turbulence? With these questions 
in mind we will investigate the structures and trends of the circulation regimes 
in the following sections.

\section{Phenomenology}
\subsection{Zonal mean circulation}\label{sec:zonal}
\begin{figure*}[!ht]
 \centering
 \subfloat{\includegraphics[width=0.48\textwidth,height=0.13\textwidth,clip=true]{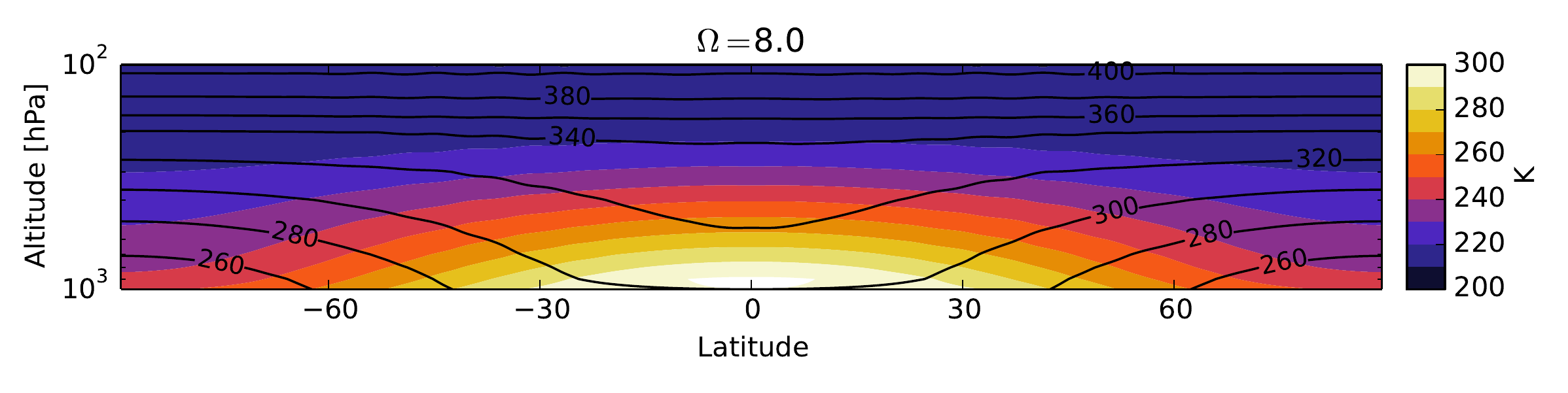}}
 \subfloat{\includegraphics[width=0.48\textwidth,height=0.13\textwidth,clip=true]{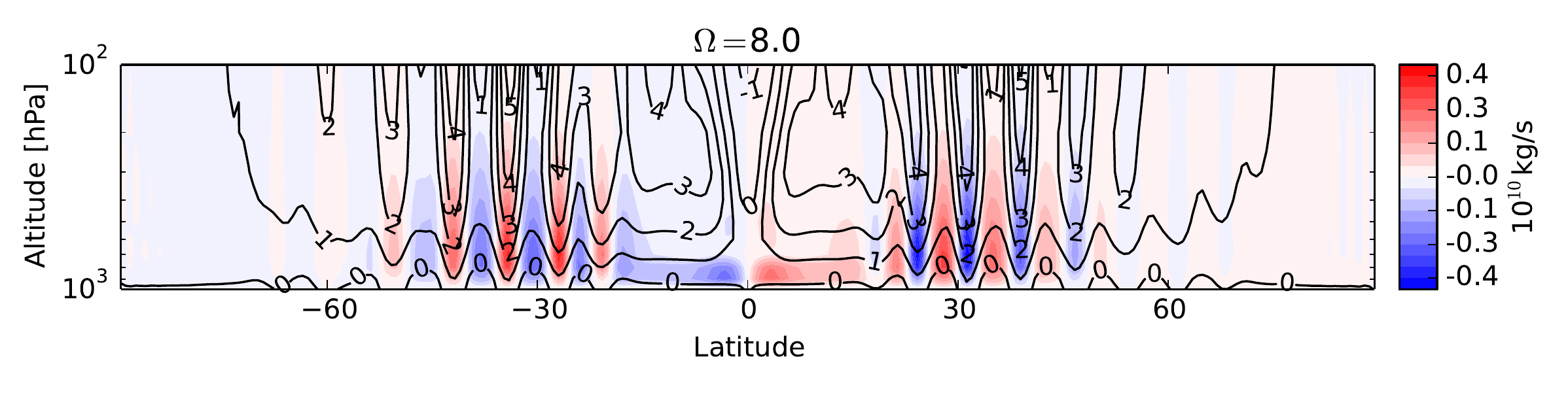}}\\
 (a) \hspace{8.5cm}(i)\\
 \subfloat{\includegraphics[width=0.48\textwidth,height=0.13\textwidth,clip=true]{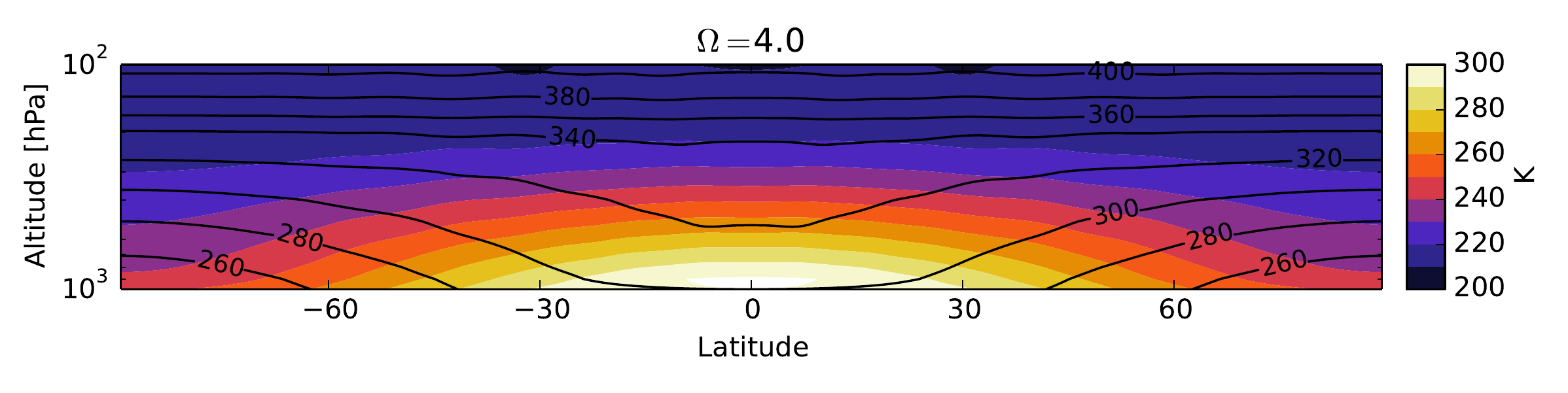}}
 \subfloat{\includegraphics[width=0.48\textwidth,height=0.13\textwidth,clip=true]{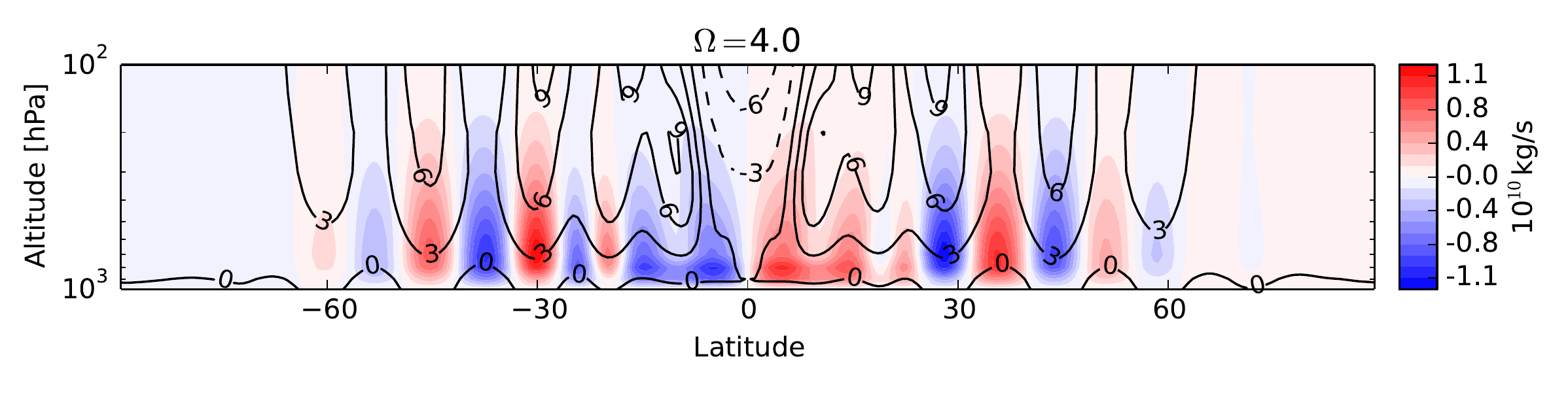}}\\
(b) \hspace{8.5cm}(j)\\
 \subfloat{\includegraphics[width=0.48\textwidth,height=0.13\textwidth,clip=true]{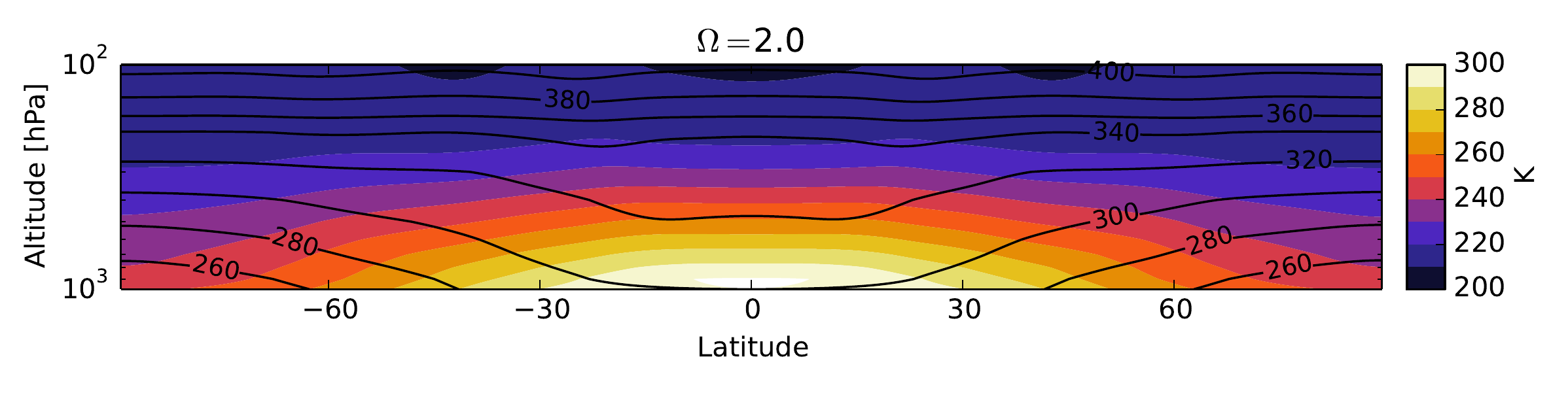}} 
 \subfloat{\includegraphics[width=0.48\textwidth,height=0.13\textwidth,clip=true]{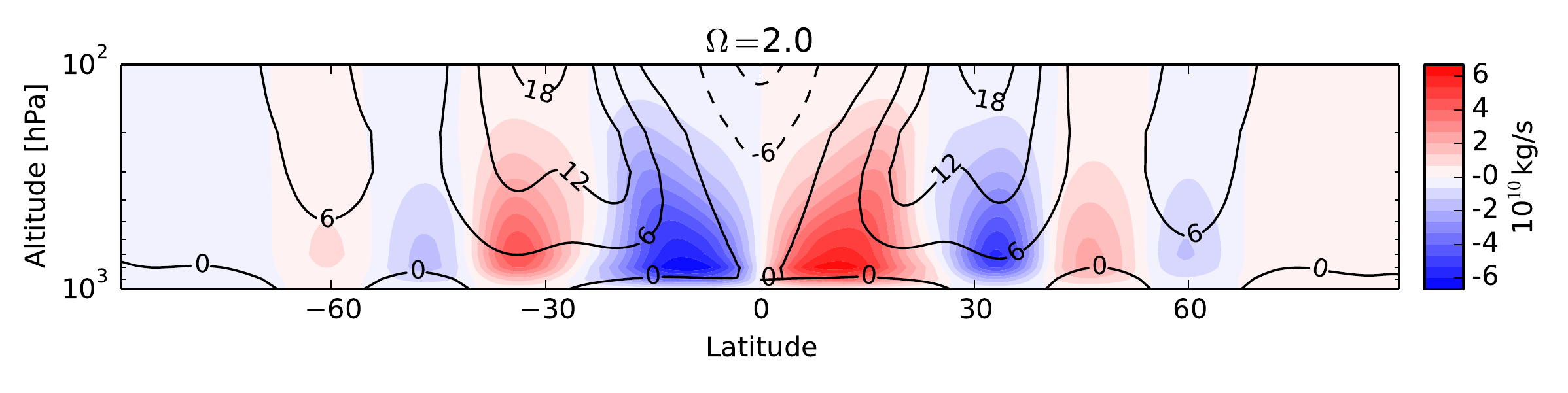}}\\
(c) \hspace{8.5cm}(k)\\
 \subfloat{\includegraphics[width=0.48\textwidth,height=0.13\textwidth,clip=true]{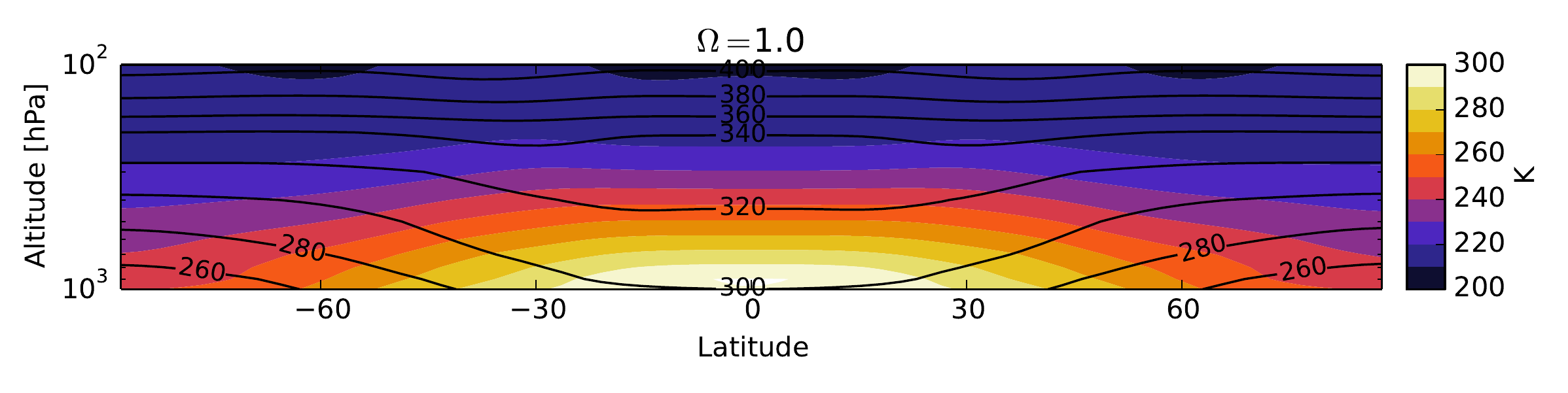}}
 \subfloat{\includegraphics[width=0.48\textwidth,height=0.13\textwidth,clip=true]{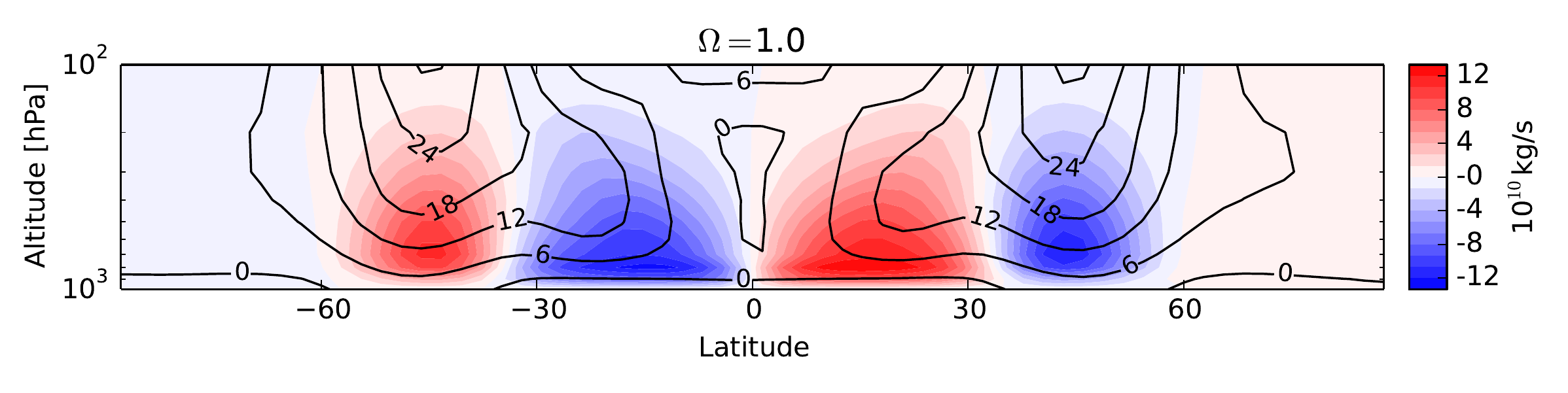}}\\
(d) \hspace{8.5cm}(l)\\
 \subfloat{\includegraphics[width=0.48\textwidth,height=0.13\textwidth,clip=true]{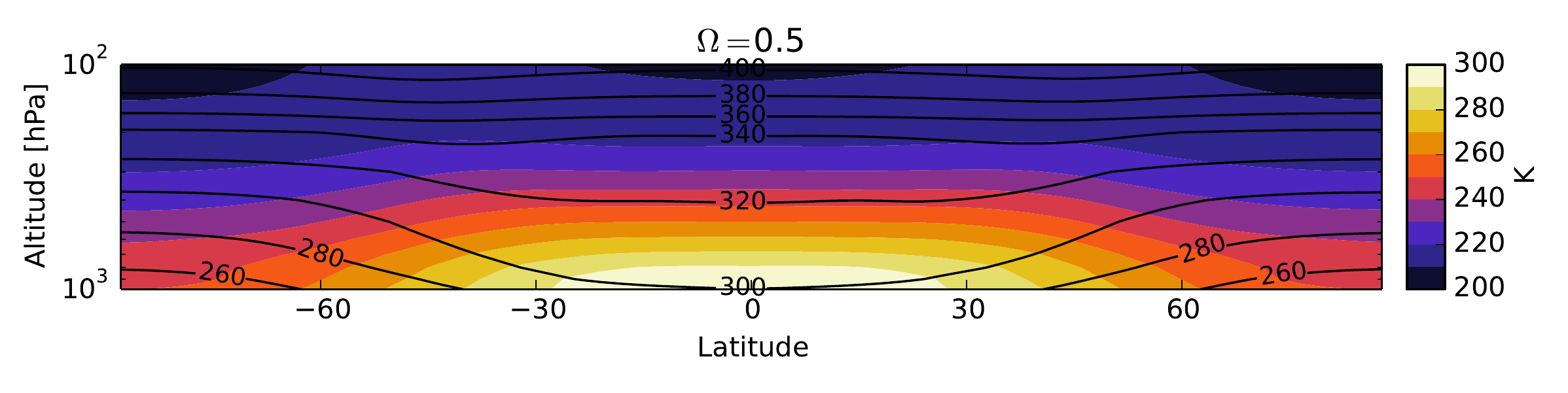}} 
 \subfloat{\includegraphics[width=0.48\textwidth,height=0.13\textwidth,clip=true]{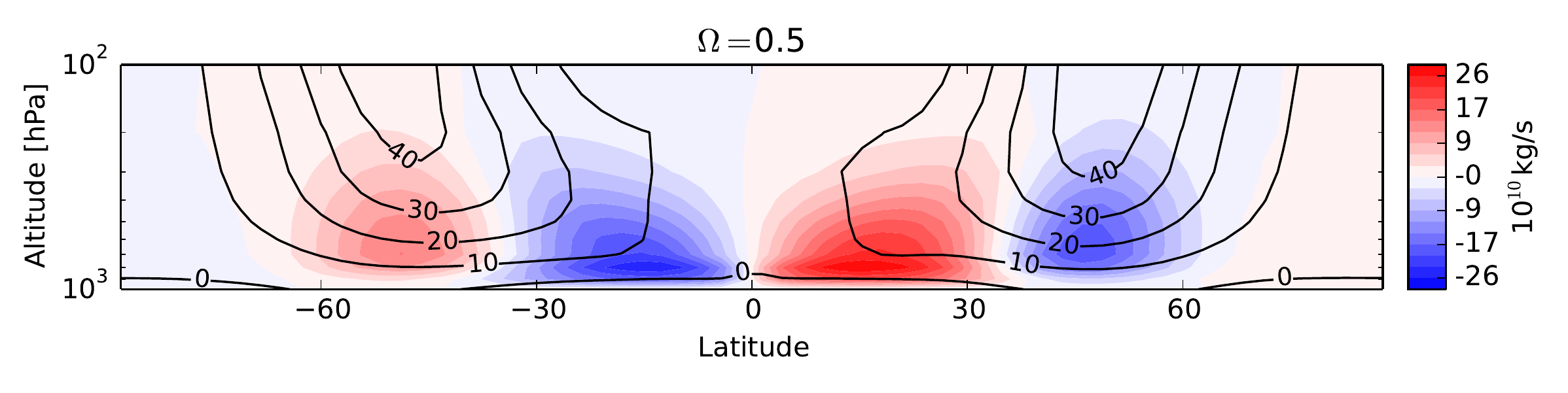}}\\
(e) \hspace{8.5cm}(m)\\
 \subfloat{\includegraphics[width=0.48\textwidth,height=0.13\textwidth,clip=true]{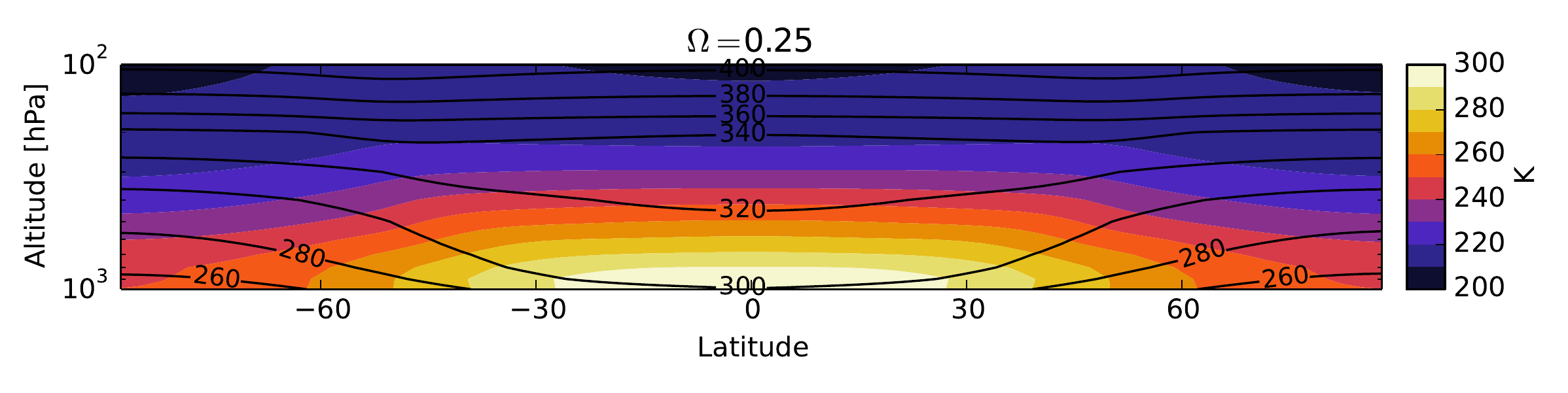}} 
 \subfloat{\includegraphics[width=0.48\textwidth,height=0.13\textwidth,clip=true]{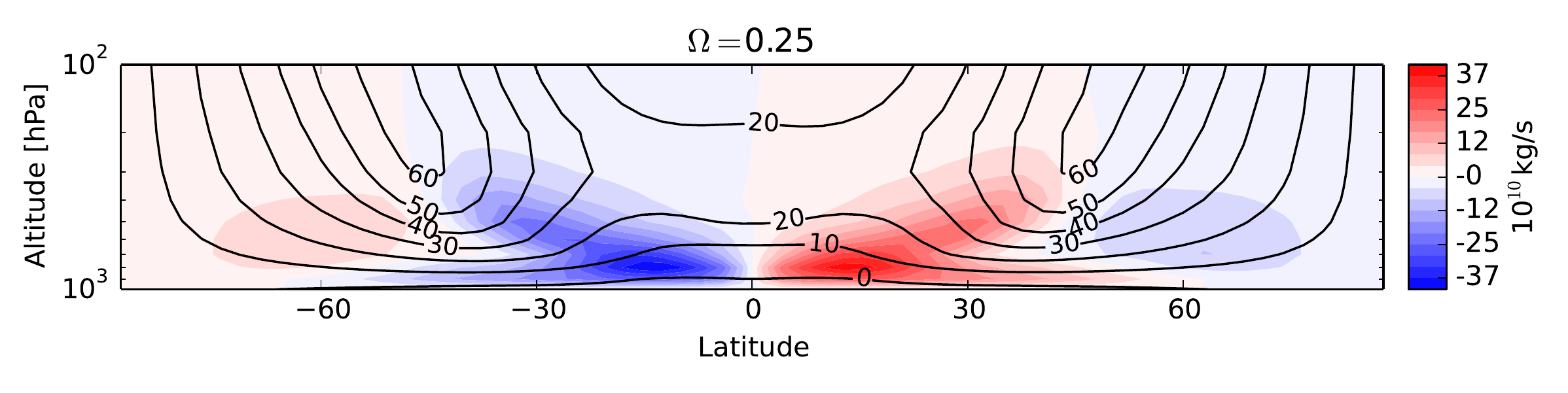}}\\
(f) \hspace{8.5cm}(n)\\
 \subfloat{\includegraphics[width=0.48\textwidth,height=0.13\textwidth,clip=true]{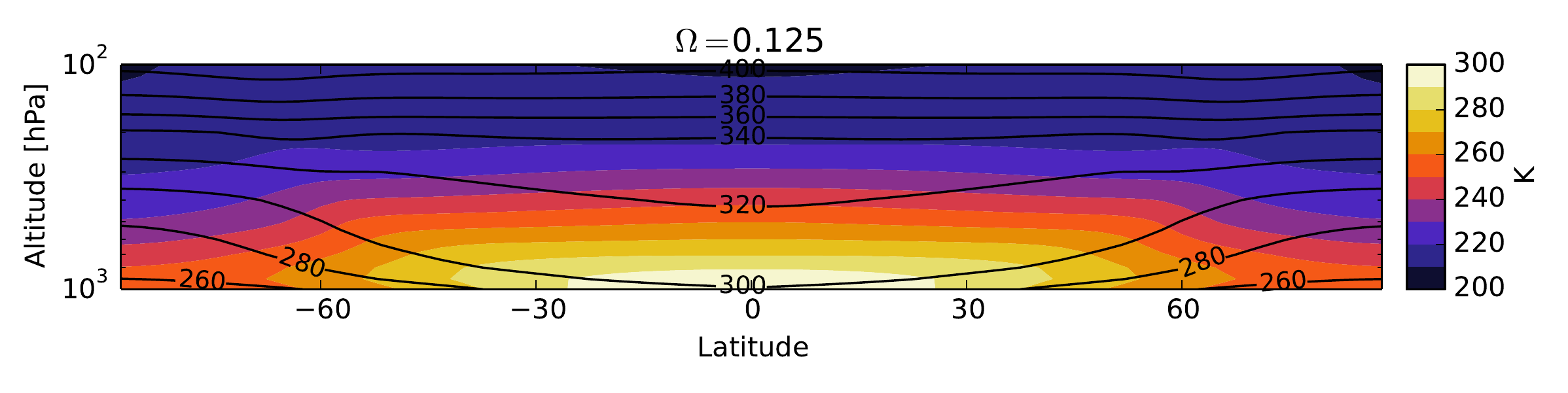}} 
 \subfloat{\includegraphics[width=0.48\textwidth,height=0.13\textwidth,clip=true]{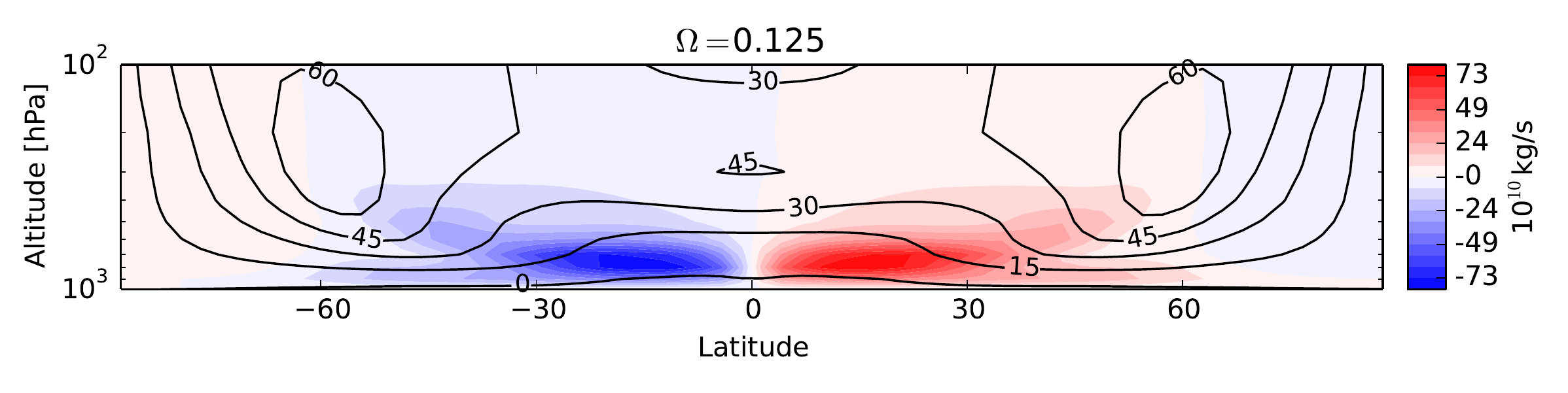}}\\
(g) \hspace{8.5cm}(o)\\
 \subfloat{\includegraphics[width=0.48\textwidth,height=0.13\textwidth,clip=true]{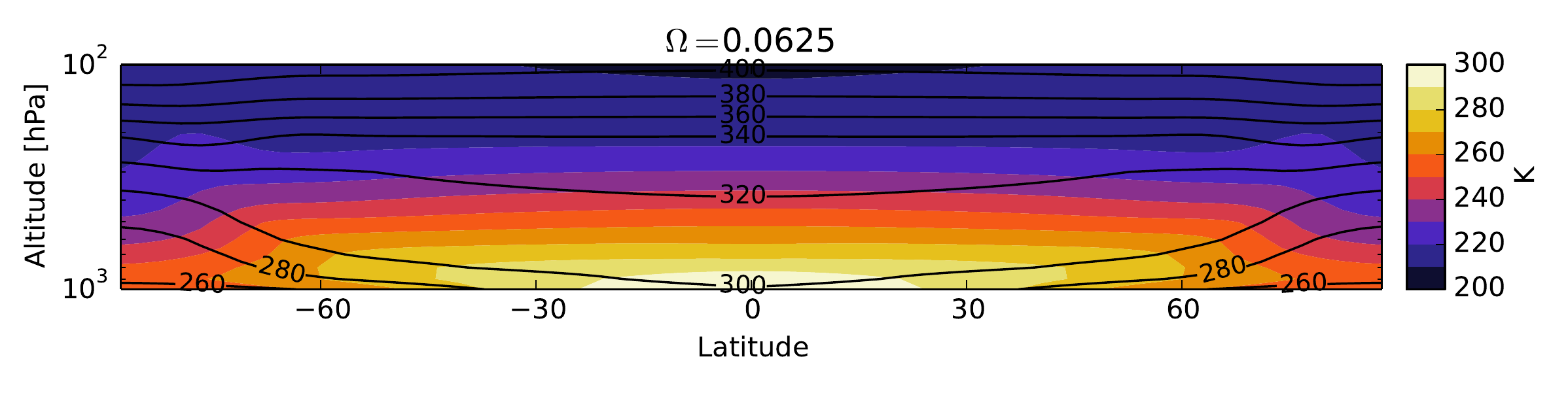}} 
 \subfloat{\includegraphics[width=0.48\textwidth,height=0.13\textwidth,clip=true]{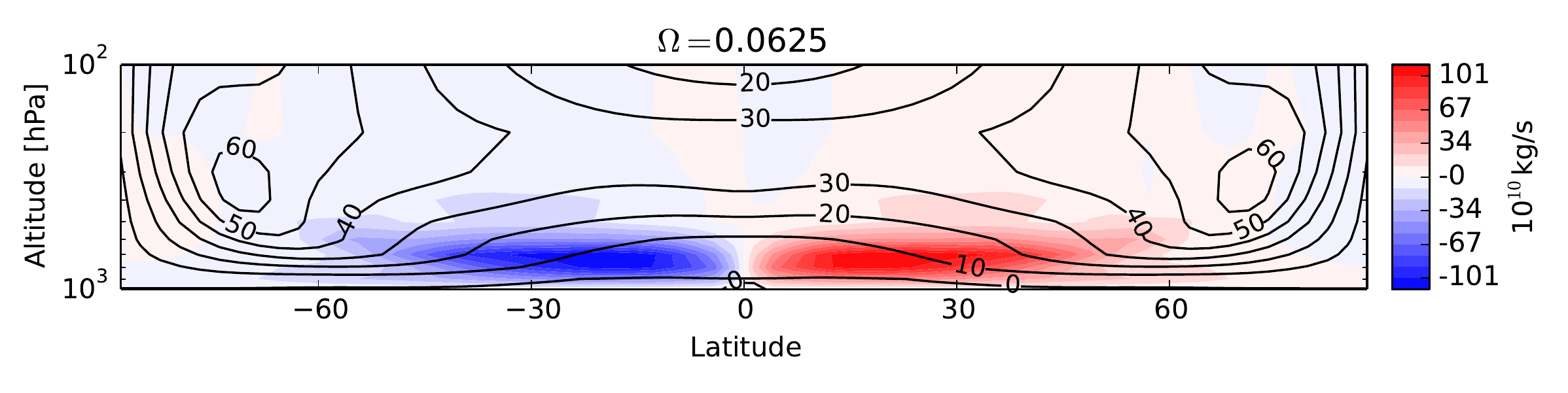}}\\
 (h) \hspace{8.5cm}(p)\\
 \caption{Zonal mean fields of temperature (absolute, $T$, and potential, $\theta$) [(a) - (h)] and zonal wind and meridional streamfunction [(i) - (p)] for $\Omega^{\ast}=8$ [(a) \& (i)], 4 [(b) \& (j)], 2 [(c) \& (k)], 1 [(d) \& (l)], 1/2 [(e) \& (m)], 1/4 [((f) \& (n)], 1/8 [(g) \& (o)] and  1/16 [(h) \& (p)]. Shading indicates absolute temperature in (a) - (h) and meridional streamfunction in (i) - (p), while contours indicate potential temperature [(a) - (h)] and zonal velocity [(i) - (p)].}
 \label{fig_ustrm_highrot_pumas}
\end{figure*}

Since the incoming solar radiation, which serves as the ultimate energy source of 
the atmospheric circulation, generally varies most in the meridional direction, the 
large-scale motion of the atmosphere has greater variations in the meridional 
direction than in the zonal direction, especially when averaging over a long time 
period compared to other dynamical timescales. Thus, analysis of zonal averages of the
meteorological variables takes a particularly significant role in the study of atmospheric 
general circulation. The zonal mean absolute and potential temperature ($T$ and $\theta$ - left column) and zonal wind ($\overline{u}$) contours (right column) for 
different $\Omega$ and $\Delta T_h = 60\ \text{K}$ are shown in 
Fig. \ref{fig_ustrm_highrot_pumas}, with the meridional 
mass streamfunction superimposed on $\overline{u}$ using colour shading.

The fastest rotating experiments ($\Omega^{\ast}>1$) are characterised by multiple parallel extratropical
jet streams. The generation and maintenance of these jets through upscale energy transfers, which 
are characteristic of highly anisotropic geostrophic turbulence, will be discussed in a further companion paper \citep{read2017}. At this stage we 
focus on describing the change of global structures---the increase of the number of jets with $\Omega$ --- and identifying the key lengthscales that determine the global structure in the zonal mean.
The characteristic latitudinal range of anisotropy associated with the latitudinal variation of Coriolis parameter $f$ can be estimated by 
the Rhines scale \citep{Rhines1975}: {
\begin{equation}
 L_{R}= \Big(\dfrac{2U}{\beta}\Big)^{1/2},
\end{equation}
in which $U$ is a characteristic horizontal wind speed. The original argument in \cite{Rhines1975} used $U_{\text{rms}}$,
which is the root mean square velocity (often taken to be the rms eddy velocity). Other work, however, has suggested that the rms zonal mean velocity, $\overline{U}$, is more appropriate in defining $L_R$ \citep[e.g.][]{dunkerton2008}. Here, we apply the thermal wind relationship to the zonal mean temperature field to estimate $\overline{U}$ (e.g. see \citet{Read2011}) which leads to:
\begin{equation}
 L_{R1}\simeq \Big(\dfrac{R\Delta \theta_h}{\Omega ^2}\Big)^{1/2} = a (\mathcal{R}o_T)^{1/2}. \label{eq:LR1}
\end{equation}
The number of eddy-driven jets expected on a planet can then be estimated as
\begin{equation}
 N_J \simeq \dfrac{\pi a}{2} \dfrac{1}{2 \pi L_R} = \dfrac{a}{4 L_R},\label{eq:nj}
\end{equation}
\noindent which leads to an estimate based on the zonal mean thermal wind of
\begin{equation}
N_{J1} \simeq \dfrac{1}{4\mathcal{R}o_T^{1/2}}. \label{eq:NJ1}
\end{equation}
This leads to an estimate that is somewhat similar to the formulation suggested by \citet{lee2005} ($N_J \propto (a /\Delta \theta_h)^{1/2}$) though with a different dependence on $a$. }

Table \ref{tab.rhines} shows the values of various key dimensionless parameters for the main series of runs discussed here, including the number of jets on a planet surface predicted by Eq. \eqref{eq:nj} {and \eqref{eq:NJ1}}. 
For experiments with very slow rotation rates ($\Omega^{\ast}<1/4$), $N_{J1}$ is smaller than $1$. This means that 
the picture of an eddy-driven jet emerging through an inverse energy cascade in the framework of geostrophic 
turbulence is no longer valid since the Rhines {wavelength, $2 \pi L_{R1}$ in this case,} exceeds the size of the planetary domain. The jets  observed in these slowly rotating simulation experiments are primarily due to the quasi-conservation of angular 
momentum within the upper branch of the extended thermally-direct Hadley cell. At $\Omega^{\ast}=1$, the 
predicted number of jets is $1.16$, which approximately corresponds to the regime of two prograde jet 
per hemisphere. For experiments with faster rotation rates ($\Omega^{\ast}>1$), Eq. \eqref{eq:NJ1} predicts 
up to $14$ jets for $\Omega^{\ast}=8$. This is {broadly} consistent with the zonal wind map shown 
in Fig. \ref{fig_ustrm_highrot_pumas} where at least 8 distinct eastward jets can be located in each hemisphere. {The multiple jets for $\Omega^\ast = 2$ and 4, however, were seen to drift slowly polewards, much as found by \citet{chemke2015b}, although this was less evident for the $\Omega^\ast = 8$ case for which the spatial resolution may have been insufficient to reveal this behaviour clearly.}


\begin{table}
 \begin{center}
  \begin{tabular}{|c|c|c|c|c|c|c|}
   \hline
$\Omega^{\ast}$ & $\mathcal{R}o_T$ & $ \mathcal{T}a_f$ & $\mathcal{T}a_R$ & $N_{J1}$ & $N_{J2}$ \\
\hline
  $1/16$ &   20.5 & 238 & 1.71 $\times 10^5$ & 0.08 & 0.49\\
  $1/8$  &  5.14 & 3.80 $\times 10^3$ & 2.73 $\times 10^6$ & 0.14 & 0.77 \\
  $1/4$ &    1.28 & 6.09 $\times 10^4$ & 4.67 $\times 10^7$ & 0.29 & 0.96 \\
  $1/2$ &   0.32 & 9.74 $\times 10^5$ & 6.98 $\times 10^8$ & 0.58 & 1.06 \\
  $1$ &  0.08 & 1.56 $\times 10^7$ & 1.12 $\times 10^{10}$ & 1.1 & 1.25 \\
  $2$ &  0.02 & 2.49 $\times 10^8$ & 1.79 $\times 10^{11}$ & 2.2 & 3.1 \\
  $4$  & 0.005 & 3.99 $\times 10^9$ & 2.86 $\times 10^{12}$ & 3.9 & 6.1 \\
  $8$ & 0.0013 & 6.38 $\times 10^{10}$ & 4.58 $\times 10^{13}$ & 7.1 & 13.9 \\
  \hline
  \end{tabular}
  \caption{Key dimensionless parameters for the baseline set of numerical simulations with $\Delta \theta_h=60 
  \text{K}$, $\tau_{ft} = 5$ Earth days and $\tau_R = 25.9$ Earth days, as defined by Equations \eqref{eq:Ro}, \eqref{eq:Taf}, 
  \eqref{eq:nj}, \eqref{eq:NJ1} and \eqref{eq:NJ2}.}
  \label{tab.rhines}
 \end{center}
\end{table}

As we can see from Figs. \ref{fig_ustrm_highrot_pumas}, the subtropical
jet stream moves to higher latitude as the rotation rate decreases, which is consistent with the 
prediction from the quasi-inviscid axisymmetric Hadley cell theory (see \citet{Held&Hou}; \citet{caballero2008}). The intensity of 
the jet stream grows stronger as the rotation rate decreases until $\Omega^{\ast}=1/8$. 
For the $\Omega^{\ast}=1/16$ run, the extratropical jet ($\sim 60\ \text{m s}^{-1}$) is actually 
weaker than 
that in the $\Omega^{\ast} = 1/8$ run ($\sim 70\ \text{m s}^{-1}$). In principle, there are two 
competing factors determining the intensity of the extratropical jet stream as the rotation rate 
decreases. The jet stream reaches a higher latitude, thus gaining more angular velocity for a given angular momentum than at 
lower latitudes. On the other hand, the reduction in rotation rate actually decreases the angular 
momentum of the whole planet, thus reducing the angular momentum obtained by the poleward moving jet
stream. For the runs from $\Omega^{\ast}=1$ to $\Omega^{\ast}=1/8$, the first factor dominates and the 
jet intensity gets stronger for smaller rotation rates. But for the $\Omega^{\ast}=1/16$ run, 
the latter factor begins to dominate and the angular momentum of the background planetary rotation reduces 
sufficiently to offset the angular momentum gain caused by the poleward motion of the jet
stream (similar results were found by \cite{Navarra2002}).

This poleward movement of the subtropical jet stream is a clear indication of the expansion of the Hadley 
cell in each hemisphere as the rotation rate decreases. This can be evaluated by the zonal mean 
meridional mass streamfunction, defined in the pressure coordinate system as (see \cite{Peixoto1992}):
$$
\Psi=\frac{2\pi a}{g}\cos{\phi}\int_0^P dp'[\overline{v}],
$$
where $a$ is the planetary radius, $[\overline{v}]$ the zonal and temporal mean meridional velocity, 
$\phi$ the latitude and $P$ the surface atmospheric pressure.

As anticipated (see Fig. \ref{fig_ustrm_highrot_pumas}(right column)), there are basically three cells in each hemisphere for the terrestrial rotation rate
($\Omega^{\ast}=1$). The positive values of $\Psi$ represent clockwise flow and the negative 
values represent counter-clockwise flow, while the magnitude reflects the strength of the overturning. The poleward edge of the overturning Hadley circulation can be estimated by the latitude of the 
boundary between the tropical cell and the adjacent mid-latitude cell. In the $\Omega^{\ast}=1$ case,
the extent of the Hadley cell in each hemisphere is roughly $30^\circ$, which is consistent with the 
observed value for the Earth. As the rotation rate decreases, the overturning Hadley cell expands
and intensifies, {much} as predicted in the previously cited \citet{Held&Hou} and \citet{caballero2008} models. In fact, at the
lowest rotation rate, $\Omega^{\ast}=1/16$, only one strong hemispherically-dominating Hadley cell is 
found in each hemisphere while the other two cells at mid- and high-latitudes disappear. Within the 
Hadley cell, baroclinicity (associated with the meridional slope of isentropic surfaces, and hence the horizontal temperature difference in the meridional direction) is 
considerably weaker compared to that in the extratropical baroclinic eddy zones. This feature is 
reflected in the zonal mean temperature cross-sections
in Fig. \ref{fig_ustrm_highrot_pumas}(left column).

{
\begin{figure}
 \centering
 \includegraphics[width=\columnwidth]{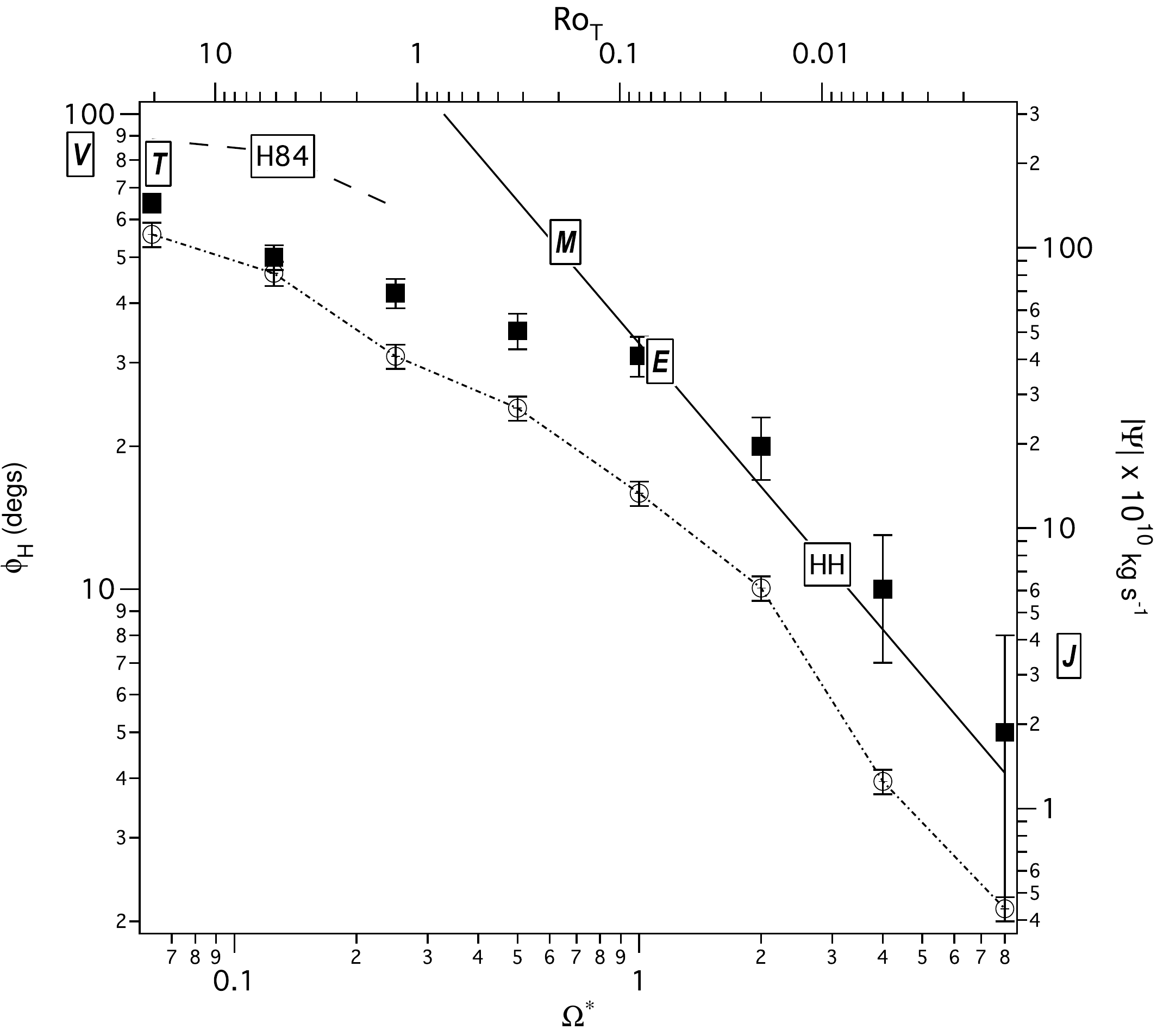}\\
 
 \caption{Variations of (a) the strength of the main Hadley streamfunction (shown as open circles) and (b) the width of the main Hadley circulation as a function of $\Omega^\ast$ and $\mathcal{R}o_T$ (shown as solid squares), estimated from the latitudinal width of the mass streamfunction contour at 10\% of the peak value in mid-troposphere in each case. The solid line shows the expected variation predicted by the simple model of \citet{Held&Hou}, scaled by $\pi/2a$, while the dashed line shows the expected variation in the slowly rotating limit by \citet{Hou1984}. The observed or modelled widths of the Hadley cells on Earth, Mars, Venus, Titan and Jupiter are indicated by their initial letters (see text). }
 \label{fig:Hadley1}
\end{figure}

 Figure \ref{fig:Hadley1} shows the variation of the actual width of the thermally-direct Hadley cell at each value of $\Omega^\ast$ and $\mathcal{R}o_T$, estimated by the angular width of the mass streamfunction contour at 10\% of the peak value (shown as filled squares). This may be compared with the latitudinal width 
\begin{equation}
Y_{HH} \simeq a \left(\dfrac{5}{3} \mathcal{R}o_T\right)^{1/2},
\end{equation}
\noindent predicted by the simple models of \citet{Held&Hou} and \citet{caballero2008} (which differs from \citet{Held&Hou} only by a factor O(1) that depends on radiative-convective factors). The length scale $\pi Y_{HH}/2a$ is shown for reference as a solid line in Fig. \ref{fig:Hadley1}. The Hadley cell width in the GCM simulations is seen to scale quite closely with the Held-Hou model for $\Omega^\ast \geq 1$, but then increases much more slowly with increasing $\mathcal{R}o_T$ as it approaches the radius of the planet itself. At the lowest values of $\Omega^\ast$, the width tends towards 80-90$^\circ$, much as predicted in the extension of the Held-Hou model by \citet{Hou1984}. 

Estimates of the Hadley cell widths for Earth, Mars, Titan, Venus and Jupiter are also shown for comparison, indicated by their initial letters. The widths for Earth and Mars were estimated following \citet{Read2011} from the ERA40 reanalysis \citep{ERA40} and the annual zonal mean results of \citet{RL2004} respectively. Widths for Venus and Titan were estimated respectively from the GCM results of \citet{Lebonnois2010} and \citet{Lebonnois2012T}, while for Jupiter we show the approximate width of the equatorial zone. These are largely consistent with the simple GCM and Held-Hou results except for Mars, whose annual mean Hadley circulation is rather broader than the Held-Hou results would suggest. However, for much of the Martian year the meridional circulation is dominated by a single large, cross-equatorial cell which extends over a wider range of latitude than at equinoxes.

The strength of the meridional Hadley circulation in the simulations is also shown in Fig. \ref{fig:Hadley1} as a function of $\Omega^\ast$ and $\mathcal{R}o_T$. This indicates a reduction in the strength of the circulation by a factor of $\sim 250$ between $\Omega^\ast = 1/16$ and 8. $|\Psi|$ follows a scaling close to $\mathcal{R}o_T^{2/5}$ for $\Omega^\ast \leq 2$ which steepens to a scaling closer to $\mathcal{R}o_T^{3/5}$ for $\Omega^\ast \geq 2$. This absolute magnitude of the mass streamfunction is of the same order as that found by \citet{kaspi2015}, although the strength of their Hadley circulations decreased somewhat less rapidly with $\Omega^\ast$. Both models show a much less steep decline in $|\Psi|$ with $\mathcal{R}o_T$ than the $\phi_H^3 \sim \mathcal{R}o_T^{3/2}$ indicated by \citet{caballero2008}. }

At higher rotation rates, the three-cell meridional circulation shrinks towards the equator and additional cells appear at high latitudes. These additional cells are essentially eddy-driven and are aligned with the positions of the multiple eddy-driven zonal jets that appear at middle and high latitudes, though retaining a thermally-direct circulation close to the equator itself in each hemisphere.



\subsection{Non-axisymmetric features --- eddies/waves}\label{sec:eddies}

For rapidly rotating planets like the Earth and Mars, the maintenance of the general circulation and the 
transport of heat and momentum depend significantly on the interactions between transient eddies and the 
mean flow. In the absence of eddies, the atmosphere of the mid- and high-latitudes would reach a state of
radiative equilibrium, which is characterised by larger meridional temperature contrasts than we observe 
here. Thermal winds induced by such strong temperature gradients would then lead to wind shears in the 
vertical direction which are baroclinically unstable. The baroclinic eddy transport of heat considerably
reduces the equator-to-pole temperature difference from that predicted by radiative equilibrium. In this
section, we will investigate the general trends of eddy activity with planetary rotation rate.





\begin{figure*}[!ht]
 \centering
 \subfloat{\includegraphics[width=0.45\textwidth,height=4cm,clip=true]{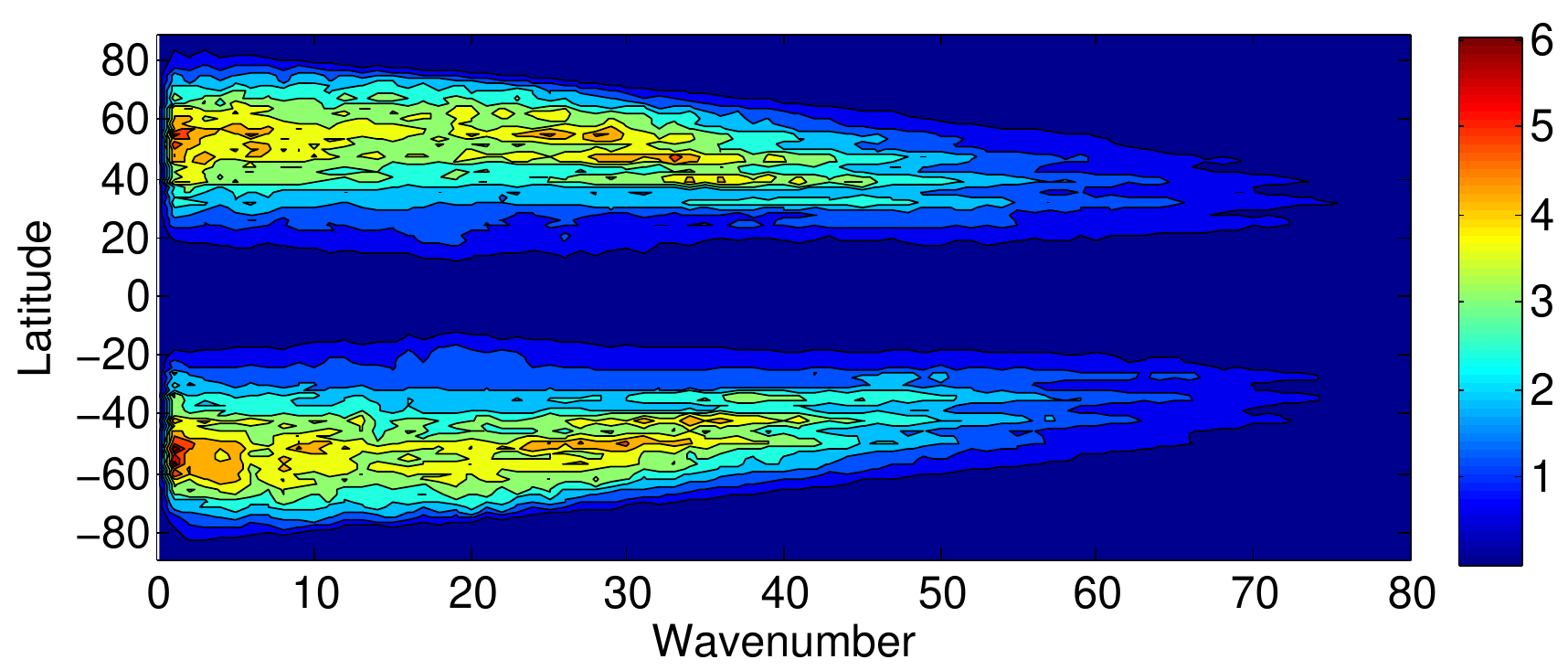}}
 \subfloat{\includegraphics[width=0.45\textwidth,height=4cm,clip=true]{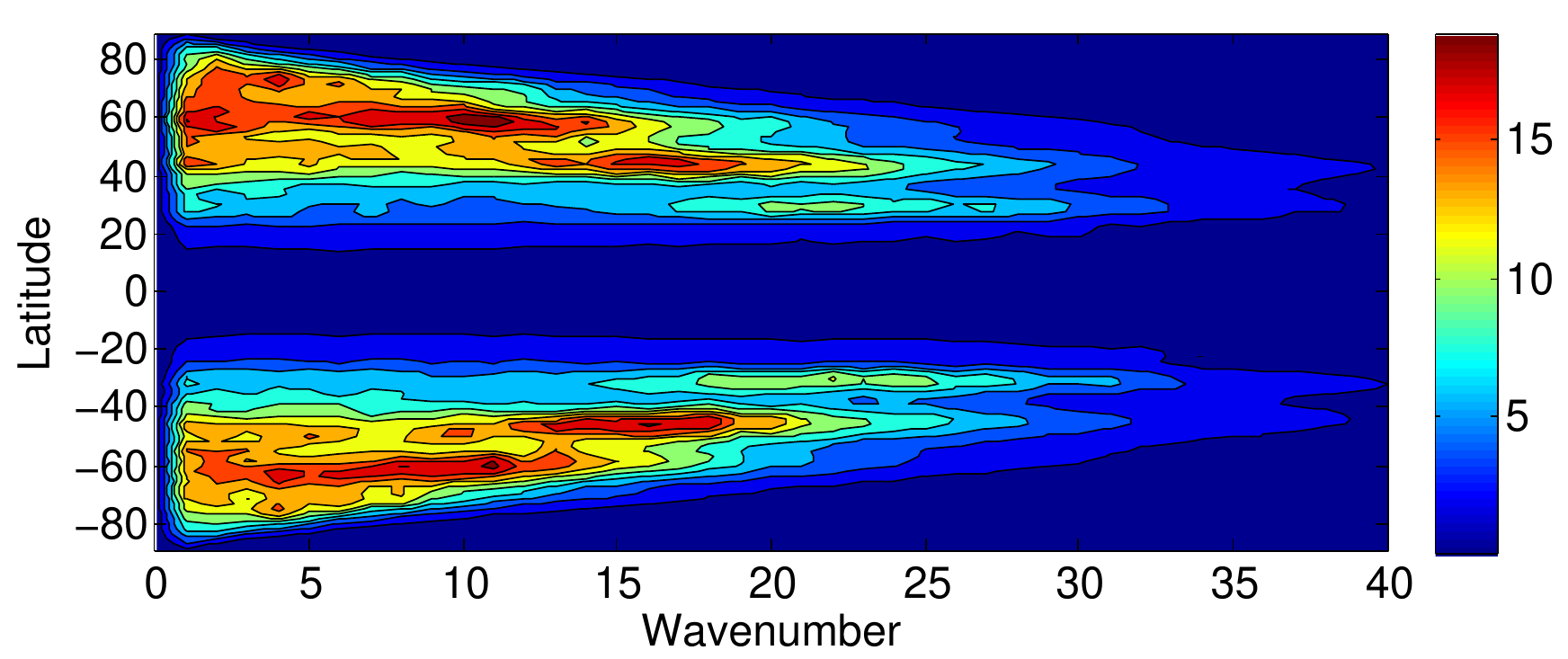}}\\
 \subfloat{\includegraphics[width=0.45\textwidth,height=4cm,clip=true]{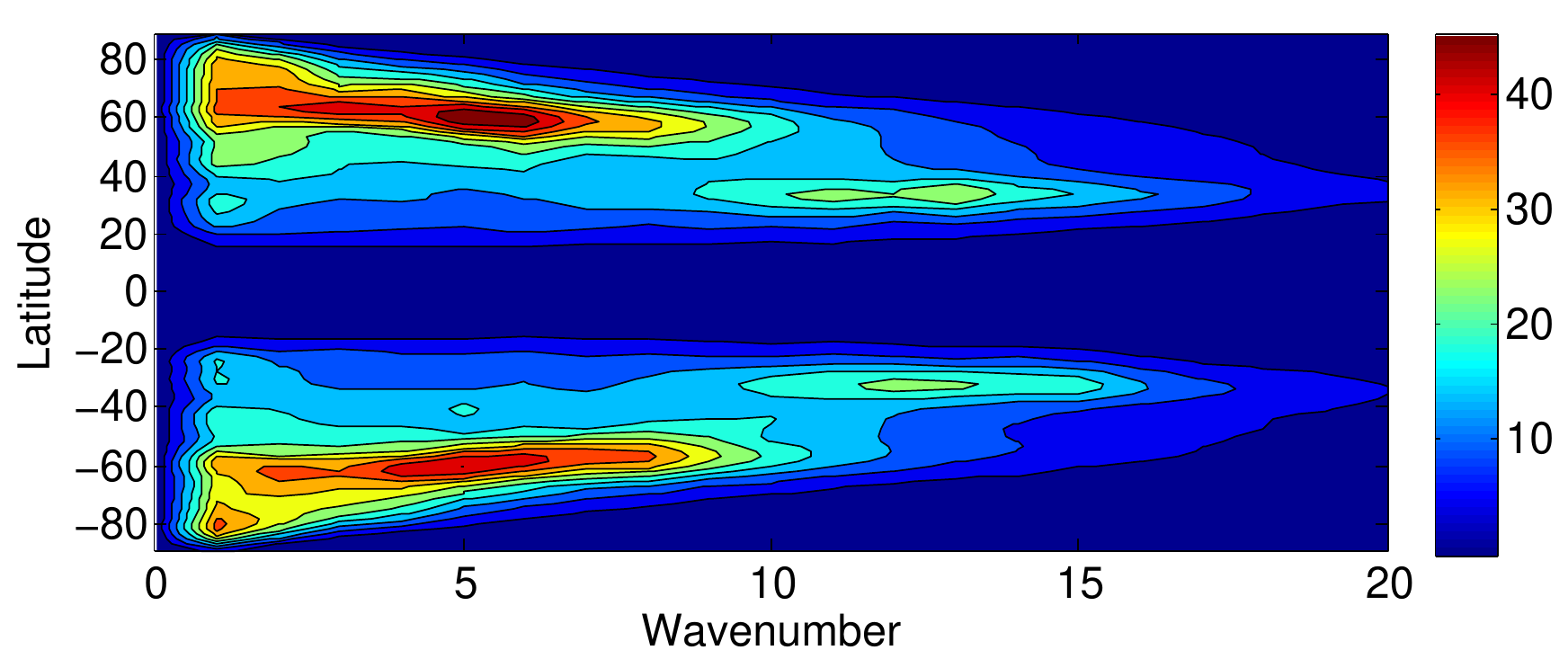}}
 \subfloat{\includegraphics[width=0.45\textwidth,height=4cm,clip=true]{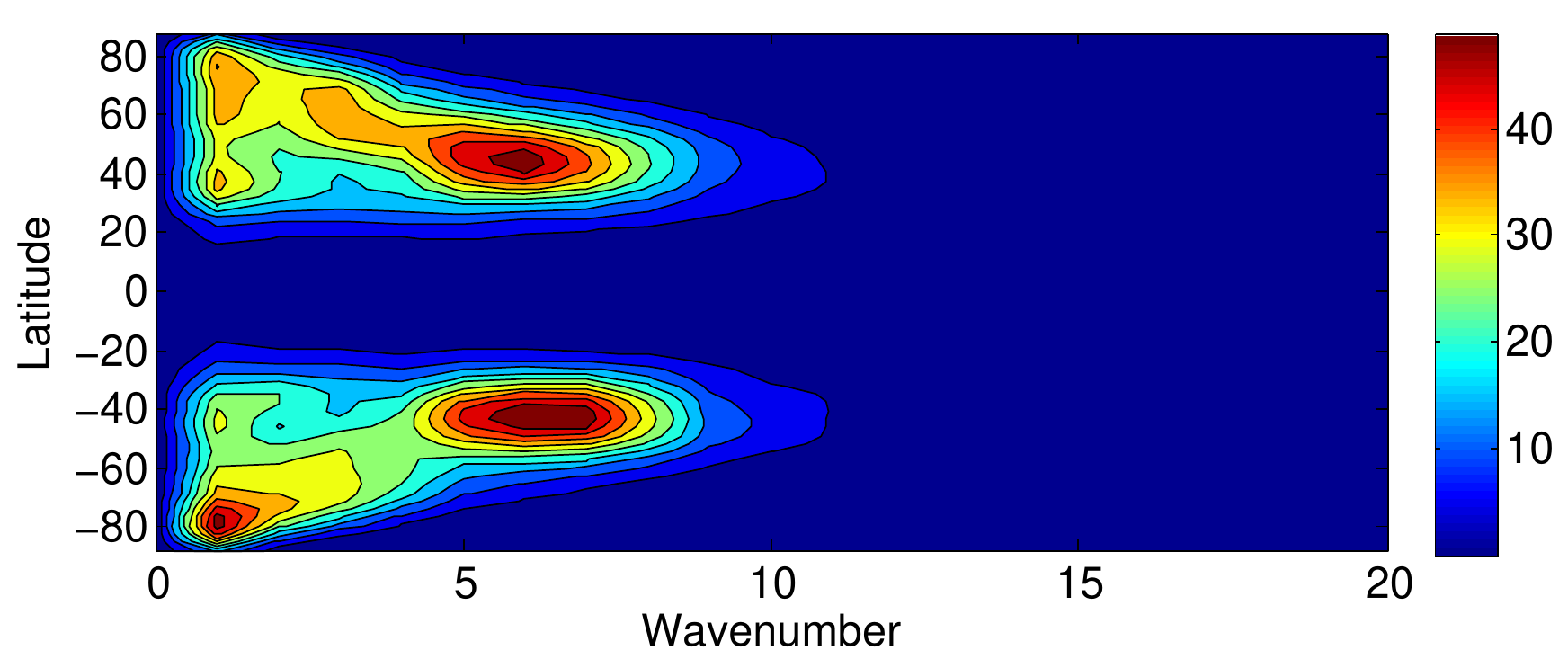}}\\
  \subfloat{\includegraphics[width=0.45\textwidth,height=4cm,clip=true]{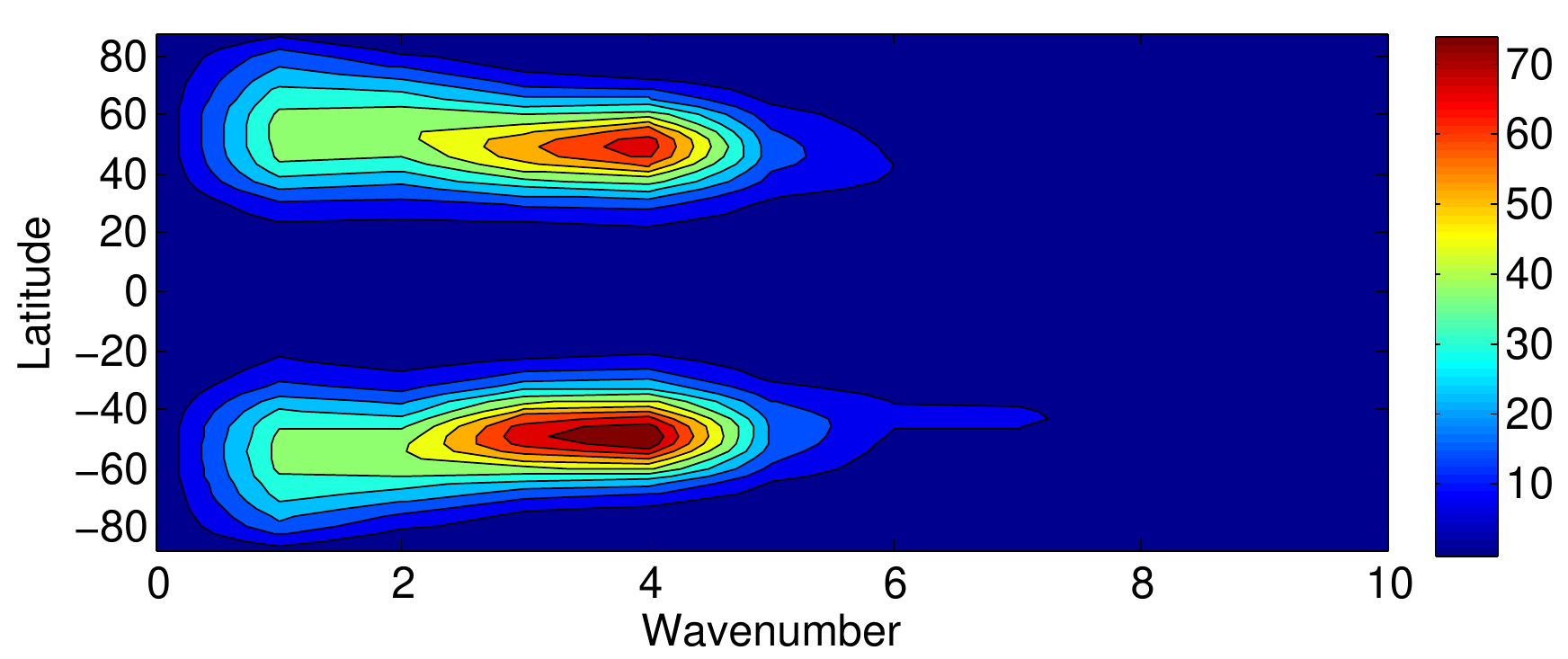}}
 \subfloat{\includegraphics[width=0.45\textwidth,height=4cm,clip=true]{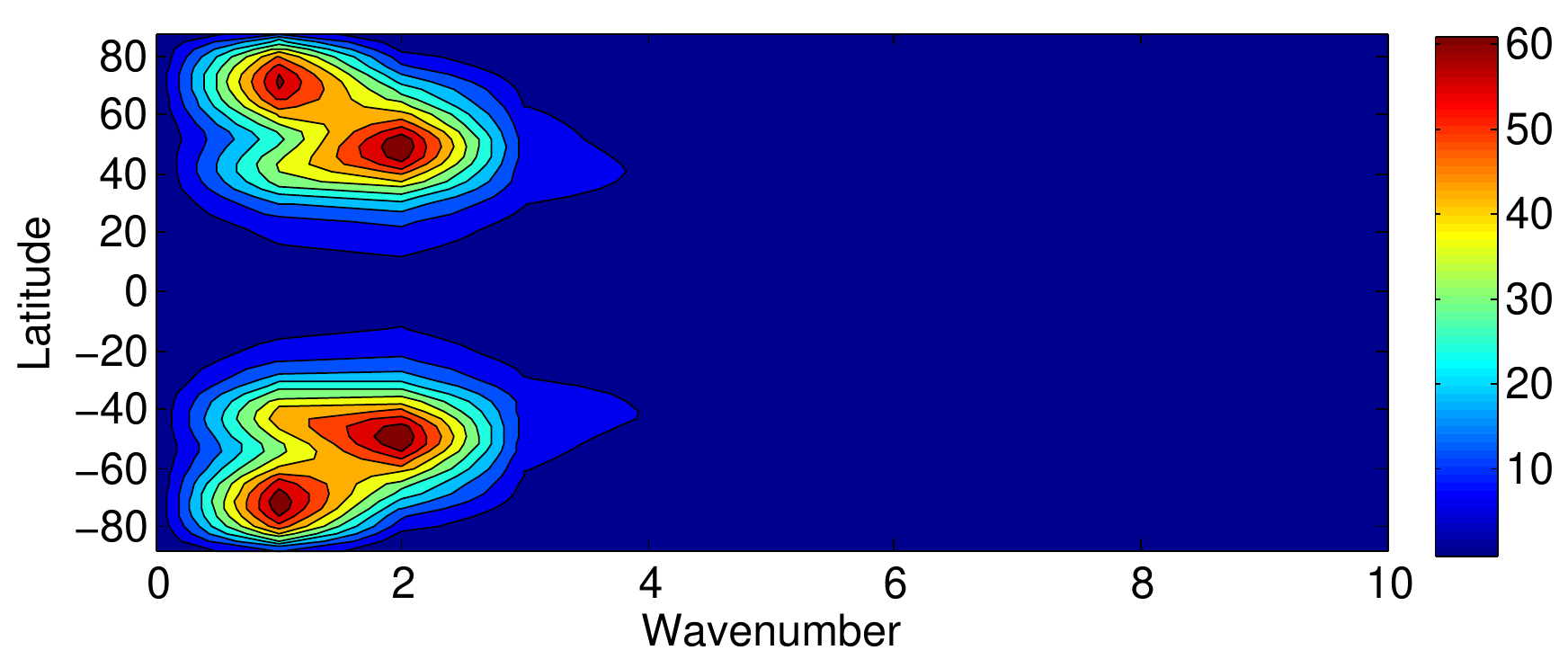}}\\
  \subfloat{\includegraphics[width=0.45\textwidth,height=4cm,clip=true]{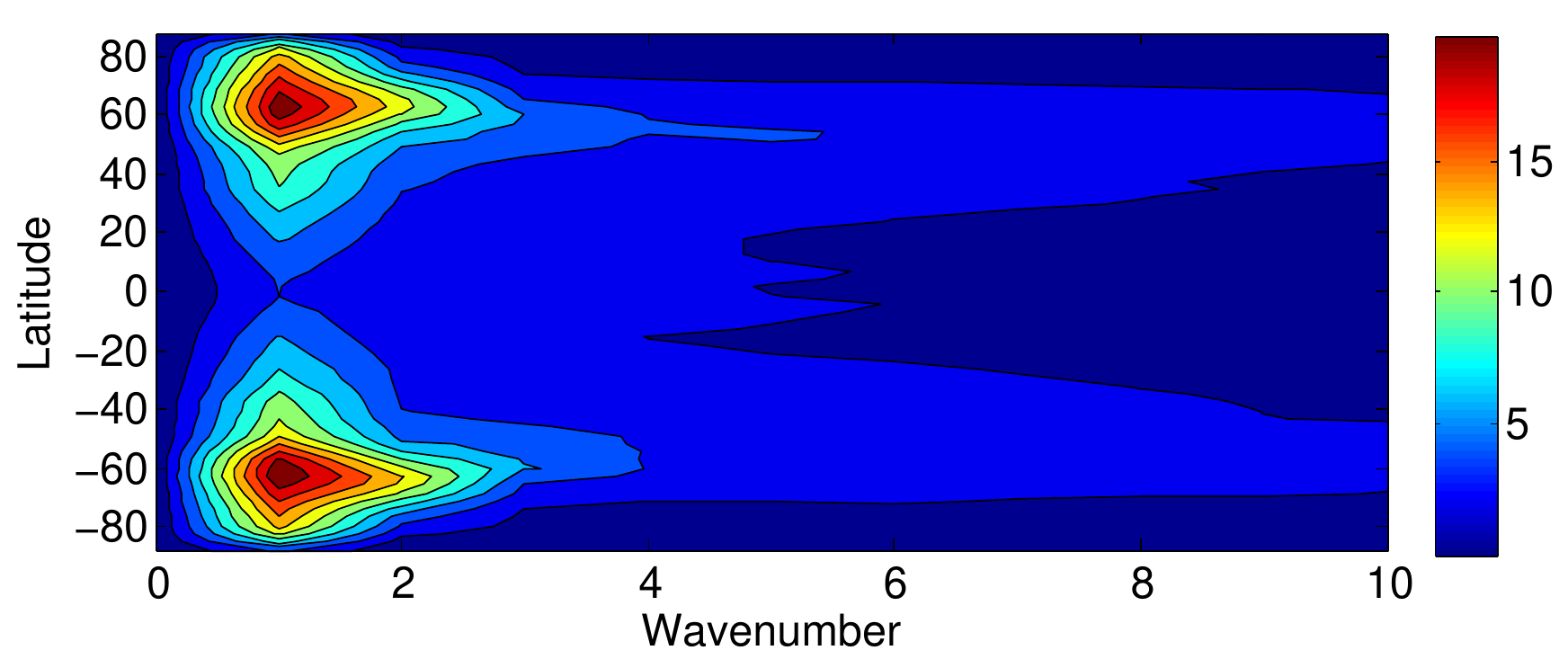}}
 \subfloat{\includegraphics[width=0.45\textwidth,height=4cm,clip=true]{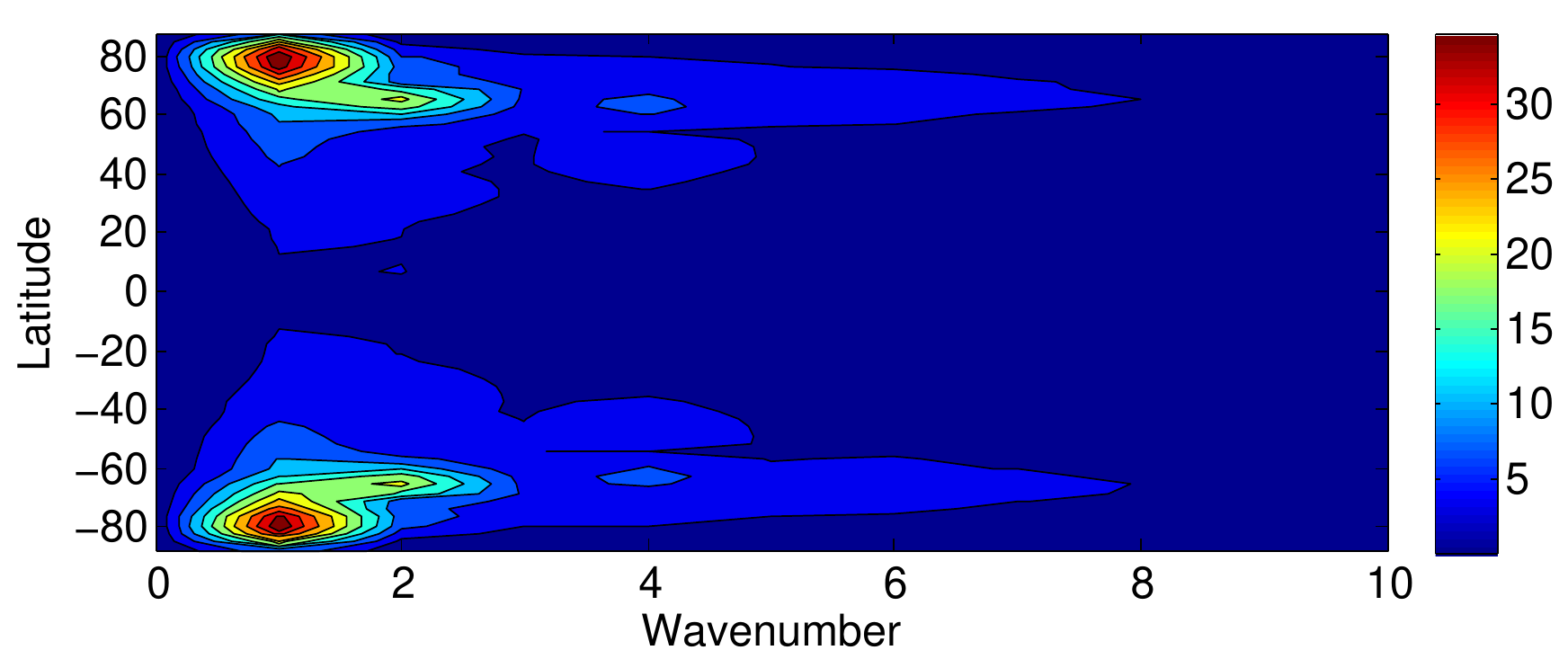}}\\
  \caption{Fourier amplitude maps of the zonal wavenumber spectra of geopotential height at 500 hPa from
 experiments at various rotation rates.: (a) $\Omega^{\ast} = 8$, (b) $\Omega^{\ast} = 4$, (c) $\Omega^{\ast} = 2$, (d) $\Omega^{\ast} = 1$, (e) $\Omega^{\ast} = 1/2$, (f) $\Omega^{\ast} = 1/4$, (g) $\Omega^{\ast} = 1/8$, $\Omega^{\ast} = 1/16$. Spectra are averaged over one model year. Unit: $\text{geopotential
 metres}$.}
 \label{fig:latfft_highrot}
\end{figure*}

Fig. \ref{fig:latfft_highrot} shows the zonal wavenumber amplitude spectra of geopotential height at the 500 hPa level as a function of latitude  from experiments with various rotation rates. As the rotation
rate decreases, the eddies become more and more confined to smaller wavenumbers, eventually converging towards $m = 1$ at the lowest values of $\Omega^\ast$. This is consistent with 
the trend with rotation of the energy-containing scales, as represented by the Rossby deformation radius and the Rhines scale. At high 
rotation rates (e.g. $\Omega^{\ast}=4$ and $8$), noticeable parallel banded structures can be seen in the 
extratropical latitudes, indicating the existence of multiple baroclinic zones, as would be expected if the
characteristic eddy length scale is significantly smaller than the planetary domain. The zonal wavenumber spectra themselves at these high rotation rates also peak at much higher wavenumbers. The Fourier amplitude 
is smaller for high rotation rate spectra as a result of the decreased eddy length scale, which leads to a 
smaller intensity of perturbations. 
The maximum strength of eddy activity is found at 
higher latitudes for experiments with lower rotation rates, which is again consistent with the trend that
Hadley cells tend to expand as $\Omega^\ast$ reduces, pushing extratropical jets (along with associated eddies) polewards as the rotation
rate decreases.


\subsection{Characteristic scales of eddies}\label{sec:turb-eddy-scale}

The preferred scale of non-axisymmetric eddies can be 
defined as the energy containing wavenumber $n_e$, based on the 
global barotropic eddy kinetic energy spectrum of the atmosphere (\cite{schneider2006}\footnote{According to 
the argument made in \cite{schneider2006}, using this $-1$ moment gives a better approximation of the observed
energy containing wavenumber than the conventional first moment, since the former is closer to the maximum of 
the KE spectra of Earth's atmosphere and their series of simulations.}):
\begin{equation}
 n_e(n_e+1) = \displaystyle\frac{\sum\limits_n E_n}{\sum\limits_n \big[n(n+1)\big]^{-1}E_n}, \label{eq:ne}
\end{equation}
in which $E_n$ is the barotropic eddy kinetic energy density at spherical harmonic wavenumber $n$, computed 
following \cite{Koshyk1999} {as a vertical, mass-weighted average} for the components with non-zero zonal wavenumber.


Linear theories of baroclinic instability (\cite{Eady1949}, \cite{Charney1947}) predict the characteristic 
eddy length scale to be the wavelength of the fastest growing unstable mode, which can be estimated by the 
first baroclinic Rossby deformation radius. But if nonlinear eddy-eddy interactions and the consequential inverse energy 
cascade are taken into account, the characteristic eddy length scale will be the scale at which the inverse 
energy cascade is inhibited\footnote{Most previous literature uses the word \lq\lq halted\rq\rq\ , which is 
argued against by \cite{Sukoriansky2007} on the grounds that some energy is transferred to lower wavenumbers, albeit anisotropically.
We follow the argument of 
\cite{Sukoriansky2007} and use
the word \lq\lq inhibited\rq\rq\ to indicate that an inverse energy cascade is not completely stopped at the Rhines scale.}.
This is shown to be characterised by the so-called Rhines scale 
(\cite{Rhines1975}), or the size of the whole domain if the Rhines scale exceeds the bound of the domain. 
It is therefore interesting to see how Rossby deformation and Rhines scales vary with the rotation rate in
our experiments and whether there is a scale separation due to the potential cascade of kinetic energy.

Following \cite{schneider2006}, the Rossby deformation radius is calculated as:
\begin{equation}
 L_D = c_D\frac{NH}{f_0}=c_D \frac{N_p\Delta p}{f_0},  \label{eq:LD}
\end{equation}
where $N_p^2 = -(\overline{\rho}_s \overline{\theta}_s)^{-1} \overline{\partial_p\theta}^s$ is the static
stability measure\footnote{Here $N_p$ is estimated using near-surface quantities following the treatment 
of \cite{schneider2006}. Using a vertically averaged value of $N_p$ does not change the result 
significantly, however.}, $\Delta p$ the pressure difference between tropopause and ground level, and $\rho_s$ is
the atmospheric density at the surface. Since it is the extratropical (baroclinic) Rossby deformation 
radius that is interesting to compare with, this deformation radius is averaged within the baroclinic 
zone whose boundaries are defined as the outermost latitudes at which the meridional heat flux
$\overline{v'\theta'}\cos{\phi}$ at the near surface level ($800\ \text{mb}$) exceeds $30\%$ of its maximum value.
The empirical constant $c_D$ is chosen to be $1.35$ following \cite{schneider2006}.

The Rhines scale is calculated as: {
\begin{equation}
 L_{R2} = \sqrt{2U/\beta}=\frac{\text{EKE}_{\text{bt}}^{1/4}}{\beta^{1/2}},   \label{eq:LR2}
\end{equation}
where $U$ is now} the rms flow speed, and $\text{EKE}_{\text{bt}}$ is the barotropic eddy kinetic energy density { (see Section \ref{sec:turb-eddy-scale} above).}
The corresponding Rossby and Rhines wavenumbers can be defined respectively as:
\begin{equation}
 n_D(n_D+1)=a^2/L_D^2,\ n_{R}(n_{R}+1) = a^2/L_{R}^2.  \label{eq:LDLR2}
\end{equation}
\noindent {This also leads to an alternative estimate of the number of jets, based on Eq \eqref{eq:nj} using $L_{R2}$ as defined in Eq \eqref{eq:LR2}, which equates to
\begin{equation}
N_{J2} = \dfrac{a}{4 L_{R2}} = \dfrac{n_R}{4}. \label{eq:NJ2}
\end{equation}
\noindent This estimate is also shown in Table \ref{tab.rhines} and is seen to differ from $N_{J1}$ by a factor $\sim 2$. $N_{J2}$ is generally larger than $N_{J1}$ and becomes greater than unity around $\Omega^\ast = 1/2$, but otherwise follows a similar trend to $N_{J1}$, at least for this set of simulations. $N_{J1}$ appears to provide a more realistic estimate of the number of jets in each hemisphere for $\Omega^\ast \geq 1$, however, similar to \citet{lee2005} although the latter is based on somewhat different physical assumptions.}

\begin{figure}
 \centering
 \includegraphics[width=0.9\columnwidth]{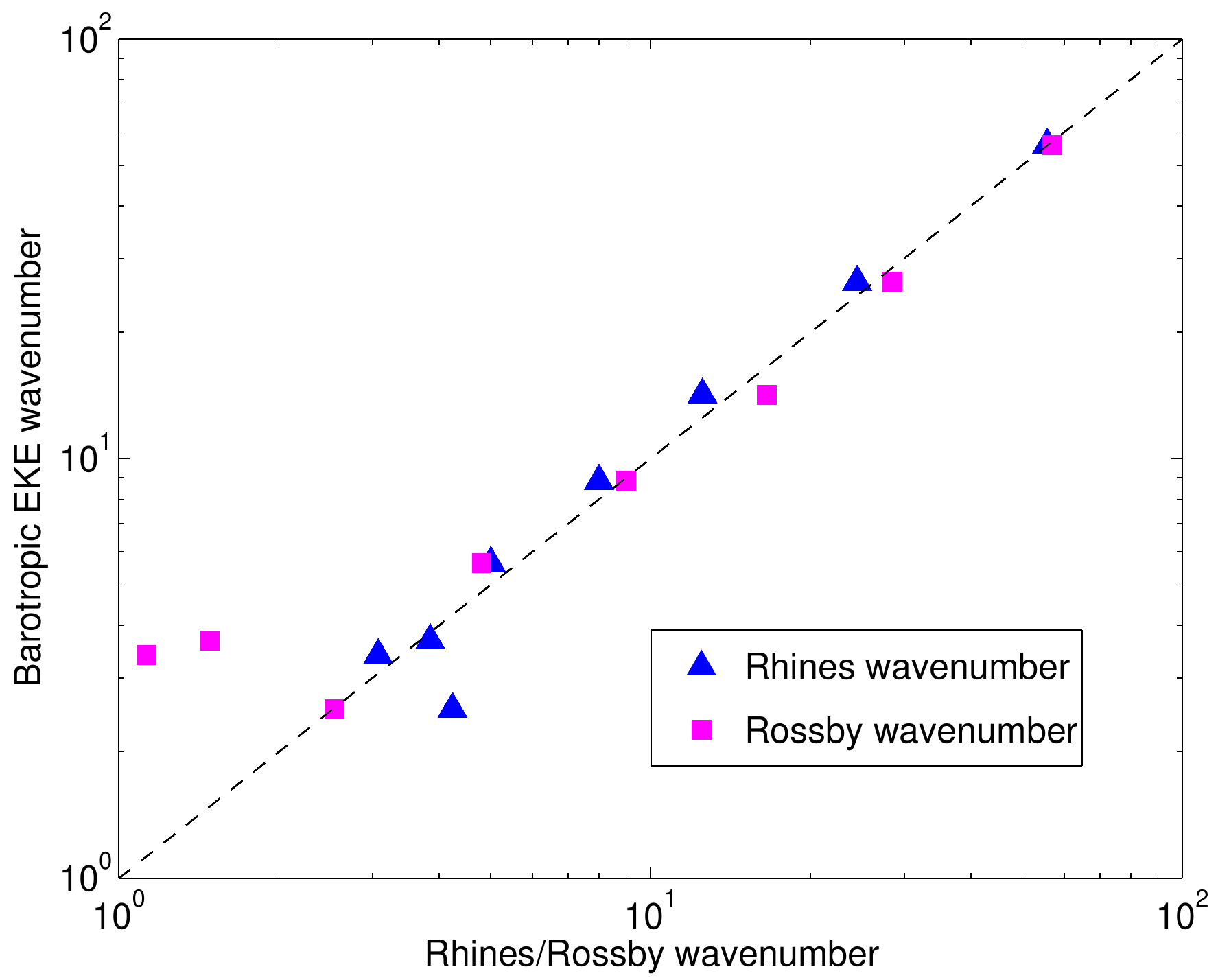}
 \caption{Energy-containing wavenumber of barotropic eddy kinetic energy ($n_e$) vs Rossby wavenumber ($n_{D}$) 
 and Rhines wavenumber ($n_{R}$) for experiments with $\Omega^{\ast}=1/16, 1/8, 1/4, 1/2, 1, 2, 4, 8$. $n_e=n_{D}$ or $n_R$ along the dashed line.}
 \label{fig:rossby_vs_ne}
\end{figure}

Fig. \ref{fig:rossby_vs_ne} show the comparison of Rossby wavenumber $n_{D}$, 
Rhines wavenumber $n_{R}$ and the energy-containing wavenumber $n_e$, as $\Omega^{\ast}$ is varied. 
It can be seen that these three wavenumbers are generally quite similar to each other in this set of simulations, though with $L_R > L_D$. This agrees with the results of \cite{schneider2006} in which they found that their simulated atmospheric 
macroturbulence tended to self-organise into a state of weak nonlinear eddy-eddy interactions with 
supercriticality $S_c\sim 1$, where supercriticality, $S_c$, is defined as 
\begin{equation}
S_c = \frac{f}{\beta H} \frac{\partial_y \overline{\theta}}{\partial_z \overline{\theta}}, \label{eq:super}
\end{equation}

\noindent a nondimensional measure of either the
slope of the isentropes near the ground \citep[see][]{schneider2006} {or the (squared) ratio between the Rossby deformation scale and the Rhines scale \citep[e.g. see][]{Held1996}}. As the Rhines wavenumber is close to the baroclinic 
Rossby wavenumber, this suggests that the inertial range for an inverse 
energy cascade via eddy-eddy interactions is likely to be largely inhibited, which is consistent with the fact that no 
well-defined $-5/3$ slope in the barotropic eddy kinetic energy spectrum is observed in our experiments (see \citet{read2017} for more discussion) {although this may be an oversimplification given the different dependence of the Rhines and Rossby lengthscales on latitude (see below).} 
According to \cite{Schneider2004a} and \cite{schneider2006}, such self-organisation is achieved through the 
adjustment of the extratropical thermal stratification by baroclinic eddies. This mode of equilibration is not allowed in the 
traditional quasi-geostrophic theories where the thermal stratification is taken to be fixed. This is consistent with the historical success of linear or weakly nonlinear theories of dynamical meteorology for 
extratropical regions. It should be mentioned, however, that some recent studies find that, under some circumstances, an Earth-like atmosphere or ocean can adjust into states with a supercriticality very different from 1 if the external forcings and planetary 
parameters are varied sufficiently (e.g. see \cite{Zurita-Gotor2008}, \cite{Jansen2012}). Thus, the results presented
here and in \cite{schneider2006} are valid probably only within a subset of the parameter space.

{ Because of the different dependences of $L_D$ and $L_R$ on latitude, the local supercriticality, $S_c(\phi)$, is also a strong function of latitude. This was explored recently by \citet{chemke2015a}, who noted that inverse energy cascades could be found over certain latitude ranges, depending upon planetary size and rotation rate. Figure \ref{fig:supercrit} shows the variation with latitude of the supercritical latitude, $\phi_S$, defined as the latitude at which the (latitude-dependent) Rhines and Rossby deformation length scales (determined according to Eqs \eqref{eq:LR2} and \eqref{eq:LD}) become equal in magnitude. This clearly shows a transition around $\Omega^\ast = 1$ between the slowly rotating regime, where $\phi_S \rightarrow 90^\circ$, and more rapidly rotating regimes where $\phi_S$ takes an intermediate value between $\sim 40^\circ$ and 20$^\circ$. Polewards of $\phi_S$, we find that $L_R > L_D$, allowing for the possibility of an inverse cascade of kinetic energy from a baroclinic injection scale around $L_D$ towards $L_R$. Equatorwards of $\phi_S$, however, $L_R < L_D$, suggesting that only forward transfers of KE are likely to occur. 

\begin{figure}
 \centering
 \includegraphics[width=\columnwidth]{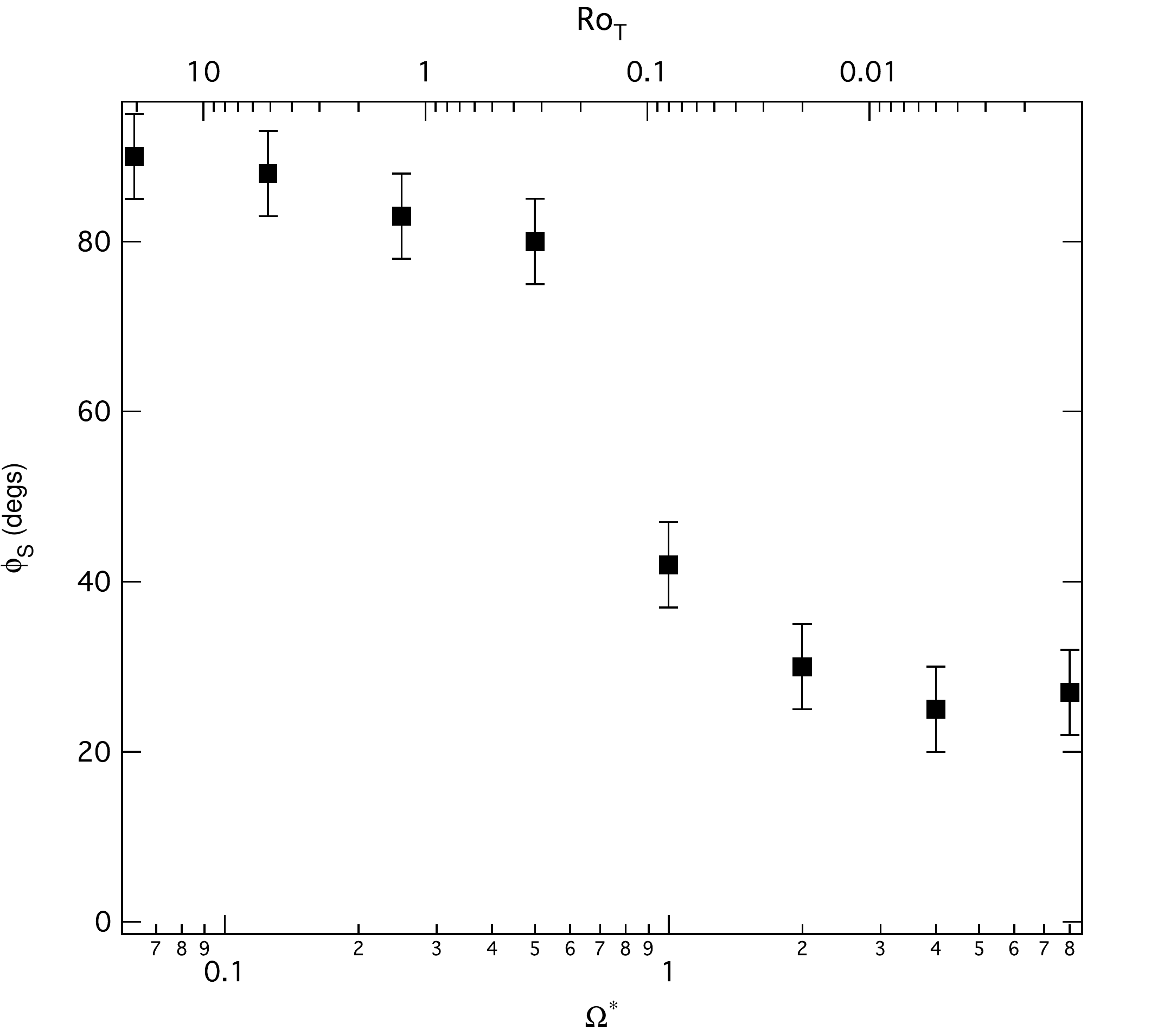}\\
 
 \caption{Variation with $|\Omega|^\ast$ and $\mathcal{R}o_T$ of the supercritical latitude, $\phi_S$, defined as the latitude where the Rhines and Rossby deformation lengthscales are equal.  }
 \label{fig:supercrit}
\end{figure}

This suggestion is further confirmed, at least for cases for which $\Omega^\ast \geq 1$, from a comparison between the Rhines and Rossby lengthscales (computed in the same way as by \citet{chemke2015a}) on the one hand and the zonal jet wavelength and kinetic energy weighted lengthscales on the other. Two examples of such a comparison for $\Omega^\ast = 4$ and 8 are shown in Figure \ref{fig:lat_scales}. The jet wavelength, $L_Z$, is shown as individual dots, representing the computed distance between successive eastward or westward extrema in $\overline{u}$, while the most energetic eddy lengthscale, $L_e$, is computed in a similar way to Eq \eqref{eq:ne} for $n_e$ but for zonal wavenumber $m$ only at each latitude point, thus,
\begin{equation}
L_e(\phi) = 2 \pi \cos \phi \left(\dfrac{\sum\limits_m E_K^{m}(\phi)/m^2}{\sum\limits_m E_K^{m}(\phi)}\right)^{1/2},
\end{equation}
\noindent where $E_K^m(\phi)$ is the eddy kinetic energy spectral density at zonal wavenumber $m$ and latitude $\phi$.

\begin{figure*}
 \centering
\subfloat{\includegraphics[width=0.45\textwidth]{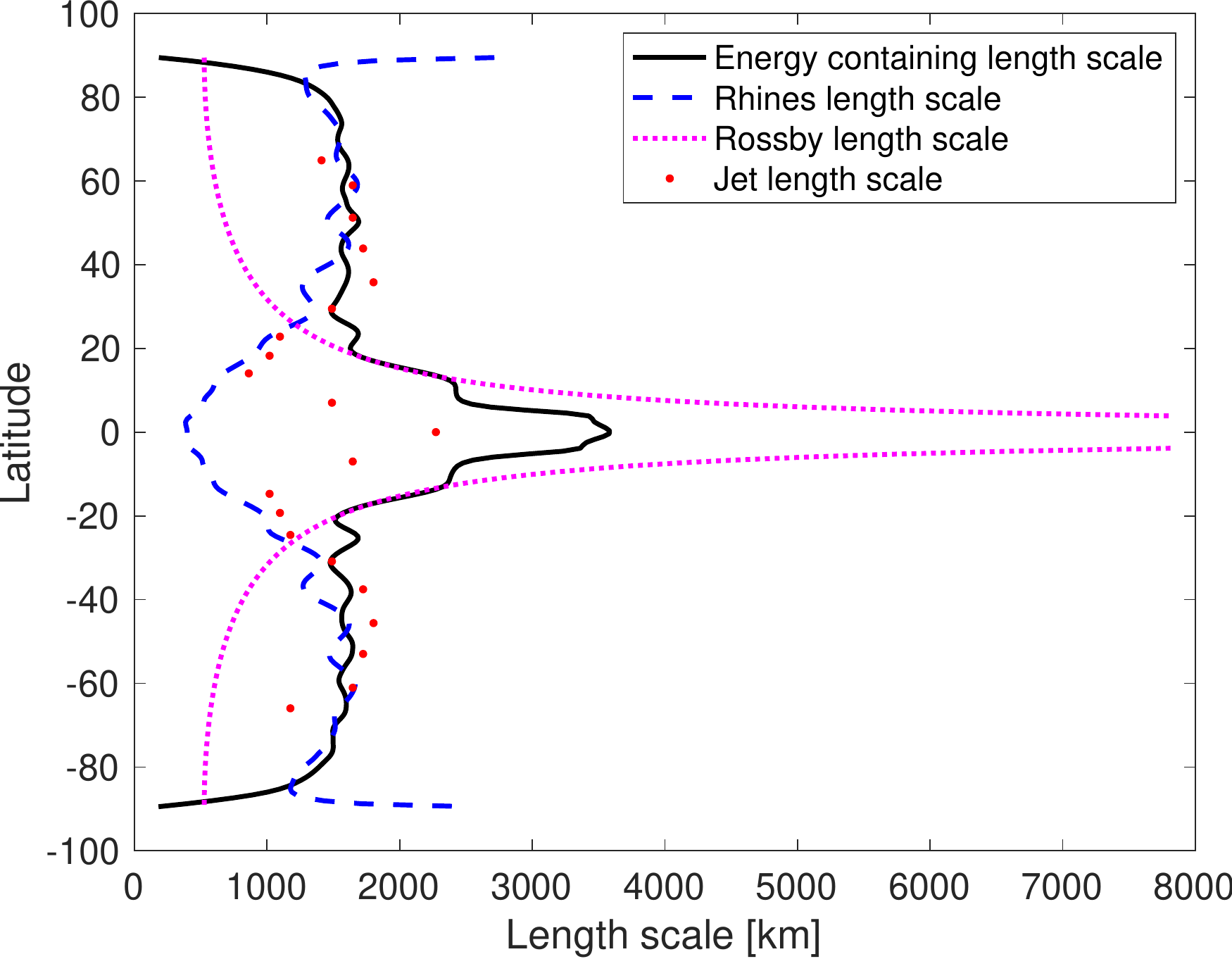}}
\subfloat{\includegraphics[width=0.45\textwidth]{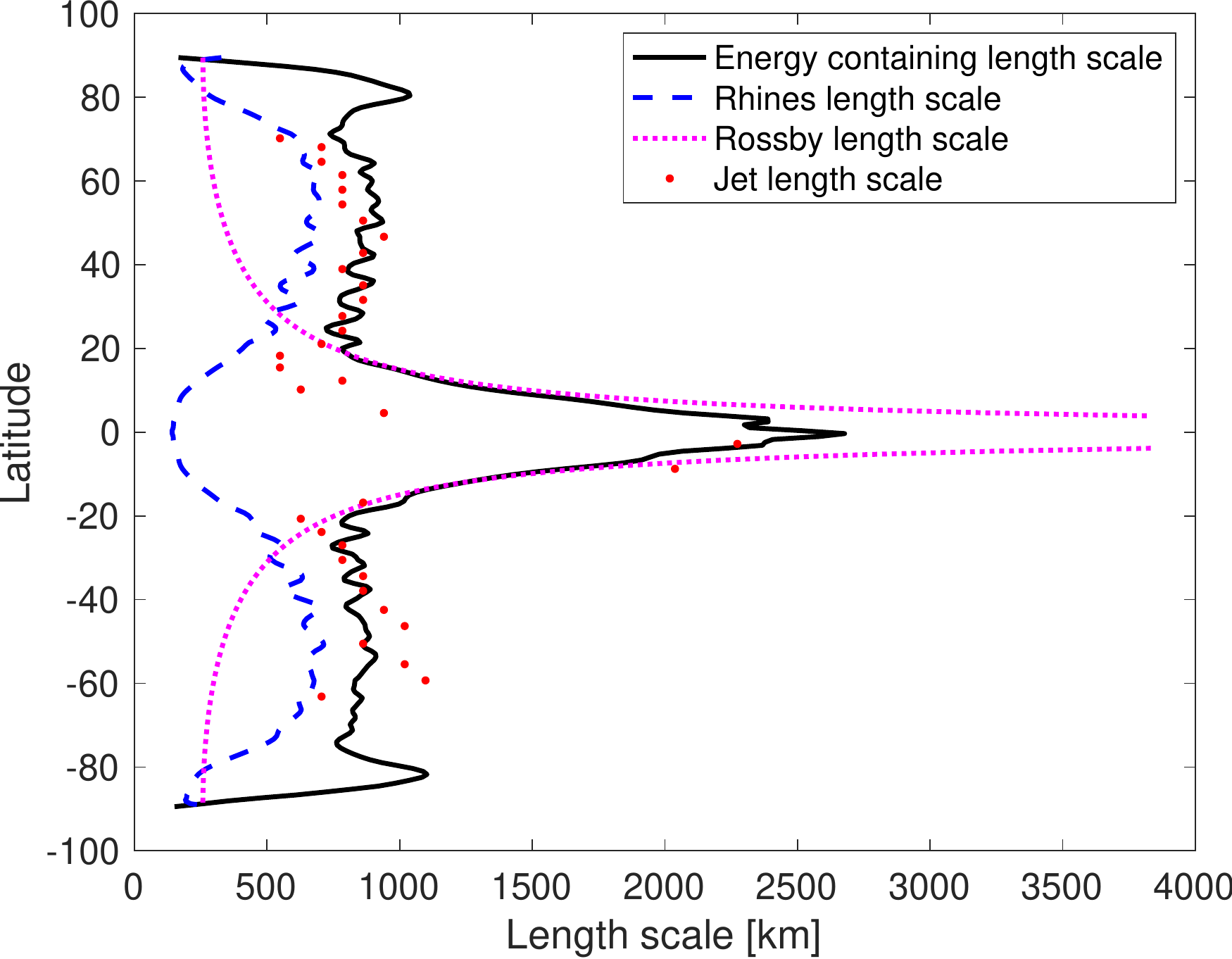}}\\
 (a) \hspace{8.5cm}(b)\\

\caption{Profiles showing the latitudinal variation of Rossby deformation and Rhines lengthscales, $L_D(\phi)$ and $L_R(phi)$, compared with the meridional wavelength of the multiple, eddy-driven zonal jets, $L_Z$, and the kinetic energy containing lengthscale, for (a) $\Omega^\ast = 4$ and (b) $\Omega^\ast = 8$. $L_e$. $L_D$ is shown as a dotted line while $L_R$ is shown as a dashed line. $L_e$ is denoted by the solid line and the zonal jet wavelength, $L_Z$, is indicated by individual dots. }
 \label{fig:lat_scales}
\end{figure*}

Fig. \ref{fig:lat_scales}(a) clearly shows both $L_Z$ and $L_e$ aligning closely with $L_R$ for $\phi > \phi_S$ (where $\phi_S \simeq 25^\circ$) for $\Omega^\ast = 4$, but with both $L_Z$ and $L_e$ diverging from $L_R$ for $\phi < \phi_S$ and following a trend that appears to parallel the behaviour of $L_D$. This behaviour is similar to what e.g. \citet{chai2014} and \citet{chemke2015a} found in their models, using different methods of radiative forcing, and suggests an important role for eddy-eddy interactions in cascading EKE to larger scales than $L_D$. A broadly similar trend is also found for $\Omega^\ast = 1$ and 2 (not shown), and for $\Omega^\ast = 8$ (see Fig. \ref{fig:lat_scales}(b)), although in this case $L_Z \simeq L_e > L_R$. This may result from the limited spatial resolution in the $\Omega^\ast = 8$ case although this should be investigated more closely in future work. At much lower rotation rates, $L_Z$ becomes ill defined (since there are no longer multiple eddy-driven jets) and $L_e$ roughly follows the trend of $L_R$, though this apparent correlation may be coincidental. Thus, the detailed links between $L_e$, $L_R$ and $L_D$ are more subtle and dependent on latitude and other parameters than the global behaviour shown in Fig. \ref{fig:rossby_vs_ne} might suggest.}

\subsection{Heat transfer}\label{sec:lorenz_heat}
The efficiency by which heat is transferred from the warm tropical regions of an Earth-like planetary atmosphere to the cooler mid- and high-latitudes is an important characteristic of the global atmospheric circulation. It largely determines how the climate varies from place to place, and in particular sets the mean temperature contrast between the tropics and polar regions. For the Earth, how this efficiency depends upon key parameters relating to the thermodynamic driving of the atmospheric circulation is important for understanding and quantifying the response of the climate system e.g. to changing amounts of greenhouse gases in the atmosphere. For other planets, heat transfer efficiency may affect their potential habitability and long term evolution, while also strongly influencing the strength of the meridional and zonal winds. The way in which horizontal heat transfer is partitioned between meridional overturning and eddies in the Earth's atmosphere has been well documented in both observations and numerical simulations for many years \citep[e.g.][]{Peixoto1974,James1995}. But for other planets in the Solar System, for which detailed observations are much sparser than for Earth, this remains a challenging question. 
Although latent heat plays an important role in the Earth's atmosphere, for a dry atmosphere (such as that of Mars or Venus) the primary form of heat energy is as sensible heat or dry static energy. The intensity by which heat energy is transferred within such a dry atmosphere depends strongly on the dynamical regime. 

From a cursory examination of the zonal mean temperature fields (see Fig. \ref{fig_ustrm_highrot_pumas}), 
the equator-pole temperature gradient becomes much smaller as the rotation rate decreases. This is especially prominent at
the middle and high levels of the atmosphere, where the isotherms become almost horizontal at low rotation
rates. As also discussed e.g. by \citep{kaspi2015}, this tendency at low rotation rates is due to the strong and extensive overturning Hadley cell in each hemisphere, which is highly efficient in smoothing out the horizontal temperature inhomogeneity. 
At higher rotation rates, however, baroclinic eddies likely play a more important role and the meridional temperature gradient is found to steepen. For $\Omega^\ast > 1$ the meridional temperature variation is seen to show staircase-like features, 
indicating the existence of multiple, zonally-parallel baroclinic zones and belts, at least superficially resembling the banded structure of the Jovian atmosphere. This is particularly apparent in the PUMA-S simulations 
for $\Omega^\ast = 2$ and 4, though with some hints of such effects even at Earth-like rotation rates.


The time-averaged meridional eddy sensible heat flux, which can be formulated as $[\overline{v^\ast T^\ast}]$, provides a measure of the intensity of meridional heat transfer by eddies. 
All of the PUMA-S simulations show that the atmosphere (especially at the lower levels) is dominated by poleward eddy heat 
transport. The strength of eddy heat flux reaches its maximum in the extratropical baroclinic regions, with 
multiple local maxima in latitude for experiments at the higher rotation rate (e.g. $\Omega^{\ast}=2,4,8$), which is
consistent with the existence of multiple baroclinic zones obtained in these experiments.

\begin{figure}
 \centering
 (a) \\
 \includegraphics[width=\columnwidth]{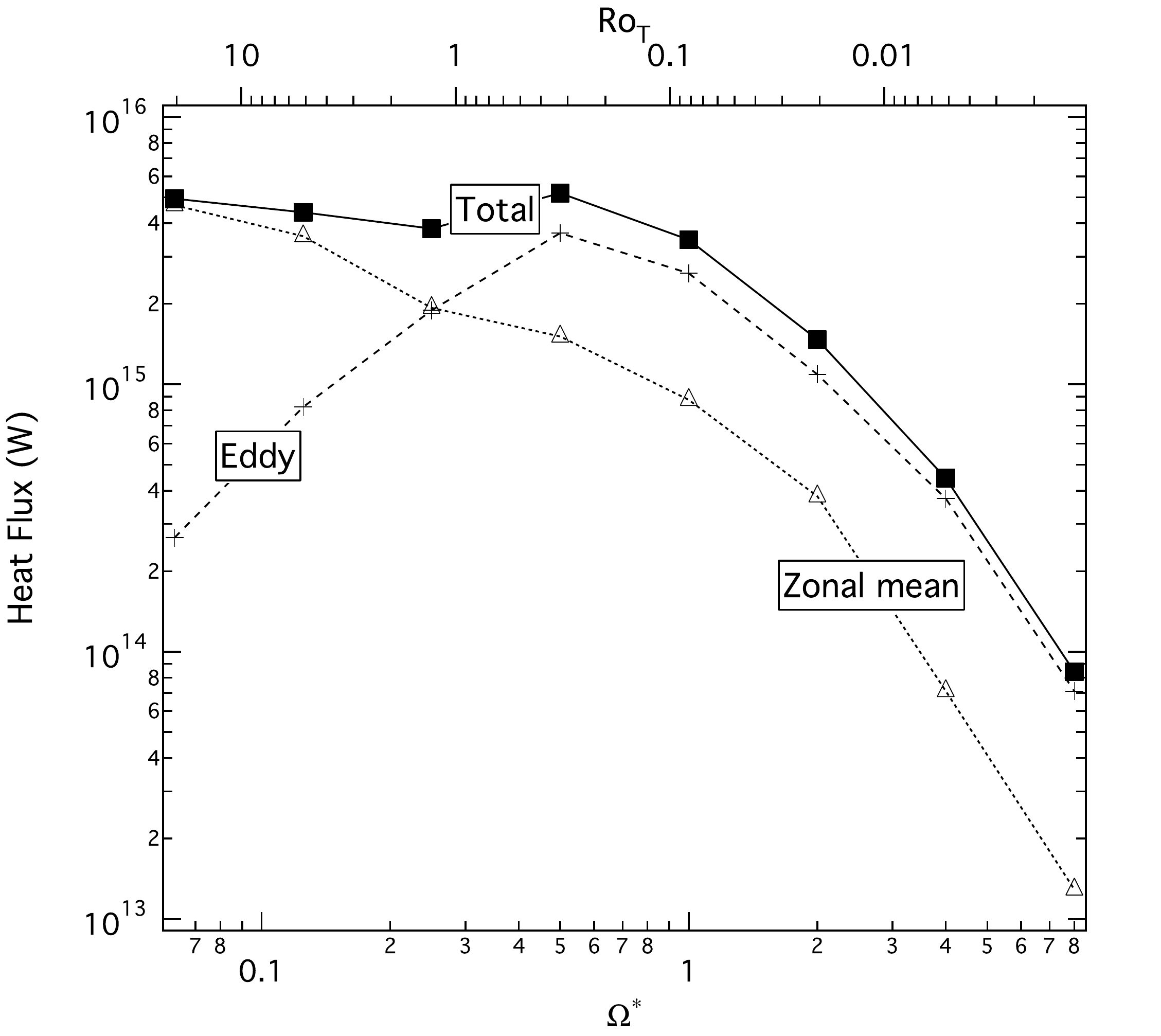}\\
 (b) \\
 \includegraphics[width=\columnwidth]{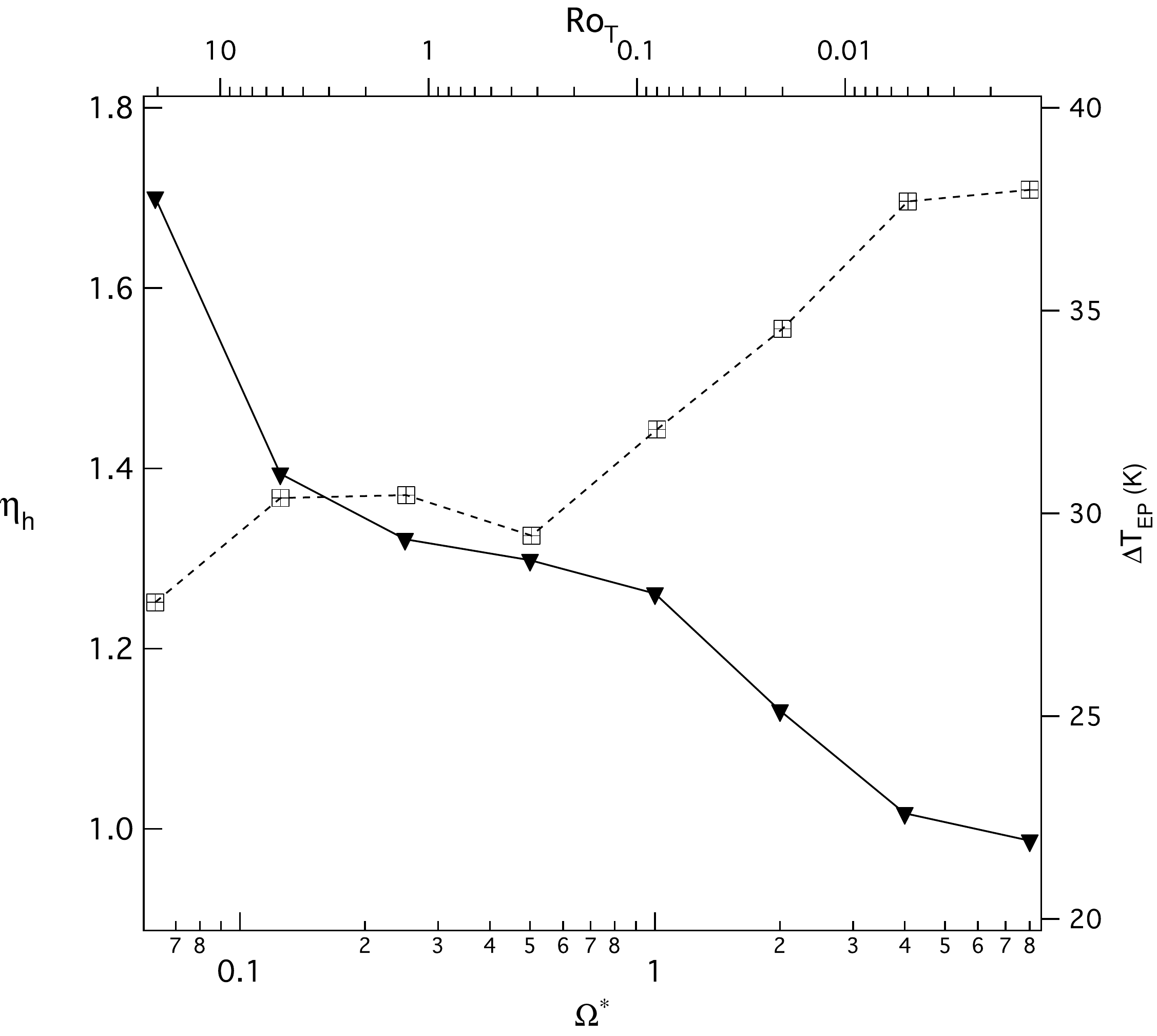}\\

 \caption{(a) Variations of peak values of meridional heat transport (in W), shown as total poleward transport (blue squares), zonal mean transport ($\int [\rho c_p \overline{v}\overline{T}] dp/g$) and eddy transport ($\int [\rho c_p \overline{v^\ast T^\ast}] dp/g$); (b) ratio of globally mass-weighted average isentropic slope in radiative-convective equilibrium (towards which the model atmosphere relaxes) to the slope obtained in fully active circulations in the latitude-height plane for experiments with different $\Omega^*$ {(shown as filled triangles connected by solid lines), plotted alongside the variation of mean equator-pole thermal contrast in the mid-troposphere, $\Delta T_{EP}$ (shown as open squares connected by a dotted line).} }
 \label{fig:pumas-theta-slope}
\end{figure}

$\overline{[v^\ast T^\ast]}$ is significant only in the extratropical regions (where the atmosphere is dominated by eddies and waves rather than quasi-axisymmetric circulation cells), and so the simple areal and mass-weighted average of $[v^\ast T^\ast]$ over the whole globe is not particularly informative. 
A clearer measure is provided by the peak value of the time-averaged heat transport, integrated (and mass-weighted) over the entire meridional plane \citep[cf][for example]{kaspi2015}. The sensitivity of the peak, vertically integrated meridional eddy heat flux $\int \rho c_p \overline{[v^\ast T^\ast]} dp/g$ to changes of rotation
rates is shown in Fig. \ref{fig:pumas-theta-slope}(a) (crosses and dashed lines), based on experiments with 
$\Delta T_{EP}=60K$ at various rotation rates, together with corresponding values of the peak zonal mean contribution $\int \rho c_p \overline{[v][T']} dp/g$ (triangles and dotted lines) and the sum of these representing the total (squares and solid lines). From these variations, it can be seen that, as rotation rate decreases from $\Omega^{\ast}=8$, the strength of the meridional eddy heat flux increases monotonically until it reaches its maximum at $\Omega^{\ast}\simeq1/2$, indicating the prominence of baroclinic eddies at this rotation rate. As the rotation rate decreases further below $\Omega^{\ast}=1/2$, the meridional eddy heat flux is seen to become less intense as the global circulation becomes more and more dominated by direct meridional overturning Hadley cells, which extend far into the high-latitudes at the slowest rotation rates. The strength of the Hadley heat transport is reflected in $\int \rho c_p \overline{[v][T']} dp/g$, which shows a monotonic decrease with $\Omega^\ast$ but evidently dominates the total heat transport at low rotation rates for which $\mathcal{R}o_T \gtrsim 1$. 

Variations in total heat transport (both meridional and vertical) lead to changes in both the equator-pole temperature contrast and the static stability, which also impacts upon the mean slope of the isotherms and isentropes within the simulated circulation. This suggests that the mean slope of the potential temperature isotherms actually reflects a measure of the total efficiency of heat transport (both horizontal and vertical) within the atmosphere. However, the slope angle by itself is not enough to quantify this efficiency since, even in pure radiative equilibrium, the isotherms would exhibit a non-zero slope because of the differential radiative heating with latitude. A more meaningful nondimensional measure of the atmospheric heat transport efficiency can therefore be defined with respect to the radiative-convective equilibrium state towards which the atmosphere tries to relax as the following:
\begin{equation}
 \eta_h = \frac{\langle\partial_y\theta_e/\partial_z\theta_e \rangle}{\langle\overline{\partial_y\theta/\partial_z\theta}\rangle}, 
\end{equation}
where $\theta$ is the potential temperature observed in the GCM results, $\theta_e$ the prescribed (equilibrium) potential temperature used by the Newtonian relaxation scheme and $\partial_z,\ \partial_y$ the vertical and meridional derivatives respectively. 
$\eta_h$ thus represents the mean ratio of the meridional slope of the isotherms relative to that of the restoration state (i.e. representing the ultimate baroclinicity of the atmosphere). The more efficient the dynamic heat transfer, the lower the realised isentrope slope. Higher values of $\eta_h$ therefore correspond to more efficient atmospheric heat transfer. This diagnostic thus provides a nondimensional measure of the efficiency of total dynamic heat transport, which redistributes heat, not only meridionally 
but also vertically. It should also be noted, however, that $\eta_h$ measures the total heat transport, by both the meridional mean circulation (e.g. Hadley cell) and the eddies. 

Fig. \ref{fig:pumas-theta-slope}(b) shows the comparison of $\eta_h$, based on the same simulations as in Fig. \ref{fig:pumas-theta-slope}(a).
It can be seen that, at the faster rotation rates, the redistribution of heat by dynamical processes tends to be inhibited, 
which includes both direct meridional overturning circulations (Hadley cells) as well as non-axisymmetric waves and 
eddies. 
Note that the downward trend is almost
arrested around $\Omega^{\ast}\simeq1/4-1$, which corresponds to where $\mathcal{R}o_T \simeq 1$. This indicates that in this parameter regime the eddy heat transport tends to compensate significantly for the decreasing efficiency of Hadley cell heat transport, thus preventing the total heat
transport from collapsing as rotation rate increases. This is similar in many respects to the findings of laboratory and numerical annulus experiments, where non-axisymmetric flows are found to sustain stronger levels of total heat transport well into the baroclinically unstable regime than would be sustained by pure axisymmetric flows as the rotation rate increases (\cite{hide1975,Read2003}). This compensating effect persists until the system enters the multiple eddy-driven jet regime, where meridional eddies are no longer able to propagate and transport heat energy uniformly from the tropics to the poles.

{A related trend is also apparent in the emergent thermal contrast between equator and poles, $\Delta T_{EP}$. This is also illustrated in Fig. \ref{fig:pumas-theta-slope}(b), indicated by the open squares connected by a dotted line, where $\Delta T_{EP}$ is defined as the temperature difference between equatorial and polar latitudes over the pressure range 600 hPa $\geq p \geq$ 400 hPa. This shows a nearly monotonically increasing trend with $\Omega^\ast$, though with an apparent pause between $\Omega^\ast = 0.2$ and 1. Such a pause is also consistent with a degree of baroclinic adjustment close to the onset of dominance of baroclinic instability over barotropic instabilties. }

\section{The regular baroclinic wave regime}\label{sec:reg-rg}
A regular baroclinic wave regime lies in the range $0.1<\mathcal{R}o_T<1.5$ within the non-axisymmetric region. 
Being favoured by intermediate rotation rates and moderate imposed meridional/radial temperature contrasts, it also resembles 
one of the major flow regimes observed in laboratory rotating annulus experiments (see e.g. \cite{hide1975}). Featuring baroclinic waves 
with one or two dominating wavenumbers and their harmonics, this regime is distinguished from irregular/chaotic
baroclinic wave regimes by its highly symmetrical regularity/predictability due to the presence of steady or (quasi-)periodic
vacillating waves. 

\begin{figure}
  \centering
\includegraphics[width=0.7\columnwidth,clip=true]{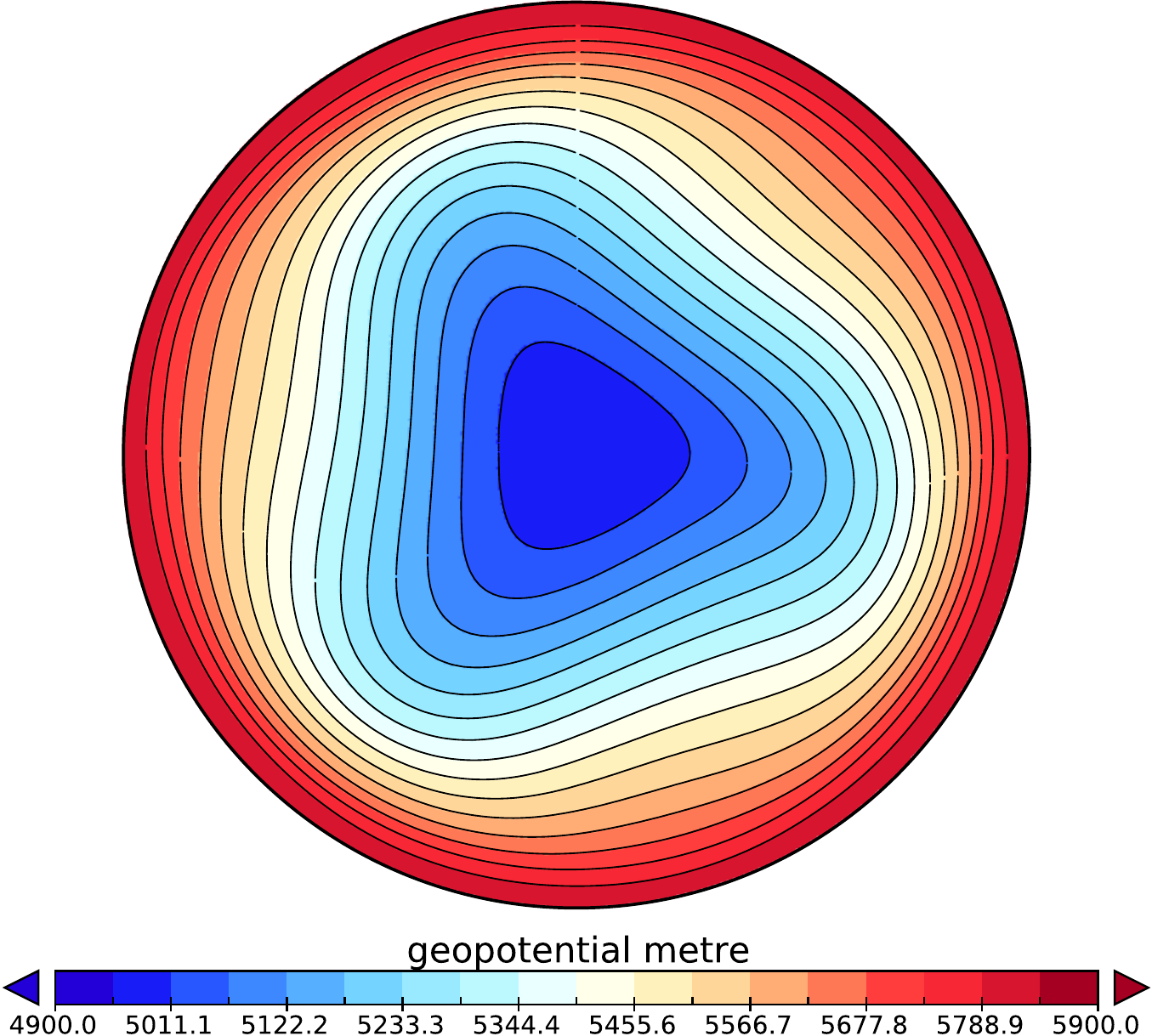}\\
 (a) \\
\includegraphics[width=0.7\columnwidth,clip=true]{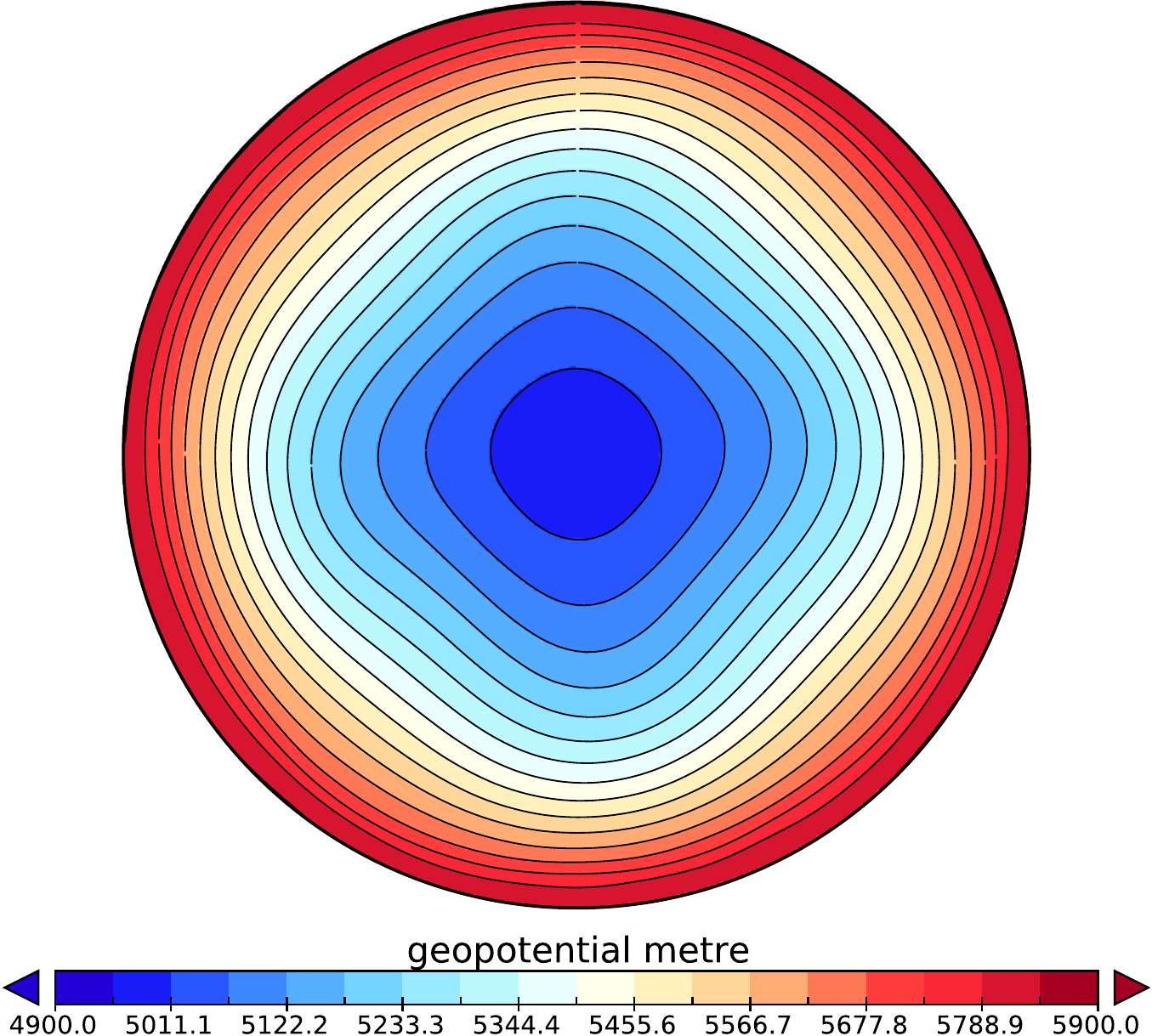} \\
 (b) 

 \caption{Snapshots of the geopotential height at the 500 hPa level viewed from the North Pole, showing (a) a wavenumber-3 
 regime 
 with $\tau_f=1.0$ day and $\tau_R=7.5$ days, and (b) a wavenumber-4 regime 
 with $\tau_f=2.0$ 
 days and $\tau_R=1.625$ days, for for experiments with a rotation rate of $1/2\Omega_E$. }
 \label{fig:wvn34_snapshot}
\end{figure}

Flows within the regular wave regime with intermediate strengths of frictional and thermal damping are typically characterised 
by one dominant wavenumber component and its harmonics.Two representative cases with dominant wavenumber 3 or 4 respectively are illustrated in Fig. \ref{fig:wvn34_snapshot}, 
which shows polar view snapshots of the geopotential height field in two experiments with a rotation 
rate of $\Omega_E/2$ and different values of the frictional and radiative damping parameters. 


In the early study by \cite{Collins1995}, a similar set of
experiments were performed across parameter space by varying the thermal and frictional damping timescales in a simplified GCM. The main 
difference between our work and \cite{Collins1995} is that parameters here were varied over a larger range in our 
experiments, and regimes dominated by either $m=2$ or $4$ were found in our experiments in addition to the $m=3$ regime noted by \cite{Collins1995}.
These circulations, with a single dominant zonal wavenumber and its harmonics, have a
latitudinally coherent structure as well over a characteristic range of latitudes corresponding to a mid-latitude `wave-guide'. Examples of this can be seen in Fig. \ref{fig:latfft_highrot}(e) and (f). 



The regular waves observed in most of our experiments are not actually perfectly steady waves that drift around the globe 
with a constant amplitude. Instead, their amplitudes undergo temporal modulations, making the waves more akin to 
the (regular or modulated) amplitude vacillation regime(s) that have been found in laboratory annulus experiments 
\citep{hide1975,Read1992,fruh1997,Young2008,read2015}. 
A Fourier spectral analysis on the timeseries of the amplitude of $m = 3$ in a simulation at $\Omega^\ast = 1/2$ revealed the time variation as a quasi-periodic modulation or modulated amplitude vacillation, with a 
well defined, dominant amplitude vacillation period of about 108 days, together with some weaker (and broader) frequency components on longer timescales. 

The trend of these regimes with respect to the strength of frictional and thermal damping is intuitive if we 
consider the growth and life cycle of baroclinic eddies. Baroclinic eddy growth tends to be suppressed under 
strong damping. 
In contrast, under weak damping the eddies can grow to larger amplitude, often leading to wave regimes with 
smaller dominant wavenumbers. With weak enough damping, more than one 
wavenumber component can grow to large amplitude and compete for dominance through wave-zonal flow interactions
(see e.g. \cite{Hart1981}, \cite{Appleby1988}), leading to a mixed 
wavenumber (or wavenumber vacillation) regime with quasi-periodic or chaotic switching of the dominant wavenumbers among 2 or 3 different wavenumber components.



\section{Discussion} \label{sec:discussion}
In this paper we have explored the range of different styles of circulation exhibited by a prototypical Earth-like planetary atmosphere over a wide range of parameters. In common with several other recent studies \citep[e.g.][]{Geisler,Williams1988a,Williams1988b,Mitchell2010,Read2011,kaspi2015}, the organisation of the circulation was shown to depend strongly on factors such as the planetary rotation rate, radius and strength of the latitudinal thermal contrast, ranging from planet-encompassing Hadley-like meridional overturning circulations with strongly super-rotating zonal winds to patterns of multiple, parallel zonal jet streams with multi-cellular meridional overturning. At intermediate parameter values, the circulation exhibits just one or two baroclinically unstable zonal jets in each hemisphere with patterns of zonally travelling waves, with a global circulation that resembles that of the Earth or Mars. 

\subsection{Dynamical similarity}
Although some of these broad trends in circulation pattern with factors such as planetary size or rotation rate have been noted previously, rather few authors have investigated the possible dependence of the observed type of circulation regime on a small number of dimensionless combinations of planetary parameters, notably the so-called thermal Rossby number, $\mathcal{R}o_T$, and various measures of mechanical and radiative damping (here characterized by various forms of Taylor number). 

An early study of such forms of dynamical similarity in the context of simple GCMs was carried out by \citet{Geisler} in a numerical model with Newtonian relaxation thermal forcing and a simple viscous representation of mechanical friction. This study identified the existence of both the regular and irregular baroclinic wave regimes also found in the present work, as well as the existence of an axisymmetric regime when viscous dissipation was sufficiently strong to damp out baroclinic instability. \citep{Geisler} also attempted to compile a rudimentary regime diagram, similar to the form presented here though for a much more restricted range of parameters. This was further explored more recently by \citet{Mitchell2010} and \citet{pinto2014} for slowly rotating planets, and more generally in the synthesis/review by \citet{Read2011}, based on the earlier model simulations by \citet{Williams1988a,Williams1988b}. But the more complete regime diagram we have presented here in Fig. \ref{fig:regdiag} gives a much more comprehensive picture as to how the form and style of the circulation regime depends upon key dimensionless parameters across the full range of possibilities. At least over the range of parameters we have been able to explore in the present study, it seems clear that the dominant dimensionless parameter determining the gross properties of the circulation regime is $\mathcal{R}o_T$, with other parameters playing a more subsidiary role, although the damping rate as measured by the radiative and frictional Taylor numbers can evidently overwhelm the effect of rotation and buoyancy forcing if it is large enough. 

The dependence of circulation regime on these dimensionless groups suggests the possibility of real dynamical similarity between the circulations on very different sizes of planet provided the rotation rate and other factors are adjusted to obtain similar values of $\mathcal{R}o_T$ and other parameters. Although not explored in detail here, this was shown graphically for slowly rotating regimes in the simple GCM study by \citep{pinto2014} (cf their Figs 4(e) and 7(a)), in which a similar circulation and wind pattern was obtained for both an Earth-sized planet rotating at $\Omega^{\ast} = 1/20$ and a planet rotating at $\Omega^{\ast} = 1$ with radius $a^{\ast} = 1/20$. However, this close similarity was only obtained if not only $\mathcal{R}o_T$ was matched but also the radiative and friction parameters too. Such similarity over much wider ranges of parameters, and for different circulation regimes, is also implicit in a comparison between the results of our simulations and those obtained in other parameter sweeping studies, such as those of \citet{Williams1988a,Williams1988b} and \citet{kaspi2015}, who used rather different parameterizations of thermal forcing and dissipation. Despite these differences between models, a very similar sequence of circulation regime with varying rotation rate was obtained in each case, indicating a fundamental dynamical similarity which is at least semi-quantitative.

\subsection{Heat transport}
The efficiency of heat transport across the planet is a key property that is well known to have an important influence on planetary climate. In common with other studies, we find that the planetary rotation rate plays an important role in determining how effectively heat can be transferred from the tropics, which experience net heating from the Sun, to higher latitudes, where net cooling is dominant. In Section \ref{sec:lorenz_heat} we confirmed that, at low rotation rates where the direct overturning Hadley circulation extends to high latitudes, horizontal advective heat transfer is very efficient, allowing potential temperature to be redistributed evenly across the planet, markedly reducing the equator-pole thermal contrast and almost eliminating the atmospheric baroclinicity by flattening isentropes across the planet. At higher rotation rates, however, Coriolis effects inhibit such direct overturning circulations and prevent such efficient transport reaching much beyond the tropics themselves. As a result, the atmosphere equilibrates to a more baroclinic state with an equator-pole thermal contrast that approaches the radiative-convective equilibrium at the highest values of $\Omega^\ast$. 

Quantification of this trend in dimensionless terms is not straightforward, however. It is conventional in dynamical studies of convection to quantify efficiency of heat transfer by the Nusselt or P\'eclet numbers, which compare the total or advective heat flow to that obtained purely by molecular conduction. But such a comparison is not appropriate or meaningful in the context of an atmosphere since, in the absence of any motion of the air, heat energy is principally transferred by radiative transfer. Since such an energy flux by radiation is set by the irradiance received from the Sun, its effectiveness is manifest in the horizontal temperature contrast obtained in radiative-convective equilibrium. In Section \ref{sec:lorenz_heat}, therefore, we quantify the effectiveness of advective heat transfer by computing the ratio between the actual slope of zonally-averaged potential temperature surfaces and the slope in radiative-convective equilibrium. This dimensionless measure clearly showed the extent to which advection reduces the slope of the isentropes {and the equator-pole thermal constrast}, especially at low rotation, but becomes much less effective at reducing this slope {and $\Delta T_{EP}$} as $\Omega^\ast \rightarrow \infty$ or $\mathcal{R}o_T \rightarrow 0$. At intermediate values of $\Omega^\ast$ around $\mathcal{R}o_T \sim 1$, near the onset of deep baroclinic instabilities, the monotonic increase of isotherm slope became inhibited. This suggests that baroclinic instability was able to maintain a higher overall efficiency of heat transport (in combination with direct meridional overturning) than would be obtained if this instability did not occur. 

Such an effect is well established in laboratory experiments on rotating, stratified flows \citep[e.g.][]{hide1975,Read2003,perez2010}, where the Nusselt number is found to remain nearly constant over a wide range of $\Omega$ until the equivalent of $\mathcal{R}o_T$ is  $<<1$. This would appear to be a form of ``baroclinic adjustment'' \citep[e.g.][]{stone1978} that reflects the tendency of baroclinic instability to alter its intensity as the zonally symmetric overturning circulation reduces in strength with increasing $\Omega$. The result is to maintain the total heat transport in the system to a more or less constant value until conditions are reached (at high values of $\Omega$) where baroclinic instability itself becomes more latitudinally intermittent and less efficient at transporting heat, which typically occurs in the laboratory when baroclinic waves become irregular and more turbulent. The effect is therefore largely confined to conditions not too far from the first onset of strong baroclinic instability.

In atmospheric models, ``baroclinic adjustment''  has been discussed in a number of contexts relating (a) to the possibility of self-organized criticality by which baroclinic instability equilibrates to adjust the structure of the atmosphere to remain close to a state of marginal instability \citep{stone1978} or (b) to the tendency of baroclinic instability to modify the stratification and height of the tropopause as it equilibrates and thereby hold the atmosphere close to a state for which the supercriticality parameter, $S_c$ (cf Eq (\ref{eq:super})), 
remains close to a value O(1) \citep[e.g.][]{schneider2006,Zurita-Gotor2007}. The former is commonly considered an oversimplification, because there is no critical point with decreasing $\Omega^\ast$ at which baroclinic instability in an atmosphere is completely suppressed. But the latter interpretation presumes that the slope of the isotherm slope tends to adopt a particular value, which the present work might suggest is only applicable to planets under conditions not too far from the present Earth. More rapidly or slowly rotating planets may be in quite a different regime, for which such a concept of ``baroclinic adjustment'' may be much less appropriate.

\subsection{Laboratory analogues}
The sequence of circulation regimes found in our study as a function of $\mathcal{R}o_T$ and frictional Taylor number bears a strong resemblance to regime diagrams obtained in laboratory experiments on rotating, stratified flows in rotating annulus experiments \citep[e.g.][]{hide1975,read2015}, at least for the transition between the regular, coherent baroclinic wave regime and the more irregular, chaotic Earth-like circulation regime. The parallels extend also to the existence of a lower axisymmetric regime at small Taylor number in both systems, at least for laboratory experiments using fluids with a large Prandtl number \citep[e.g.][]{hide1975}. Significant differences between the ``classical'' rotating annulus experiments (with strictly horizontal endwall boundaries) and the GCM simulations obtained here and elsewhere are found both at very low rotation rates (high values of $\mathcal{R}o_T$) and at very high rotation rates (low values of $\mathcal{R}o_T$). At very small values of $\mathcal{R}o_T$, planetary vorticity gradients in a spherical atmosphere dominate in setting the most energetic length scales and lead to the break-up of the flow into multiple parallel baroclinic zones and zonal jets which grow in strength to dominate over the small-scale eddies. In the laboratory, however, the decreasing Rossby deformation radius leads to increasingly turbulent eddy-dominated flows unless strongly sloping topography is present along the upper and/or lower boundary. But laboratory experiments with conically sloping upper and/or lower endwall boundaries do show a similar tendency to the GCM simulations in producing zonally dominated flows and multiple baroclinic zones and eddy-driven zonal jets \citep[e.g.][]{mason1975,bastin1998,Wordsworth2008}. At low rotation rates, however, laboratory experiments exhibit an upper axisymmetric limit to baroclinically unstable flows for $\mathcal{R}o_T >$ some limiting value O(1), in contrast to the GCM simulations that develop ever smaller and more intense circumpolar vortices that become barotropically unstable.  One might speculate that this difference is due to the presence of the inner cylinder in the rotating annulus experiments, that limits how small the ``circumpolar vortex'' in such experiments can become, thereby suppressing the barotropically unstable super-rotating regime at large values of $\mathcal{R}o_T$. However, it would be of interest to explore other configurations in the laboratory without a solid inner cylinder to determine whether a barotropically unstable vortex develops close to the rotation axis. 

{The trends in zonally symmetric and eddy heat transport, discussed in Section \ref{sec:lorenz_heat}, also show some similarities with what has been found in the laboratory \citep[e.g. see][and references therein]{hide1975,Read2003,read2015}. For such laboratory systems, the heat transport due to the axisymmetric overturning circulation is found to decrease as $\Omega$ increases, with Nusselt number scaling as 
$\Omega^{-3/2}$ associated with the combined effects of $\Omega$ on thermal wind velocity scales and the thickness of the Ekman layers. Meanwhile, eddy heat transfer increases with $\Omega$ following the onset of baroclinic instability but saturates at a maximum value, such that the overall combined heat transfer decreases in a similar way to the axisymmetric contribution as $\sim \Omega^{-3/2}$ \citep{Read2003}. A similar trend with $\Omega^\ast$ is apparent for our GCM simulations in Fig. \ref{fig:pumas-theta-slope}(a), with overall heat transport being held nearly constant until $\Omega^\ast \sim 1$, beyond which it decreases at a rate that appears to tend towards $\sim \Omega^{-3/2}$. }

\subsection{The regular wave regime}
The existence of a regular wave regime, dominated by a temporally coherent, spatially almost monochromatic travelling baroclinic wave, as confirmed in the present series of GCM simulations, is a remarkable and in some ways somewhat surprising occurrence, given our experience with the atmosphere of the Earth. But such a regime has an important analogue in the laboratory, where it has been observed and studied for many years \citep{hide1975}. The apparent absence of such a regime in early laboratory experiments with an open cylinder or ``dishpan'' \citep{Fultz1959} led to a suggestion among some theoreticians \citep[e.g.][]{davies1959} that a rigid inner cylinder was necessary to observe such regular waves. More recent work, however, showed that regular waves could be observed in open cylinder experiments \citep[e.g.][]{hidemason1970,spence1977} but required precise and stable experimental control of the external parameters. But some scepticism persisted for a while that such a regime would not be found on the scale of a planetary atmosphere until features such as the comparatively coherent baroclinic waves on Mars or Saturn's north polar hexagon wave were discovered \citep{Ryan1978,Barnes1981,godfrey1988}. 

Linear theories of baroclinic instability and eddy growth 
have been successful in delineating the initial development of unstable disturbances on 
the background axisymmetric flow and predicting the location of the transition boundary between axisymmetric and
non-axisymmetric flow regimes within the regime diagram (\cite{hide1975}). But for the regular baroclinic wave
regime, as the eddies/disturbances grow to amplitudes that are non-negligible compared to the background flow, 
linear theories begin to fail and nonlinear eddy-eddy and eddy-mean flow 
interactions need to be taken into account. As an analytical first attempt, weakly nonlinear theories have been
proposed since the 1960s \citep[e.g.][]{Pedlosky1970,Pedlosky1971,lovegrove2001,lovegrove2002} which assumes the flow to be just weakly 
supercritical to baroclinic instability so that only a small range of wavenumbers is unstable and grows 
relatively slowly compared to the wave phase propagation. Although these idealised models are able to predict the 
existence of vacillating or steady waves under appropriate conditions of viscous dissipation, there have been
frequent debates as to whether these weakly nonlinear models are applicable to the fully developed baroclinic 
instability (e.g. see \cite{Boville1981}, \cite{Esler2012}). The challenge of the nonlinearity hinders the 
development of a complete analytical understanding of the equilibration process of these regular 
finite-amplitude baroclinic waves. But insights can 
be obtained from numerical modelling studies, such as the present one based
on the fully nonlinear primitive equations.

The regular baroclinic wave regime found in our study also compare reasonably favourably to the planetary scale baroclinic 
waves that have been observed in the Martian atmosphere. Quasi-periodic surface pressure variations (especially
in the northern winter hemisphere) have been observed by in-situ measurement (Viking landers) which can be 
attributed to large scale baroclinic wave activity (\cite{Ryan1978}, \cite{Barnes1981}). A highly regular 
regime dominated by zonal wavenumber 3-4 could be inferred from these observations, indicating that they are
quite akin to what we have observed in our idealised GCM experiments. Compared with Earth's irregular and chaotic
atmosphere, therefore, the Martian atmosphere may be be classifiable as lying within a regular baroclinic wave regime. 
The relatively large value of $\mathcal{R}o_T$ compared with the Earth (due to smaller planetary size) and the much stronger radiative and frictional damping of the Martian atmosphere than on Earth \citep[e.g.][]{Nayvelt1997} are therefore likely to be the principal factors responsible for the occurrence of these regular baroclinic waves.

\ack 
PLR, FT-V and RMBY acknowledge support from the UK Science and Technology Research Council during the course of this research under grants ST/K502236/1, ST/K00106X/1 and ST/I001948/1. 
We are also grateful to an anonymous referee for his/her constructive comments on an earlier version of this paper.

\balance

\bibliographystyle{wileyqj}
\bibliography{atmosphere_EB,mars,refs}\label{refs}
\end{document}